\documentclass[ journal]{IEEEtran}

\usepackage{amsmath,amssymb,latexsym,bm}
\usepackage{paralist}
\usepackage{graphicx}
\usepackage{float}
\usepackage{algorithmic}
\usepackage{algorithm}
\usepackage{cases}
\usepackage{subfigure}
\usepackage{caption2}
\usepackage{epstopdf}
\usepackage{makecell,multirow,diagbox}


\usepackage{xcolor}
\usepackage{tikz}
\usetikzlibrary{spy}
\usepackage{wrapfig}
\usetikzlibrary{arrows}
\tikzstyle{block}=[draw opacity=0.7,line width=1.4cm]
\tikzstyle{process} = [rectangle, minimum width=3cm, minimum height=1cm, text centered, draw=black, fill = yellow!50]

\usepackage[colorlinks=true]{hyperref}
\hypersetup{urlcolor=blue, citecolor=red,linkcolor=blue,}
\begin{document}
\title{
	Statistical Image Reconstruction Using Mixed Poisson-Gaussian Noise
	Model for X-Ray CT
}
\author{Qiaoqiao Ding, Yong Long\IEEEauthorrefmark{1}, Xiaoqun Zhang\IEEEauthorrefmark{1} and Jeffrey A. Fessler%
	\IEEEcompsocitemizethanks{
		\IEEEcompsocthanksitem Xiaoqun Zhang and Qiaoqiao Ding are supported in part by Chinese 973 Program (Grant No. 2015CB856000) and National Youth Top-notch Talent program in China.
		Yong Long is supported in part by the SJTU-UM Collaborative Research Program, Shanghai Pujiang Talent Program (15PJ1403900) and NSFC (61501292).
		Jeffrey A. Fessler is supported in part by NIH grant U01 EB018753.
		\IEEEcompsocthanksitem Q. Ding and X. Zhang \IEEEauthorrefmark{1} (e-mail: xqzhang@sjtu.edu.cn) are with School of Mathematical Sciences and Institute of Natural Sciences, Shanghai
		Jiao Tong University. 800, Dongchuan Road, Shanghai, China, 200240
		\IEEEcompsocthanksitem Y.  Long\IEEEauthorrefmark{1} (e-mail: yong.long@sjtu.edu.cn)  is with the University of Michigan-Shanghai Jiao Tong University Joint Institute, Shanghai Jiao Tong University, 800 Dongchuan Road, Shanghai, China, 200240
		\IEEEcompsocthanksitem J. A. Fessler  is with the Department of Electrical Engineering and Computer Science, University of Michigan, Ann Arbor, MI 48109.
	}
}
\maketitle
\thispagestyle{empty}
	
\begin{abstract}
Statistical image reconstruction (SIR) methods for X-ray CT produce high-quality and accurate images, while greatly reducing patient exposure to radiation. When further reducing X-ray dose to an ultra-low level by lowering the tube current, photon starvation happens and electronic noise starts to dominate, which introduces negative or zero values into the raw measurements. These non-positive values pose challenges to post-log SIR methods that require taking the logarithm of the raw data, and causes artifacts in the reconstructed images if simple correction methods are used to process these non-positive raw measurements. The raw data at ultra-low dose deviates significantly from Poisson or shifted Poisson statistics for pre-log data and from Gaussian statistics for post-log data. This paper proposes a novel SIR method called MPG (mixed Poisson-Gaussian). MPG models the raw noisy measurements using a mixed Poisson-Gaussian distribution that accounts for both the quantum noise and electronic noise. MPG is able to directly use the negative and zero values in raw data without any pre-processing. MPG cost function contains a reweighted least square data-fit term, an edge preserving regularization term and a non-negativity constraint term. We use Alternating Direction Method of Multipliers (ADMM) to separate the MPG optimization problem into several sub-problems that are easier to solve. Our results on 3D simulated cone-beam data set and synthetic helical data set generated from clinical data indicate that the proposed MPG method reduces noise and decreases bias in the reconstructed images, comparing with the conventional filtered back projection (FBP),  penalized weighted least-square (PWLS) and shift Poisson (SP) method for ultra-low dose CT (ULDCT) imaging.
\end{abstract}
\section{Introduction}
X-ray Computed Tomography (CT) provides high-resolution images of anatomical structures for diagnosis and management of human diseases.
For example, CT has a tremendous impact on cancer diagnosis and treatment.
Studies have indicated that current CT usage may be responsible for 1.5\%-2\% of all cancers in the U.S. \cite{brenner2007computed}.
Significantly lowering radiation dosages from CT has become a growing concern both in the public and professional societies.
Ultra-low dose CT (ULDCT) scans that still provide suitable image quality could significantly alleviate potential damage caused by radiation and open new clinical applications using CT scans.

Developing CT image reconstruction methods that could reduce patient radiation exposure while maintaining high image quality is an important area of research.
Statistical image reconstruction (SIR) methods \cite{thibault2007three}  improve the ability to produce high-quality and accurate images, while greatly reducing patient exposure to radiation.
Further reducing dose to an ultra-low level could be achieved by reducing the number of projection views,  causing aliasing artifacts due to under-sampled sinograms when the number of views is too small \cite{long13aou}.
Lowering the X-ray tube current is an alternative approach, but it causes photon starvation and electronic noise starts to denominate \cite{whiting06pop}.
This approach of reducing dose introduces negative and zero values into the raw data and consequently causes artifacts and bias in the CT images reconstructed by methods \cite{nuyts13mtp} based on post-log sinograms obtained from pre-processing of raw data.

Most SIR methods assume standard or shifted Poisson (SP) distributions for pre-log data or assume Gaussian statistics for post-log data.
The CT measurements at ultra-low photon counts deviate significantly from Poisson or Gaussian statistics.
For ULDCT imaging, the logarithm cannot be directly taken on the raw measurements because of negative or zero values due to electronic noise in the data acquisition systems (DAS).
To take the logarithm of noisy measurements, simple methods, such as replacing the negatives with a small positive value or replacing them with their absolute values, corrupt the true
statistical nature of the raw data and introduce bias in reconstructed images \cite{thibault2006recursive}.
Wang et al. \cite{wang2006penalized} filtered noisy measurements using an adaptive trimmed mean filter (ATM) \cite{hsieh1998adaptive}, and then replaced non-positive values in filted measurements with a small positive value to enforce the logarithm transform that is applied on positive numbers.
The ATM filter dynamically adjusted its parameters to adapt to the local noise characteristics of the CT projection measurements \cite{hsieh1998adaptive}.
Thibault et al. \cite{thibault2006recursive} proposed a recursive filter which preserves the local mean while pre-processing noisy measurements.
Before applying the recursive filter, the method in \cite{thibault2006recursive} used a non-linear function to map any real valued noisy measurements to strictly positive values.
Poisson distribution models the number of events which should be non-negative.
The SP model \cite{la2006monotonic,la2006penalized} added a positive value associated with the variance of electronic noise to the raw CT data,
but the shifted data may still have negative or zero values for ULDCT imaging.
Compound Poisson (CP) distribution \cite{la2006monotonic,elbakri2003efficient} that takes the polyenergetic X-rays and Poisson light statistics in the scintillator of energy-integrating detector into consideration has the potential to accurately model the measurement statistics in ULDCT imaging.
However, the CP model has a complicated likelihood that hinders its direct use in SIR methods.
Furthermore, electronic readout noise leads to a distribution that is even more complicated than a CP model.

This paper proposes a new SIR method whose data-fit term considers the mixed Poisson-Gaussian (MPG) distribution model for CT measurements \cite{foi2008practical,li2015reweighted}.
The proposed MPG method is able to directly process negative or zero valued raw CT measurements that contain (some, albeit limited) information about the scanned object.
We apply Alternating Direction Method of Multipliers (ADMM, also known as split
Bregman method \cite{GoldsteinOsher2009}) to solve the MPG reconstruction problem.
We focus on $\ell_1$ regularization in this paper, but the ADMM optimization method can be applied to MPG with any regularization, such as the q-GGMRF regularization \cite{thibault2007three}.
We apply the proposed MPG method to ULDCT image reconstruction, and our experimental results show the MPG method reconstructs images with improved quality in terms of noise, artifacts and bias, comparing with the FBP, PWLS and SP method.

This paper is organized as follows. Section \ref{method} mathematically formulates the MPG method for X-ray CT reconstruction as a Penalized-Likelihood (PL) cost function and solves it using ADMM. Section \ref{Comparison} reviews the PWLS and SP method and compares the MPG method with them. Section \ref{result} presents experimental results. Finally, Section \ref{conclusion} concludes.
\section{MPG Model}
\label{method}
\subsection{Measurement Model}
\label{method,model}
Quantum noise and electronic noise are the two major noise sources in clinical X-ray CT scanners using current integrating detectors \cite{LinComparison,xu2009electronic}.
 Electronic noise can be modeled as a Gaussian random variable with mean $m$ and variance $\sigma^2$.
The offset mean $m$ of background signals such as dark current can be estimated using blank measurements prior to each scan and subtracted from the measured intensity \cite{xu2009electronic,HsiehCTBook},  so we assume $m=0$ hereafter.
For the case of normal clinical exposures, the X-ray CT  measurements $z_i$ are often modeled as the sum of a Poisson distribution representing photon-counting statistics and an independent Gaussian distribution representing additive electronic noise, \emph{i.e.},
\begin{eqnarray}
z_i=i+\eta_i
  \label{e,zi}
\end{eqnarray}
where
$y_i\sim \mathrm{Poisson}(\bar{y}_i(\bm{x}))$ and $\eta_i\sim \mathrm{N}(0,\sigma^2)$. $y_i$ denotes the number of X-ray photons incident on detector for the $i$th
ray where $i = 1,\cdots,N_d$, and $N_d$ is the number of rays. $\sigma$ denotes the standard deviation of electronic noise which has been converted to photon units \cite{LinComparison}.

For a monoenergetic source, we model the mean  of X-ray photons as \cite{Gualtieri99orderedsubsets}:
\begin{eqnarray}
  \bar{y}_i=\bar{y}_i(\bm{x})\triangleq I_i\exp(-[\bm{A}\bm{\bm{x}}]_i)
  \label{elemodel}
\end{eqnarray}
where $\bm{x}$ denotes the attenuation map, and its $j$th element $x_j$ is the average linear attenuation coefficient in the $j$th voxel for $j =1,\cdots,N_p$, where $N_p$ denotes the number of voxels. $\bm{A}$ is the $N_d \times N_p$ system matrix with entries $a_{ij}$, and $[\bm{A}\bm{x}]_i=\sum_{j=1}^{N_p}a_{ij}x_{j}$ denotes the line integral of the attenuation map $\bm{x}$ along the $i$th X-ray.
 We treat each $I_i$ as known nonnegative quantities,
where $I_i$ is the incident X-ray intensity incorporating X-ray source illumination and the detector efficiency. Although the measurement model in \eqref{elemodel} ignores beam-hardening effects \cite{joseph1978method,elbakri2002statistical}, polyenergetic measurement models that account for the source spectrum and energy-dependent attenuation will be employed in our future work.
\subsection{Penalized Weighted Least Square for Poisson-Gaussian Mixed Noise}
\label{method,noise}
We adopt the reweighted least square method \cite{li2015reweighted,green1984iteratively} to develop a tractable likelihood function for the mixed Poisson-Gaussian measurement model.
Assuming $y_i$ and $\eta_i$ are independent, we have
\begin{align}
\mathrm{E}[z_i] = \mathrm{E}[y_i]=\bar{y}_i
\label{Mean}
\end{align}
and
\begin{align}
\mathrm{Var}[z_i] =  \mathrm{Var}[y_i] + \mathrm{Var}[\eta_i]=  \bar{y}_i + \sigma^2.
\end{align}
The key to the proposed method
is that we approximate $z_i$  with a normal distribution, \emph{i.e.}, $z_i \sim N(\bar{y}_i,\bar{y}_i + \sigma^2)$. The Probability Density Function (PDF) of $z_i$ is
\begin{align}
  P(z_i;\bm{x})=\frac{1}{\sqrt{2\pi (\bar{y}_i(\bm{x})+\sigma^2)} }e^{-\frac{(z_i-\bar{y}_i(\bm{x}))^2}{2(\bar{y}_i(\bm{x})+\sigma^2)}}
\end{align}
In this paper $e^{(\cdot)},\log{(\cdot)},\sqrt{\cdot}$ and division are all point-wise operations. The corresponding approximate negative log-likelihood
for independent measurements $z_i$ has the form

\begin{align}
  \bar{L}(\bm{x})&=-\sum_{i=1}^{N_d}\log(P(z_i;\bm{x}))
 \nonumber\\
  &\equiv\frac{1}{2}\|\bm{z}-\bm{\bar{y}(\bm{x})}\|^2_{\bm{W}(\bm{x})} +\frac{1}{2}\langle\log{(\bm{\bar{y}(\bm{x})}+\bm{\sigma^2})},\bm{1}\rangle,
\end{align}
where $\equiv$ means  ``equal to within irrelevant constants independent of $\bm{x}$", the image-dependent diagonal weight matrix $\bm{W}(\bm{x})$ is
\begin{eqnarray}
  \bm{W}(\bm{x})=\mathrm{diag}\bigg{\{}\frac{1}{\bar{y}_i(\bm{x})+
  \sigma^2}\bigg{\}},
\end{eqnarray}
 $\bm{z}\in \mathbb{R}^{N_d} $ and $\bm{\bar{y}}(\bm{x})\in\mathbb{R}^{N_d} $ have elements of $z_i$ and $\bar{y}_i(\bm{x})$ respectively,
 $\bm{\sigma}^2\in\mathbb{R}^{N_d}$ and $\bm{1}\in\mathbb{R}^{N_d}$ have every element equal to $ \sigma^2$ and $1$ respectively, and $\langle\cdot,\cdot\rangle$ is inner product.

 We estimate the attenuation map $\bm{x}$  from the noisy measurements $\bm{z}$ by minimizing a Penalized-Likelihood (PL) cost function as follows:
\begin{align}
  \hat{\bm{x}}&=\arg\min_{\bm{x}} \Psi (\bm{x})\label{PL1}\\
  \Psi (\bm{x})&\triangleq \bar{L}(\bm{x})+R(\bm{x})+\chi_B(\bm{x}),\label{PL2}
\end{align}
where $\chi_B$ is the charactistic  function of the nonnegativity constraint set
 $B=\{\bm{x}: x_j\geq 0, \forall j\}$.
   \begin{eqnarray}
\chi_B(\bm{x})=\left\{\begin{array}{ccc}
 0  ,                 &\bm{x}\in B,\\
+\infty    ,                    &\bm{x}\notin B.
\end{array}\right.
\end{eqnarray}
The regularization term $R(\bm{x})$ is
\begin{eqnarray}
   R(\bm{x})=\lambda\sum_{r=1}^{N_r}\beta_r\psi([\bm{C}\bm{x}]_r)
 \label{e,r}
\end{eqnarray}
where the regularization parameter $\lambda$ controls the noise and resolution
tradeoff, $\beta_r$ is the spatial weighting in the $r$th direction \cite{fessler1996spatial},
$\psi(\cdot)$  is a potential function, $\bm{C}\in \mathbb{R}^{N_r\times N_p} $
is a  finite-differencing matrix
and $[\bm{C}\bm{x}]_r=\sum_{j=1}^{N_p}C_{rj}x_j$.
The proposed method MPG can work with any potential function, such as Huber function and generalized Gaussian \cite{bouman1993a}.
This paper focuses  on $l_1$ norm regularization, \emph{i.e.}, $\psi_r(t)=|t|$.
We incorporate $\beta_r$ into the finite-differencing matrix $\bm{C}$,
and rewrite the regularization term $R(\bm{x})$ as
\begin{align}
R(\bm{x})=\lambda \|\bm{C} \bm{x}\|_1.
\label{Regur}
\end{align}
\subsection{Optimization Method}
\label{method,Optimization}
We develop an optimization algorithm based on Alternating Direction Method of Multipliers (ADMM) to solve \eqref{PL1} which is difficult to optimize directly.

\subsubsection{Equivalent Reconstruction Problem}
\label{method,Equivalent}
Introducing auxiliary variables $\bm{u}\in \mathbb{R}^{N_d},\bm{v}\in \mathbb{R}^{N_r},\bm{w}\in \mathbb{R}^{N_p}$, we rewrite the MPG problem  \eqref{PL1} as the following equivalent constrained problem:
\begin{align}
\min_{\bm{x},\bm{u},\bm{v}}&\frac{1}{2}\|\frac{\bm{z}-Ie^{-\bm{u}}}{\sqrt{Ie^{-\bm{u}}+\bm{\sigma}^2}}\|^2_{2}+\frac{1}{2}\langle\log{(Ie^{-\bm{u}}+\bm{\sigma}^2)},\bm{1}\rangle\nonumber\\
&+\lambda\|\bm{v}\|_1+\chi_c(\bm{w})\nonumber\\
\mathrm{s.t.} &~~~\bm{u}=\bm{A}\bm{x},
                   \bm{v}=\bm{C}\bm{x},
                   \bm{w}=\bm{x}.
\label{constrain}
\end{align}
To simplify, we reformulate \eqref{constrain}  as the following constrained problem, where the constraints are written as a linear transform,
 \begin{eqnarray}
&\min_{\bm{x},\bm{u},\bm{v}, \bm{w}}\mathcal{D}(\bm{u})+\lambda\|\bm{v}\|_1+\chi_c(\bm{w}) \nonumber\\
&~~~~~~~~\mathrm{s.t.} ~~~\bm{P}\bm{x}=(\bm{A}\bm{x}, \bm{C}\bm{x},
\bm{x})^{T}=(\bm{u},\bm{v}, \bm{w})^{T}
\label{model}
\end{eqnarray}
where
\begin{align}
\mathcal{D}(\bm{u})=\frac{1}{2}\|\frac{\bm{z}-Ie^{-\bm{u}}}{\sqrt{Ie^{-\bm{u}}+\bm{\sigma}^2}}\|^2_{2}+\frac{1}{2}\langle\log{(Ie^{-\bm{u}}+\bm{\sigma}^2)},\bm{1}\rangle.
\end{align}
\subsubsection{Alternating Direction Method of Multipliers}
\label{method,SplitBregman}
We use ADMM to  solve the optimization problem in \eqref{model}. For  a penalty parameter $\mu_1,\mu_2,\mu_3 > 0$, the augmented Lagrange function of the optimization problem \eqref{model} is defined as:
\begin{align}
\mathcal{L}(\bm{x},\bm{u},&\bm{v},
\bm{w},\bm{b})=\mathcal{D}(\bm{u})+\lambda\|\bm{v}\|_1+\chi_c(\bm{w})\nonumber\\
+&\langle \bm{b},\bm{P}\bm{x} -(\bm{u},\bm{v}, \bm{w})^T\rangle
+\frac{\mu_1}{2}||\bm{A}\bm{x}-\bm{u}||_2^2\nonumber\\
+&\frac{\mu_2}{2}||\bm{C}\bm{x}-\bm{v}||_2^2+\frac{\mu_3}{2}||\bm{x}-\bm{w}||_2^2
\label{Lagrange}
\end{align}
where $\bm{b}=(\bm{b}_1,\bm{b}_2,\bm{b}_3)^T$, $\bm{b}_1\in\mathbb{R}^{N_d},\bm{b}_2\in \mathbb{R}^{N_r}, \bm{b}_3\in \mathbb{R}^{N_p}$
have the same size as $ \bm{A}\bm{x}, \bm{C}\bm{x}, \bm{x}$ respectively.
For ease of notations, we set
\begin{align}
\frac{1}{2}\|\bm{P}\bm{x}& -(\bm{u},\bm{v}, \bm{w})^T\|_{\bm{\mu}}^2\nonumber\\
:=&\frac{\mu_1}{2}||\bm{A}\bm{x}-\bm{u}||_2^2
+\frac{\mu_2}{2}||\bm{C}\bm{x}-\bm{v}||_2^2+\frac{\mu_3}{2}||\bm{x}-\bm{w}||_2^2.
\end{align}
 Given $\bm{x}^{(0)},\bm{s}^{(0)}$ and $\bm{b}^{(0)}$, ADMM updates the sequence
$(\bm{x}^{(j)},\bm{u}^{(j)},\bm{v}^{(j)}, \bm{w}^{(j)},\bm{b}^{(j)})$ using
\begin{subequations}
 \begin{numcases}
{}\bm{x}^{(j+1)}=\langle \bm{b}^{(j)},\bm{P}\bm{x} -(\bm{u}^{(j)},\bm{v}^{(j)}, \bm{w}^{(j)})^T\rangle\nonumber\\
~~~~~~~~~~~~~+\frac{1}{2}\|\bm{P}\bm{x} -(\bm{u}^{(j)},\bm{v}^{(j)}, \bm{w}^{(j)})^T\|_{\bm{\mu}}^2,\label{equationa}\\
(\bm{u}^{(j+1)},\bm{v}^{(j+1)},\bm{w}^{(j+1)})\nonumber\\
~~~~~~~=\arg\min_{\bm{u},\bm{v},\bm{w}}\mathcal{L}(\bm{x}^{(j+1)},\bm{u},\bm{v},\bm{w},\bm{b}^{(j)}),\label{equationb}\\
 \bm{b}^{(j+1)} =\bm{b}^{(j)} + \bm{\mu}(\bm{P}\bm{x}^{(j+1)}-(\bm{u}^{(j+1)},\bm{v}^{(j+1)},\bm{w}^{(j+1)})^T).\nonumber\\\label{equationc}
 \end{numcases}
 \label{equationall}
 \end{subequations}
\subsubsection{Algorithms for subproblems}
\label{method,Algorithm}

Firstly, we solve \eqref{equationa} to obtain image update $\bm{x}^{(j+1)}$.
Since \eqref{equationa} is quadratic and differentiable on $\bm{x}$, its solution has an analytical expression:
\begin{align}
\bm{x}^{(j+1)*}&=\bm{G}^{-1}\big{[}\mu_1\bm{A}^{T}(\bm{u}^{(j)}-\bm{b}_1^{(j)})
+\mu_2\bm{C}^{T}(\bm{v}^{(j)}-\bm{b}_2^{(j)})\nonumber\\
&+\mu_3(\bm{w}^{(j)}-\bm{b}_3^{(j)})\big{]}
\label{solution1}
\end{align}
where $\bm{x}^{(j+1)*}$ represents the exact solution  and $\bm{G}=\mu_1\bm{A}^{\top}\bm{A}+\mu_2\bm{C}^{\top}\bm{C}+\mu_3 \bm{I}$ is nonsingular when $\mu_3 > 0$ because $\bm{I}$ is positive definite and $\bm{A}^{\top}\bm{A}$  and $\bm{C}^{\top}\bm{C}$ are  semidefinite positive.
Although \eqref{solution1} is an exact analytical solution, it is impractical to store and invert $\bm{G}$ exactly due to its huge size for CT reconstruction. We use the conjugate gradient (CG) method \cite{burden2011numerical} to obtain an approximate update $\bm{x}^{(j+1)}$.

Secondly, we solve \eqref{equationb}   separately for $\bm{u},\bm{v},\bm{w}$ and in parallel as follows
\begin{align}
\bm{u}^{(j+1)}=&\arg\min_{\bm{u}}\mathcal{D}(\bm{u})+\langle \bm{b}_1^{(j)},A\bm{x}^{(j+1)}-\bm{u}\rangle\nonumber\\
&+\frac{\mu_1}{2}\|A\bm{x}^{(j+1)}-\bm{u}\|_2^2 ,\label{sub1}\\
\bm{v}^{(j+1)}=&\arg\min_{\bm{v}}\lambda\|\bm{v}\|_1+\langle \bm{b}^{(j)}_2, \bm{C}\bm{x}^{(j+1)}-\bm{v}\rangle\nonumber\\
&+\frac{\mu_2}{2}\|\bm{C} \bm{x}^{(j+1)}-\bm{v}\|_2^2 ,\label{sub2}\\
\bm{w}^{(j+1)}=&\arg\min_w \chi_c(\bm{w})+\langle \bm{b}^{(j)}_3, \bm{x}^{(j+1)}-\bm{w}\rangle\nonumber\\
&+ \frac{\mu_3}{2}\|\bm{x}^{(j+1)}-\bm{w}\|_2^2.\label{sub3}
\end{align}
Subproblem \eqref{sub1} is a smooth, differentiable, nonconvex and separable problem. Many methods, such as Newton's method \cite{WrightNO}, can be used  to
solve the subproblem \eqref{sub1}. Minimization with respect to $\bm{v}$ in \eqref{sub2} is the proximal operator of the $\ell_1$ norm.
We update each $v_j$ separately using soft-thresholding, \emph{i.e.},
 \begin{align}
\bm{v}^{(j+1)}=\mathcal{S}_{\frac{\lambda}{\mu_2}}\left(\bm{C}\bm{x}^{(j+1)}+\frac{\bm{b}^{(j)}_2}{\mu_2} \right),\label{Subsolb}
\end{align}
where $\mathcal{S}$ denotes the soft-thresholding operator. Subproblem \eqref{sub3} is the projection on the set $B$.
Let $\mathcal{P}$ denote the projection operation, and then we can obtain,
 \begin{align}
 \bm{w}^{(j+1)}=\mathcal{P}_B\left( \bm{x}^{(j+1)}+\frac{\bm{b}^{(j)}_3}{\mu_3} \right)=\mathrm{max}\left(\bm{x}^{(j+1)}+\frac{\bm{b}^{(j)}_3}{\mu_3},0\right)
.\label{Subsolc}
\end{align}

Thirdly, the dual variable $\bm{b}$ is updated straightforwardly as given in \eqref{equationc}. 
 We can numberically check the primal and dual residual for the ADMM updates \eqref{equationall} 
  as the stopping criteria  \cite{goldstein2014fast,boyd2004convex}
\begin{eqnarray}
r^{(j)}=
\left(
\begin{array}{ccc}\bm{u}^{(j)}- \bm{Ax}^{(j)}\\
\bm{v}^{(j)}-\bm{C}\bm{x}^{(j)} \\
\bm{w}^{(j)}-\bm{x}^{(j)}
\end{array}
\right),
\label{PrimalRes}
\end{eqnarray}
\begin{eqnarray}d^{(j)}=
\left(\begin{array}{ccc}
\mu_1\bm{A}^{T}(\bm{u}^{(j)}-\bm{u}^{(j-1)})\\
\mu_2\bm{C}^{T}(\bm{v}^{(j)}-\bm{v}^{(j-1)})\\
\mu_3(\bm{w}^{(j)}-\bm{w}^{(j-1)})
\end{array}\right).
\label{DualRes}
\end{eqnarray}

Algorithm  \ref{algMPG} summarizes the optimization algorithm of the proposed MPG method.
	\begin{algorithm}[H]
	\caption{MPG Algorithm}
	\label{algMPG}
	\begin{algorithmic}
		\STATE \footnotesize{ \textbf{Input.} $\bm{x}^{(0)}$, $\lambda$, $\mu_1$, $\mu_2$, $\mu_3$. }
		\STATE Initial $\bm{u}^{(0)}=\bm{A}\bm{x}^{(0)}$,$\bm{v}^{(0)}=\bm{C}\bm{x}^{(0)}$, $\bm{w}^{(0)}=\bm{x}^{(0)}$, $\bm{b}^{(0)}=(\bm{b}_1^{(0)},\bm{b}_2^{(0)}, \bm{b}_3^{(0)})=0$, $\rm{Maxiter}$, $\rm{tol}$,  $j=1$.
		\WHILE{$\|r^{(j)}\|>\rm{tol}$, $\|d^{(j)}\|>\rm{tol}$, $j<\rm{Maxiter}$}
		\STATE Solve for $\bm{x}^{(j+1)}$
		by applying CG iterations to \eqref{solution1}.
		\STATE Solve for $\bm{u}^{(j+1)}$ by \eqref{sub1}.
		\STATE Solve for $\bm{v}^{(j+1)}$ using \eqref{Subsolb}.	
		\STATE Solve for $\bm{w}^{(j+1)}$ using \eqref{Subsolc}.
		\STATE $\bm{b}_1^{(j+1)}=\bm{b}_1^{(j)}+\mu_1(\bm{A}\bm{x}^{(j+1)}-\bm{u}^{(j+1)})$.
		\STATE $\bm{b}_2^{(j+1)}=\bm{b}_2^{(j)}+\mu_2(\bm{C}\bm{x}^{(j+1)}-\bm{v}^{(j+1)})$.
		\STATE $\bm{b}_3^{(j+1)}=\bm{b}_3^{(j)}+\mu_3(\bm{x}^{(j+1)}-\bm{w}^{(j+1)})$.
		\STATE $j=j+1$.
		\ENDWHILE	
	\end{algorithmic}
\end{algorithm}
\section{MPG compared with SP and PWLS}
\label{Comparison}
The penalized weighted least-squares (PWLS) reconstruction method \cite{thibault2007three,sauer1993a,fessler2000statistical} is a widely used
post-log reconstruction method for CT. The shifted poisson (SP) method is a commonly used pre-log reconstruction method. This section briefly reviews the PWLS and SP method, and compares the proposed MPG method with them.
\subsection{The SP Method}
\label{Sec:SP}
The SP method shifts noisy CT measurement $z_i$ by the variance of electronic noise $\sigma^2$, and models the shifted measurement $\widetilde{z}_i=z_i+\sigma^2$ using a Poisson distribution, \emph{i.e.},
\begin{align}
\widetilde{z}_i\sim{\rm{Poisson}}(\bar{y}_i(\bm{x})+\sigma^2).
\label{SP}
\end{align}
The shifted measurement $\widetilde{z}_i$ has equal mean and variance of $\bar{y}_i(\bm{x})+\sigma^2$. The PDF of $\widetilde{z}_i$ is
\begin{align}
P( \widetilde{z}_i:\bm{x})
=\frac{(\bar{y}_i(\bm{x})+\sigma^2)^{(z_i+\sigma^2 )}}{(z_i +\sigma^2)!}
e^{- (\bar{y}_i(\bm{x})+\sigma^2)}.
\end{align}
The corresponding negative log-likelihood for independent measurements $z_i$ is
\begin{align}
\bar{L}_{SP}(\bm{x})&\equiv \langle\bar{y}(\bm{x})+\bm{\sigma}^2,\bm{1}\rangle-
\langle \bm{z} +\bm{\sigma}^2, \log(\bar{y}(\bm{x})+\bm{\sigma}^2\rangle.
\end{align}
With the same regularization \eqref{Regur} and the characteristic function for non-negativity constraint used in the MPG model \eqref{PL2}, the SP reconstruction problem can be written as
\begin{align}
\hat{\bm{x}}_{SP} = \arg\min_{\bm{x}} & \bar{L}_{SP}(\bm{x})+ \lambda\|\bm{C}
\bm{x}\|_1+\chi_B(\bm{x}).
\label{SPCost}
\end{align}
We apply ADMM to solve the SP reconstruction problem \eqref{SPCost}, as described in Supplementary Material \ref{APP:SP}. 
\subsection{PWLS Reconstruction}
\label{Sec:PWLS}
PWLS is a post-log reconstruction method that requires one to take the logarithm of the noisy measurements $z_i$.
To obtain line integrals $\hat{p}_i$, a small positive value $\varepsilon$ is typically used to  replace non-positive and zero measurement elements \cite{LinComparison}, \emph{i.e.},
\begin{align}
\hat{p}_i=\log\left( \frac{I_i}{\max(z_i,\varepsilon)}\right).
\label{set1}
\end{align}
The statistical weight $w_i$ in PWLS that considers electronic noise is \cite{thibault2006recursive,LinComparison},
\begin{align}
 w_i=\frac{\max(z_i,\varepsilon)^2}{\max(z_i,\varepsilon)+\sigma^2}.
 \label{set2}
 \end{align}
 With the same regularization \eqref{Regur} and the characteristic function for nonnegativity constraint used in the MPG model \eqref{PL2}, the PWLS reconstruction problem can be written as follows,
\begin{align}
\hat{\bm{x}}_{PWLS} =
\arg\min_{\bm{x}}\frac{1}{2}\|\bm{A}\bm{x}-\hat{\bm{p}}\|^2_{\bm{W}}+\lambda\|\bm{C}\bm{x}\|_1+\chi_B(\bm{x})
\label{PWLSCost}
\end{align}
where $\bm{W}=\mathrm{diag}(w_i)$ and $\hat{\bm{p}}\in \mathbb{R}^{N_d}$ has elements of $\hat{p}_i$.
We apply the ADMM algorithm proposed in \cite{ramani2012a} to solve the PWLS reconstruction problem \eqref{PWLSCost}.
\subsection{Comparison}
For PWLS reconstruction, the logarithm simply cannot be directly taken on noisy measurements for low dose CT imaging because the measurements may have negative and zero values.
To take the  logarithm, it is necessary to correct the non-positive values in measurements.
Due to correction of non-positive values and nonlinearity of logarithm, estimating statistical weights for post-log sinogram is a challenging problem.
Both correction of non-positive measurements and unmatched weights can introduce bias in the reconstructed images.

The SP model \eqref{SP} requires the shifted measurements $\widetilde{z}_i$ to be nonnegative, which may not be satisfied for ULDCT imaging \cite{nuyts13mtp}.
The SP model \eqref{SP} uses a Poisson distribution with mean and variance of $ \bar{y}_i(\bm{x})+\sigma^2$ to model the shifted measurements $z_i+\sigma^2$, \emph{i.e.},
\begin{align}
\mathrm{Poisson}(\bar{y}_i(\bm{x})+\sigma^2)\sim\mathrm{Poisson}(\bar{y}_i(\bm{x}))+N(0,\sigma^2)+\sigma^2.
\label{PoiGau}
\end{align}
For the two independence Poisson distributions $\mathrm{Poisson}(\bar{y}_i(\bm{x}))$ and $\mathrm{Poisson}(\sigma^2)$, the sum of them is a Poisson distribution, \emph{i.e.},
\begin{align}
\mathrm{Poisson}(\bar{y}_i(\bm{x}))+\mathrm{Poisson}(\sigma^2)=\mathrm{Poisson}(\bar{y}_i(\bm{x})+\sigma^2).
\end{align}
The SP model is equivalent to using a Poisson distribution $\mathrm{Poisson}(\sigma^2)$ to model the shifted electronic noise $N(0,\sigma^2)+\sigma^2$ in \eqref{PoiGau} that is a Gaussian distribution.
Comparing with the original Poisson + Gaussian distribution \eqref{e,zi} that has a mean of $y_i(\bm{x})$ calculated in \eqref{Mean}, the SP model has a larger signal mean $y_i(\bm{x}) + \sigma^2$ which increases with the increase of electronic noise variance $\sigma^2$.
For low dose CT imaging where photon starvation happens and electronic noise dominates, the SP model needs to correct negative values in shifted measurements, which introduces bias in the reconstructed images.
The proposed MPG model has the same signal mean as the original Poisson + Gaussian distribution.
The MPG method directly reconstruct images from noisy measurements even if there are non-positive values, without introducing bias through correcting measurements.
\section{Results}
\label{result}
We evaluate the proposed method,  MPG,  using XCAT phantom \cite{segars2008realistic} and synthetic sinogram data
from a clinical CT scan, and compare its performance with those of the FBP, PWLS and SP method.
Both SP and MPG reconstruct images from uncorrected pre-log data and require knowledge of electronic noise variance on a CT scanner.
This kind of pre-log data and electronic noise variance value are proprietary to CT venders, especially for ULDCT imaging.
We generated pre-log measurements using a CT volume reconstructed from clinical data at regular dose, and added electronic noise at different levels to produce synthetic ULDCT sinogram data.
Some elements of ULDCT measurements $\bm{z}$ were non-positive.
The proposed MPG method can directly use these measurements in reconstruction without any pre-processing.
We generated sinogram and weight used by FBP and PWLS according to equation \eqref{set1}, \eqref{set2} in section \ref{Sec:PWLS}.
For the SP method, we replaced negative shifted measurements $\widetilde{z}_i<0$ with $\widetilde{z}_i=0$.
We used FBP reconstructions to initialize PWLS reconstructions, and initialized the
SP method and the proposed MPG method with PWLS reconstructions.

\subsection{Evaluation }
To compare various methods quantitatively for the XCAT phantom experiments, we calculated the Root Mean Square Error (RMSE) and Signal Noise Ratio (SNR)
of reconstructions in a region of interest (ROI).
RMSE in (modified) Hounsfield units (HU) , where air is $0$ HU, is defined as
\begin{align}
\mathrm{RMSE}=\sqrt{\frac{\sum_{j=1}^{N}(\hat{x}_j-x_j)^2}{N}}
\end{align}
where $x_j$ and $\hat{x}_j$ denotes the $j$-th voxel of the true image and reconstructed image respectively, and $N$ is the number of voxels in the ROI. SNR is defined as
\begin{align}
\mathrm{SNR}=10\log_{10} \frac{\sum_{j=1}^{N}(\hat{x}_j-x_j)^2}{\sum_{j=1}^{N} (x_j-\bar{x})^2}
\end{align}
where  $\bar{x}$ is the mean of  $N$ voxels of the groundtruth in the ROI,  \emph{i.e.},
$\bar{x}=\frac{\sum_{i=1}^{N} x_i}{N}$.

\subsection{XCAT Phantom Results}
\begin{figure}[ht]
\begin{center}
\subfigure[XCAT Phantom]{\label{TrueA}
\includegraphics[width=.45\linewidth, height=.45\linewidth]{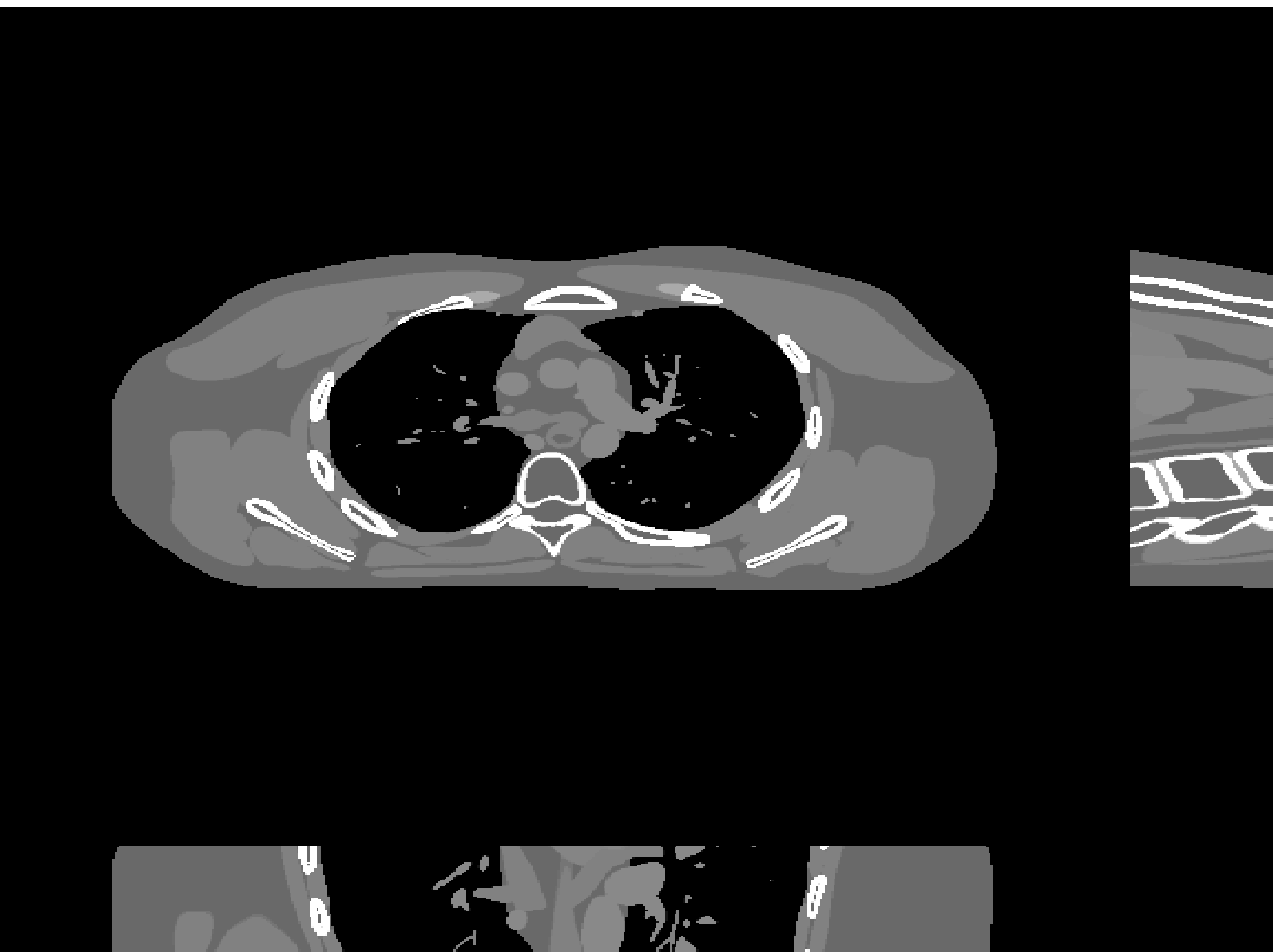}}
\subfigure[Clinical Phantom]{\label{TrueB}
\includegraphics[width=.45\linewidth, height=.45\linewidth]{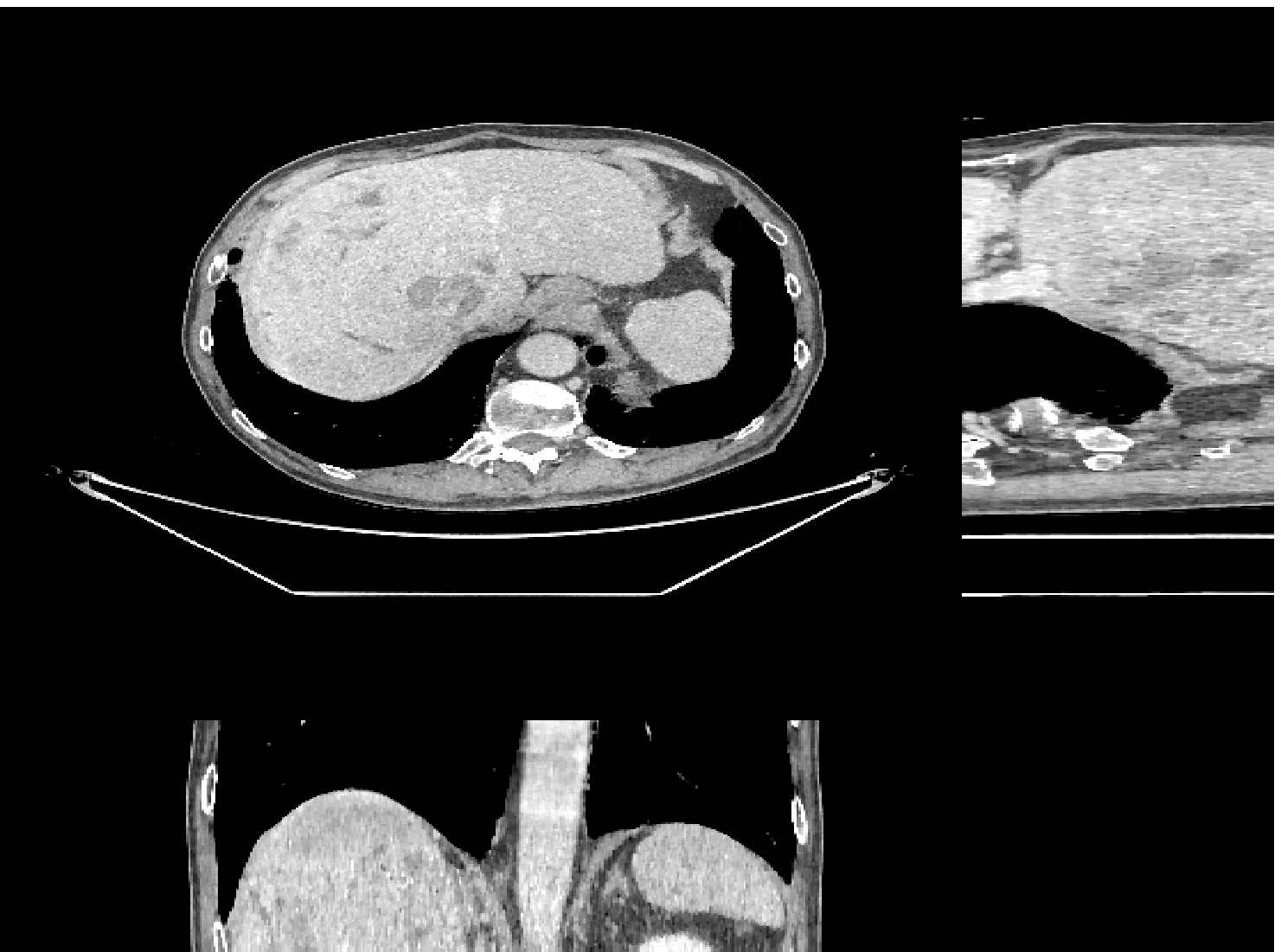}}
\end{center}
\caption{True images of the XCAT phantom and clinical data. The top left, the bottom and the right of each image are the axial plane, coronal plane and sagittal plane respectively.
The images are displayed with a window of [800,1200] HU.}
\label{fig:True}
\end{figure}

We simulated an axial cone-beam CT scan using a $1024\times1024\times100$ XCAT phantom with $\Delta_x=\Delta_y=0.4883$ mm and $\Delta_z=0.625$ mm.
We generated a $888\times64\times984$ noisy sinogram with GE LightSpeed cone-beam geometry corresponding to a monoenergetic source with $I_i= 10^4$ and $I_i=5\times10^3$ incident photons per ray.
For $I_i=10^4$ incident photons per ray, we set the standard deviation of electronic noise $\sigma$ to be $\{20,30,40,50,60,70,100\}$ photons per projection ray \cite{xu2009electronic,rui2015ultra-low}.
For the lower dose case, $I_i=5\times10^3$ incident photons per ray, we set the standard deviation of electronic noise $\sigma$ to be $\{50,60,70,100\}$ photons per projection ray.
We reconstructed $512\times 512 \times100$ volumes with a coarser grid, where $\Delta_x=\Delta_y=0.9766$ mm and $\Delta_z=0.625$ mm.
A ROI for 3D reconstruction consisted of the central $64$ of $100$ axial slices and circular regions in each slice.
Figure \ref{TrueA} shows central slices of the true XCAT phantom along three directions.

\begin{figure*}
\begin{center}
\begin{tabular}{c@{\hspace{2pt}}c@{\hspace{2pt}}c@{\hspace{2pt}}c@{\hspace{2pt}}c@{\hspace{2pt}}c@{\hspace{2pt}}c@{\hspace{2pt}}c@{\hspace{2pt}}c@{\hspace{2pt}}c}
\put(-40,50){$\sigma^2=50^2$}&
\includegraphics[width=.22\linewidth, height=.22\linewidth]{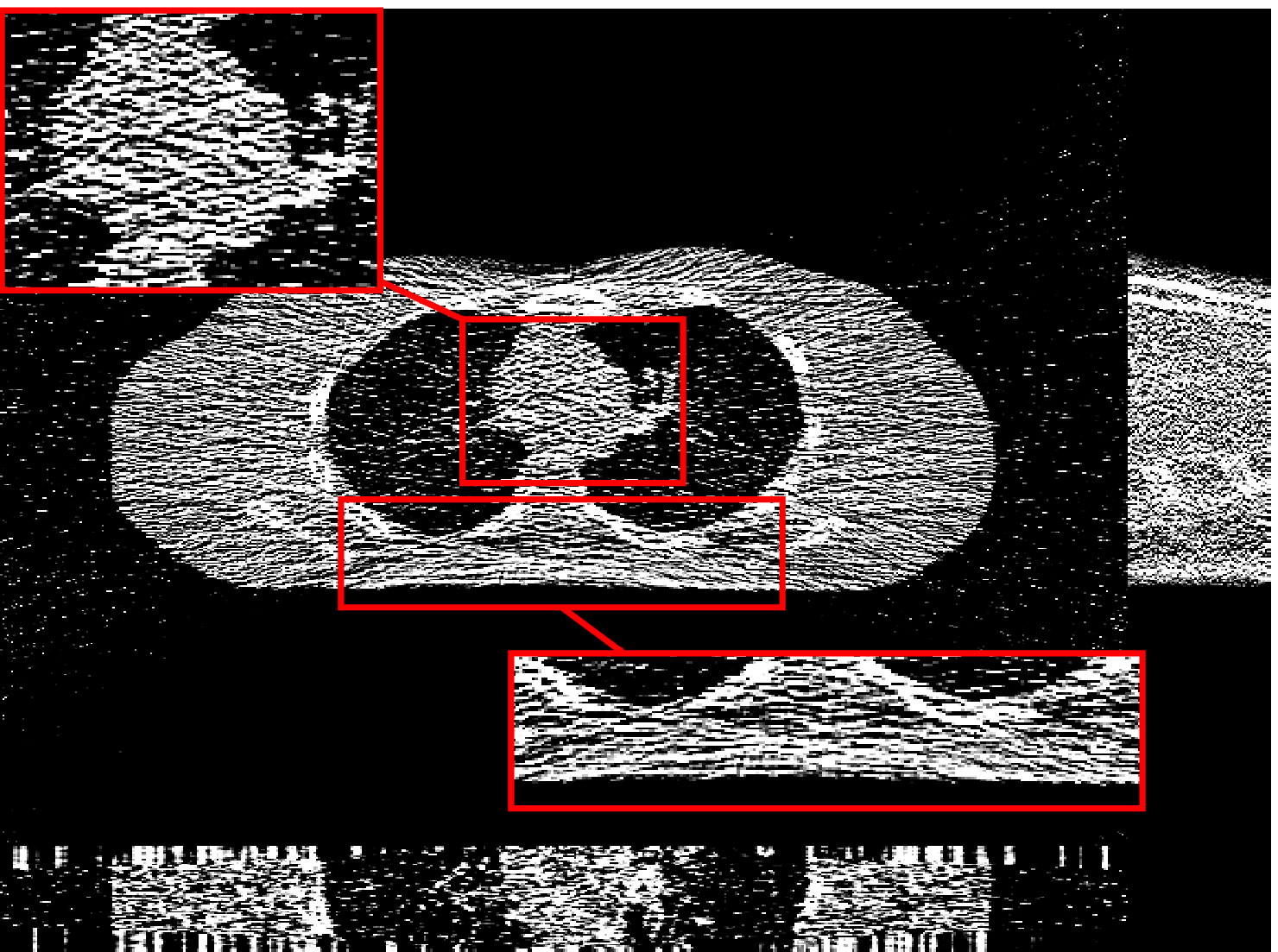}&
\includegraphics[width=.22\linewidth, height=.22\linewidth]{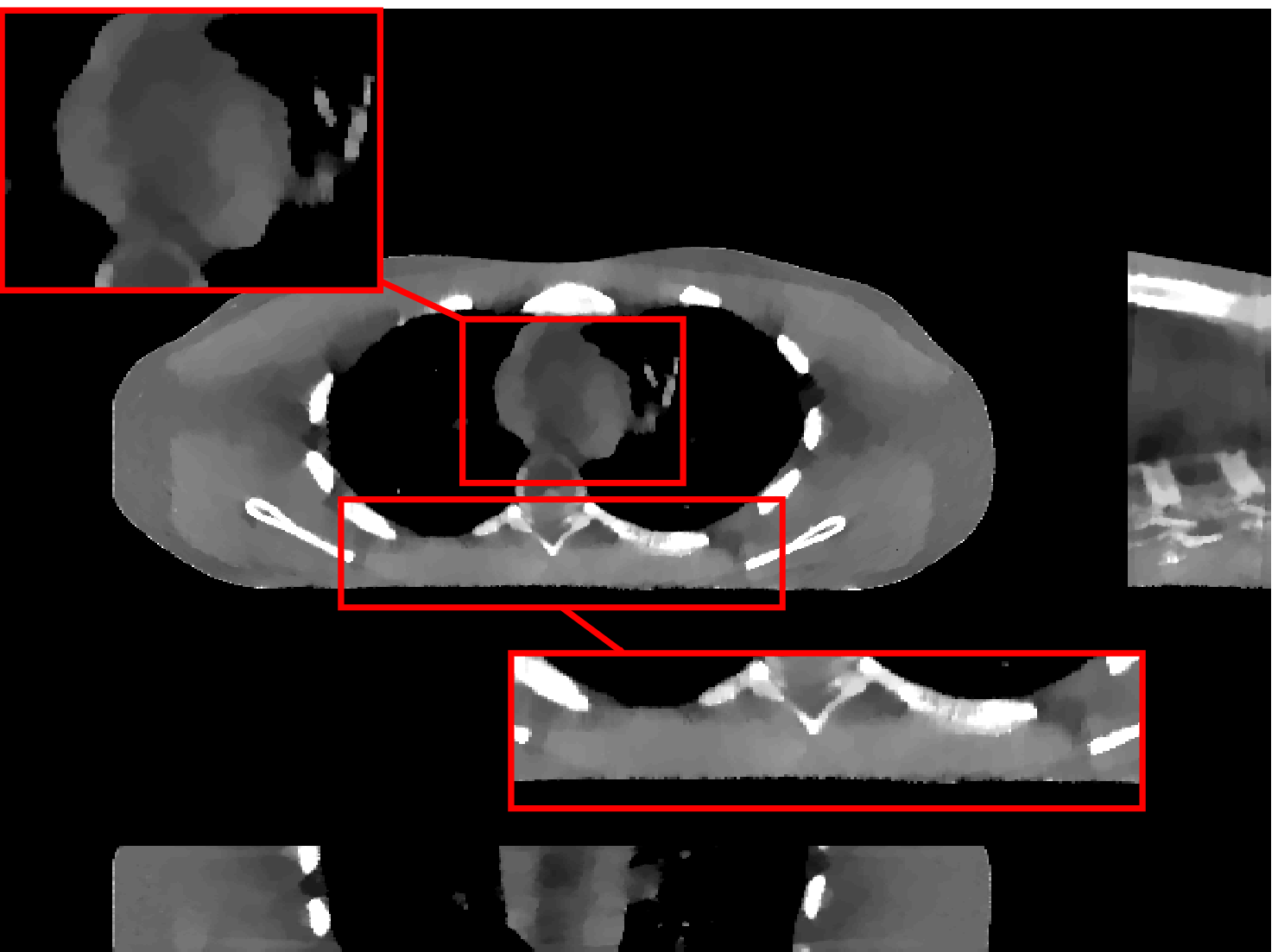}&
\includegraphics[width=.22\linewidth, height=.22\linewidth]{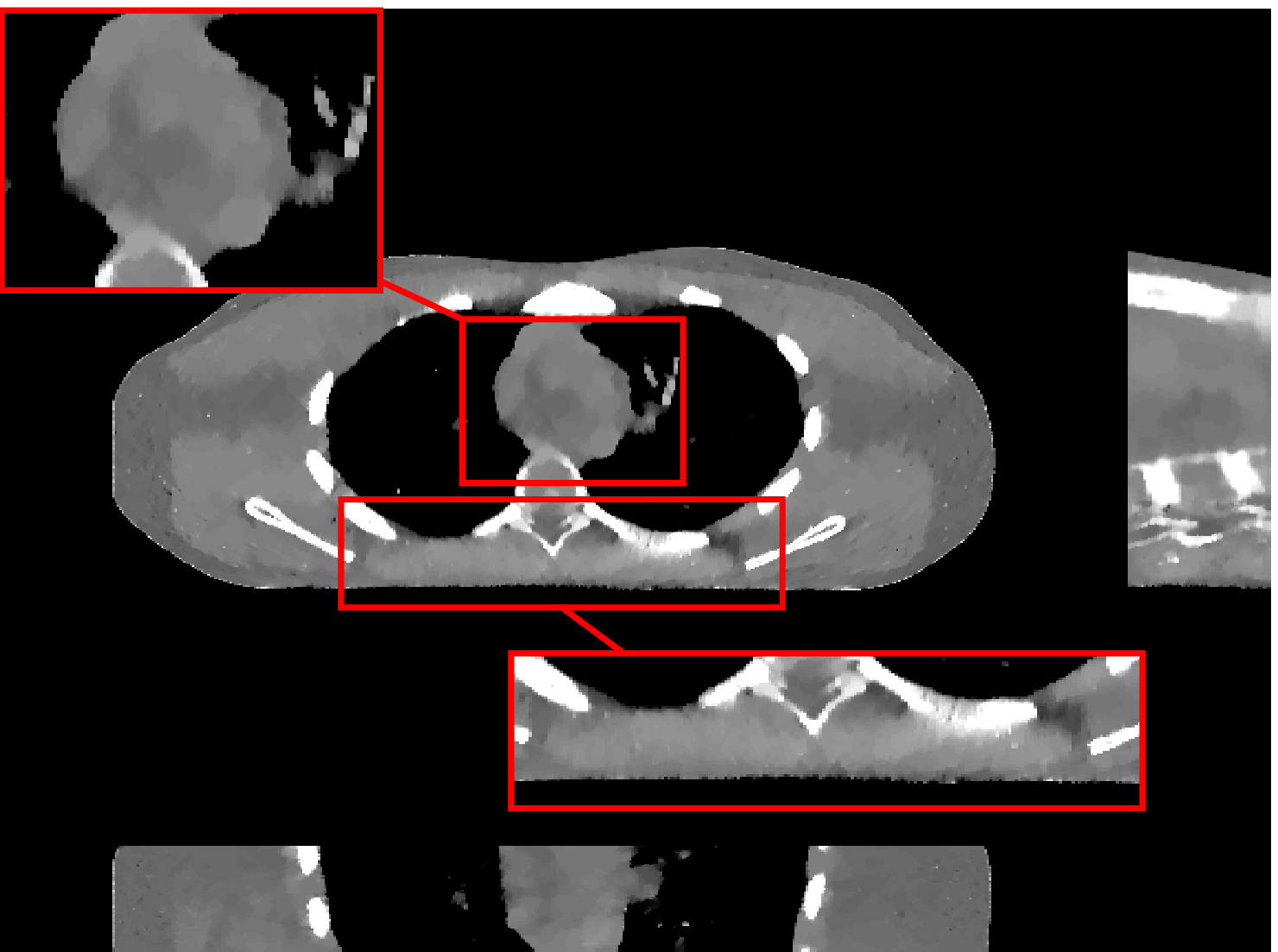}&
\includegraphics[width=.22\linewidth, height=.22\linewidth]{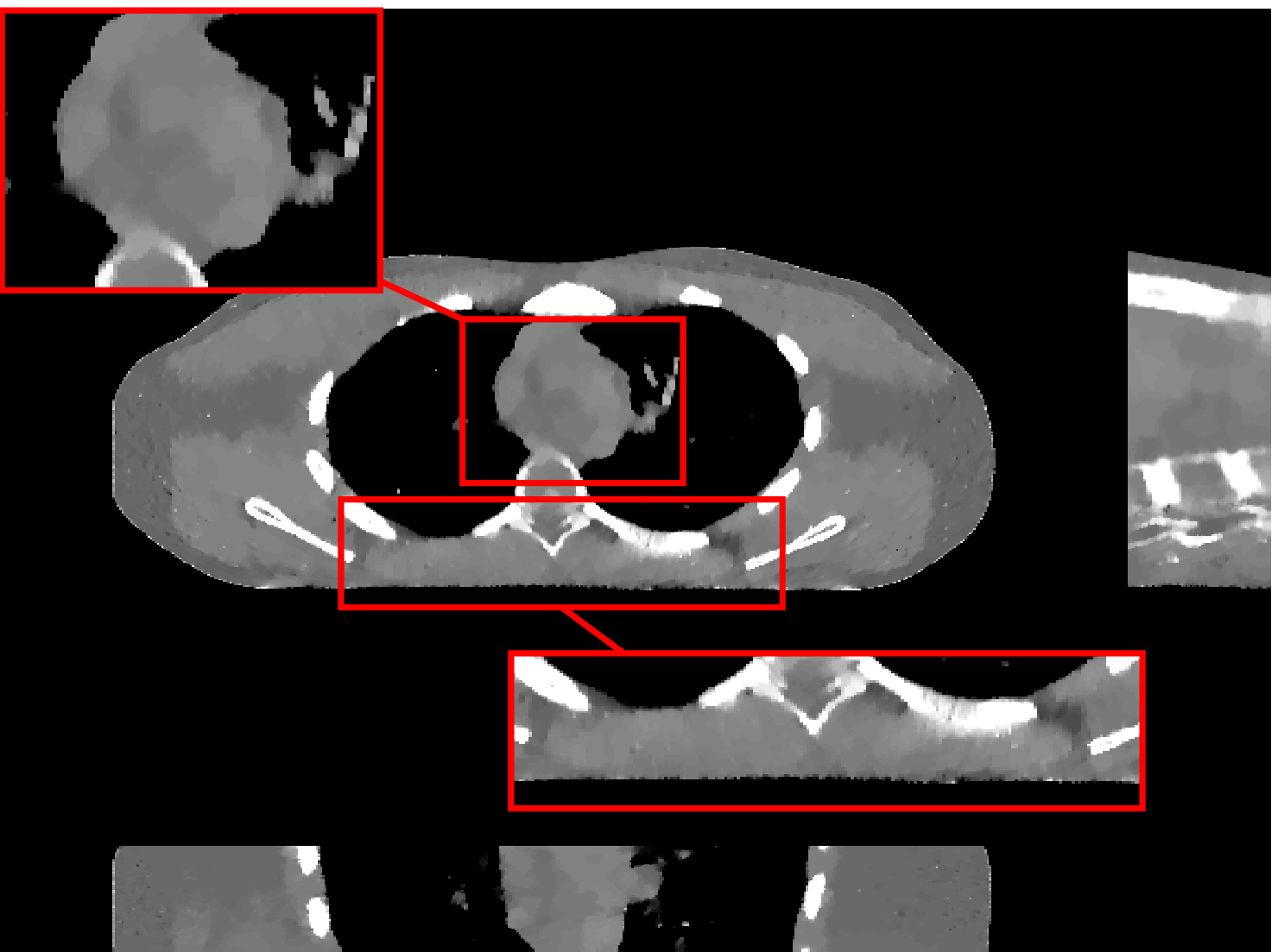}\\
\put(-40,50){$\sigma^2=60^2$}&
\includegraphics[width=.22\linewidth, height=.22\linewidth]{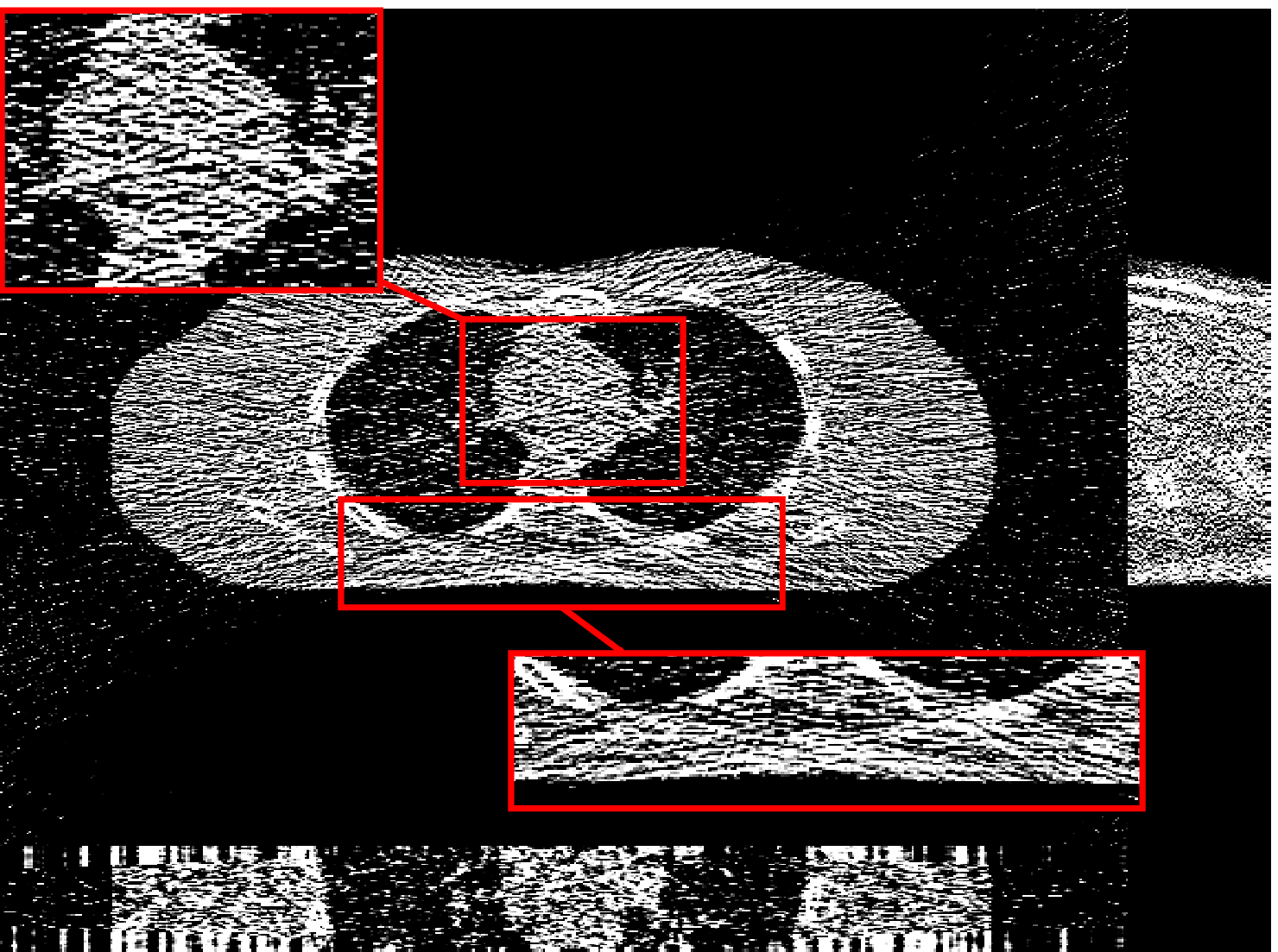}&
\includegraphics[width=.22\linewidth, height=.22\linewidth]{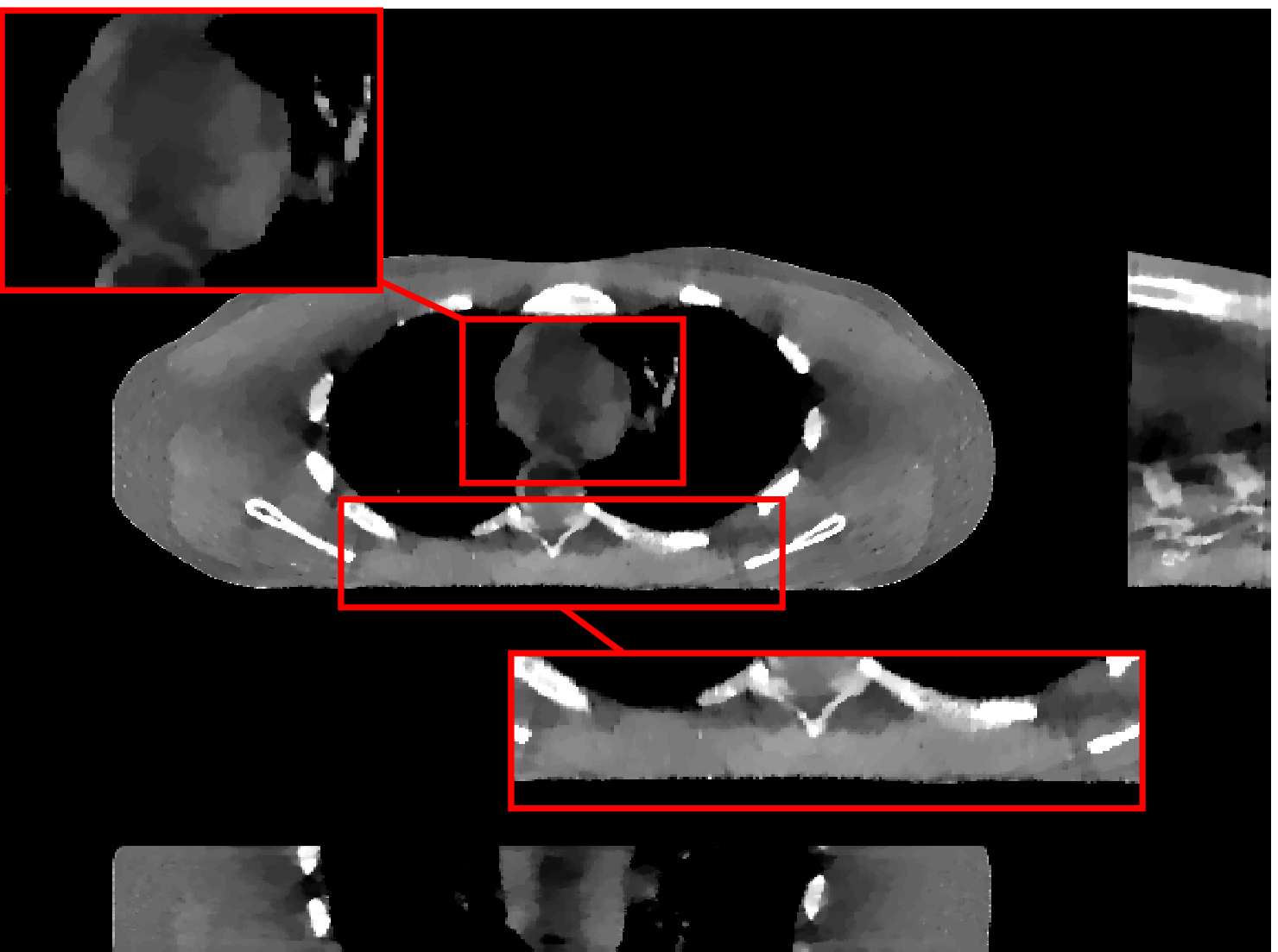}&
\includegraphics[width=.22\linewidth, height=.22\linewidth]{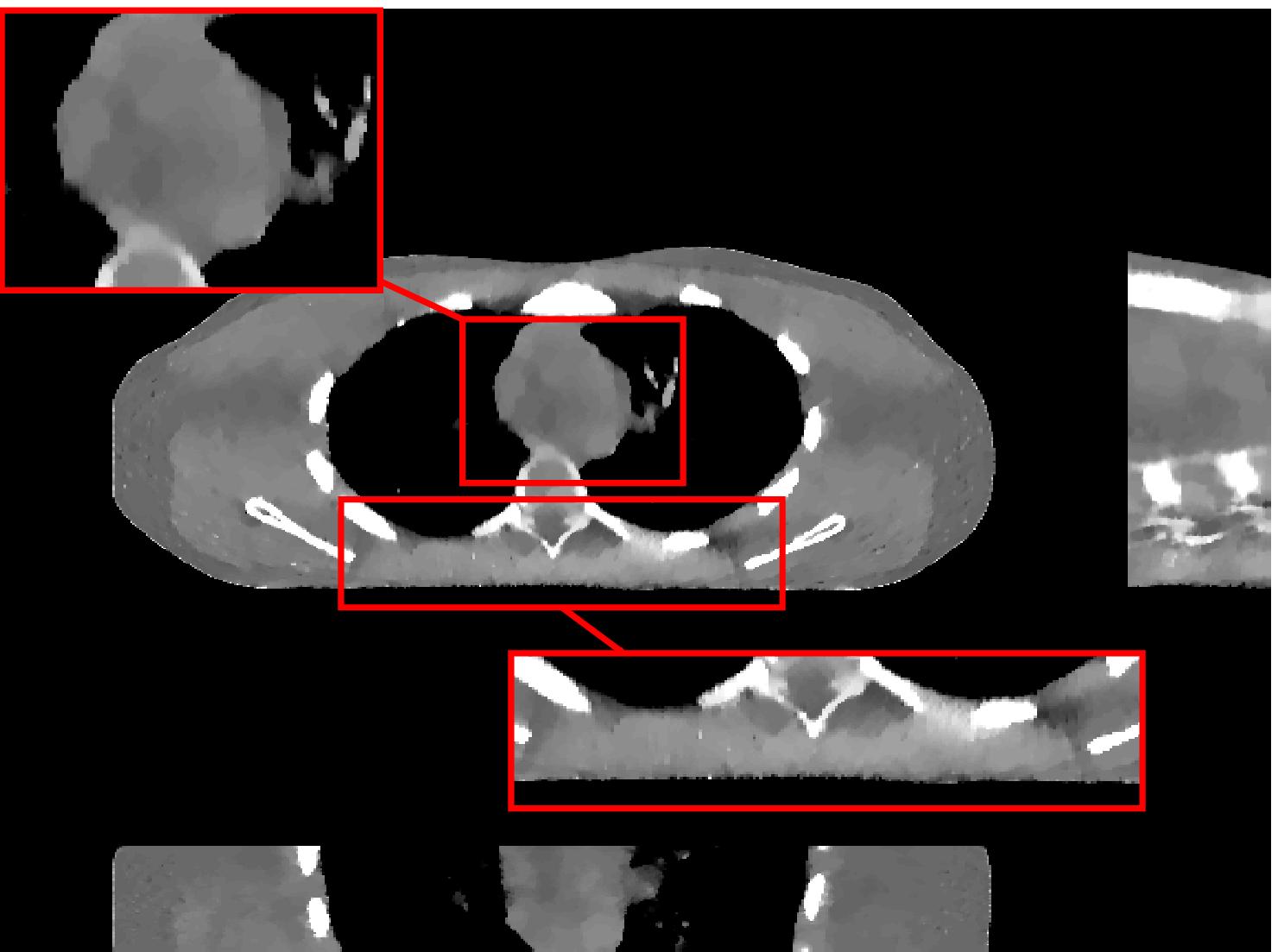}&
\includegraphics[width=.22\linewidth, height=.22\linewidth]{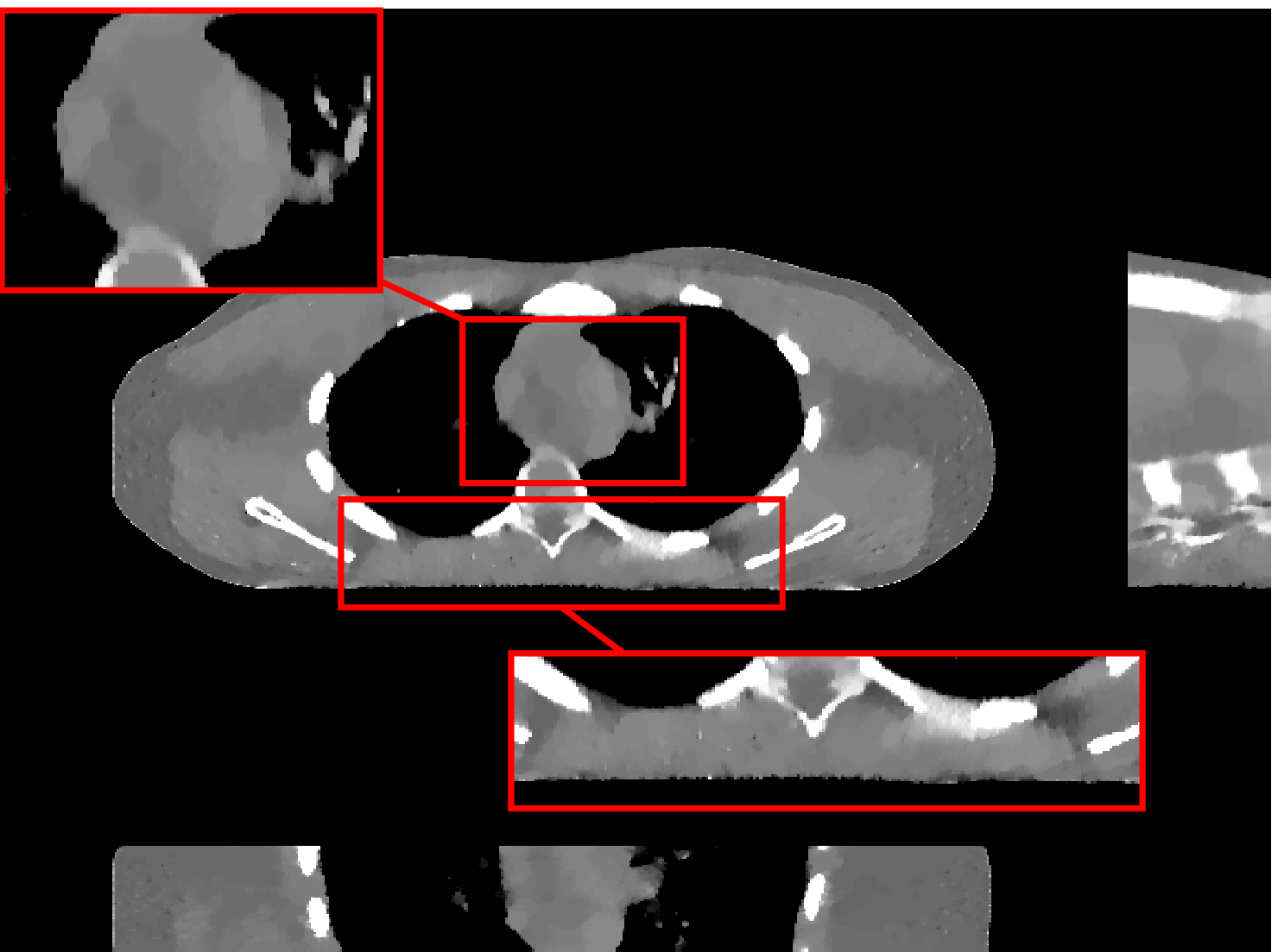}\\
\put(-40,50){$\sigma^2=70^2$}&
\includegraphics[width=.22\linewidth, height=.22\linewidth]{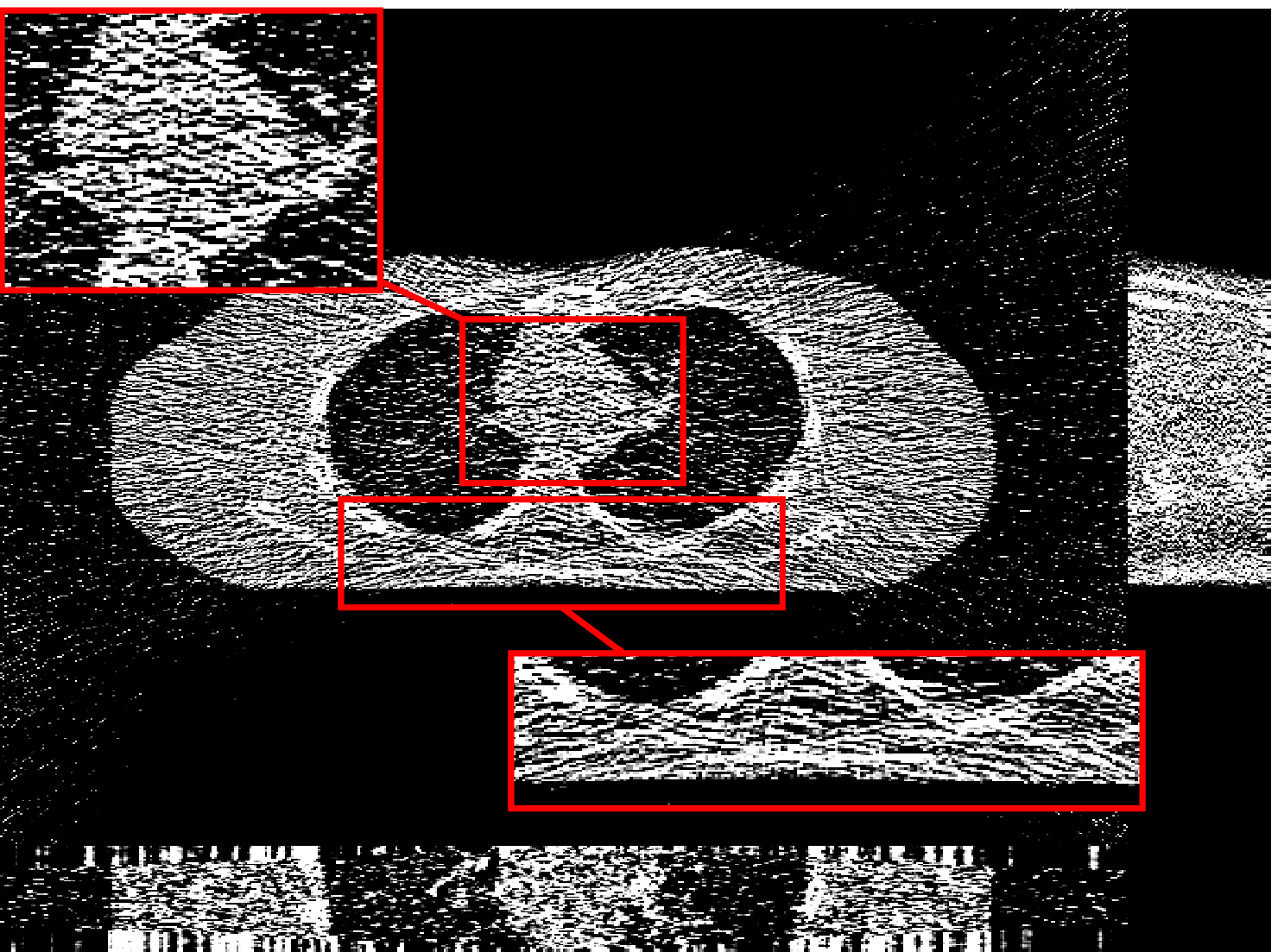}&
\includegraphics[width=.22\linewidth, height=.22\linewidth]{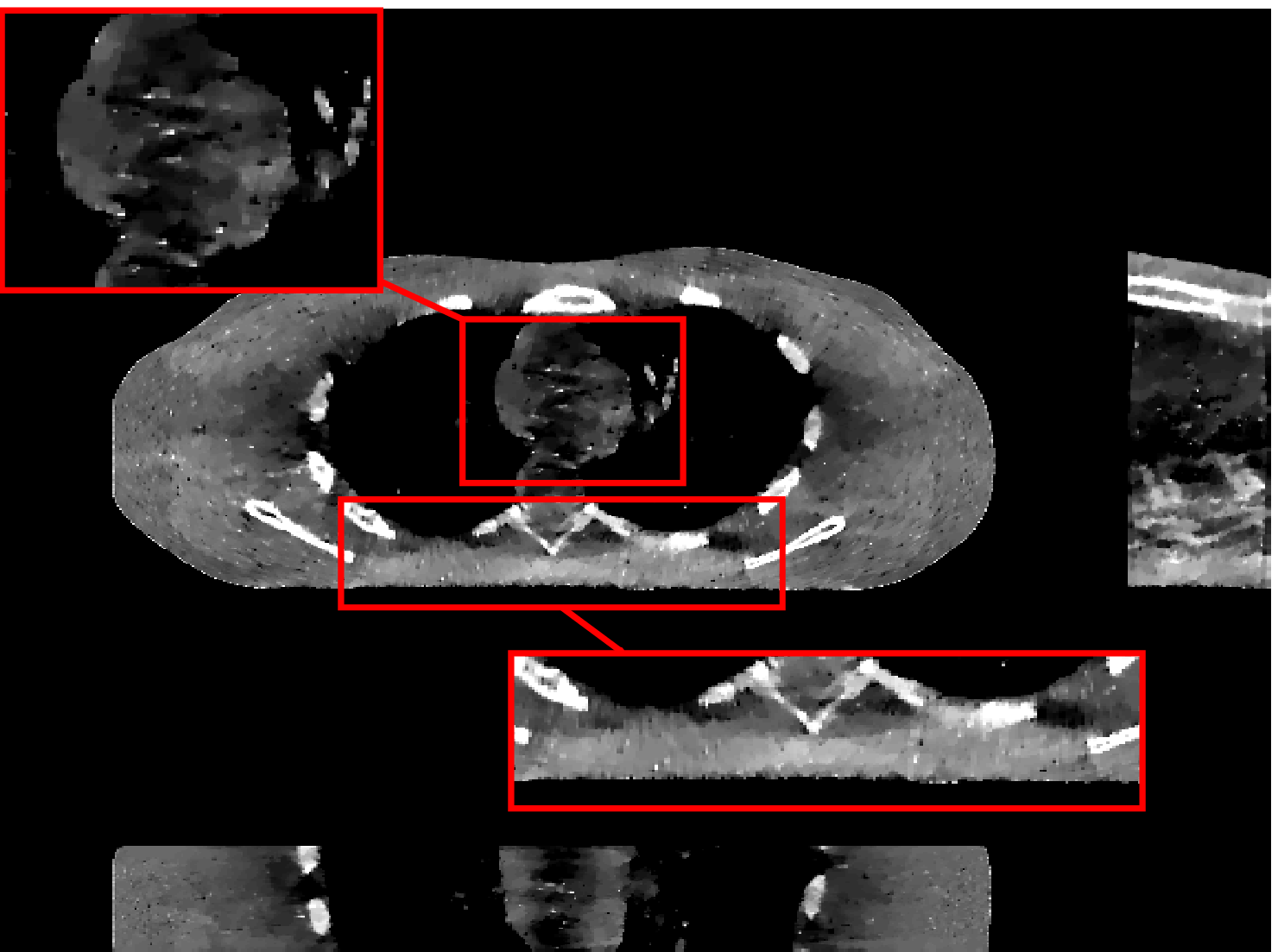}&
\includegraphics[width=.22\linewidth, height=.22\linewidth]{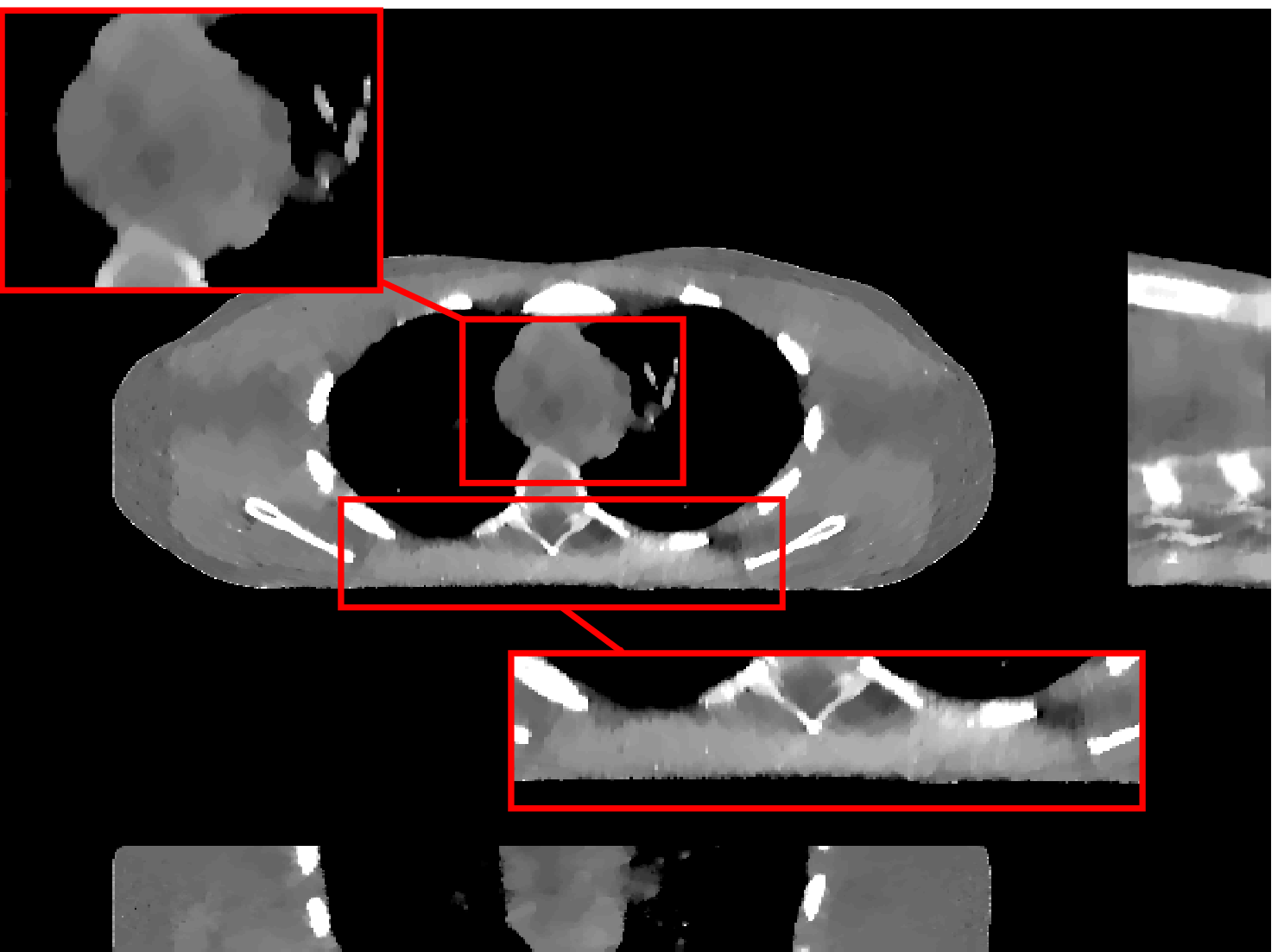}&
\includegraphics[width=.22\linewidth, height=.22\linewidth]{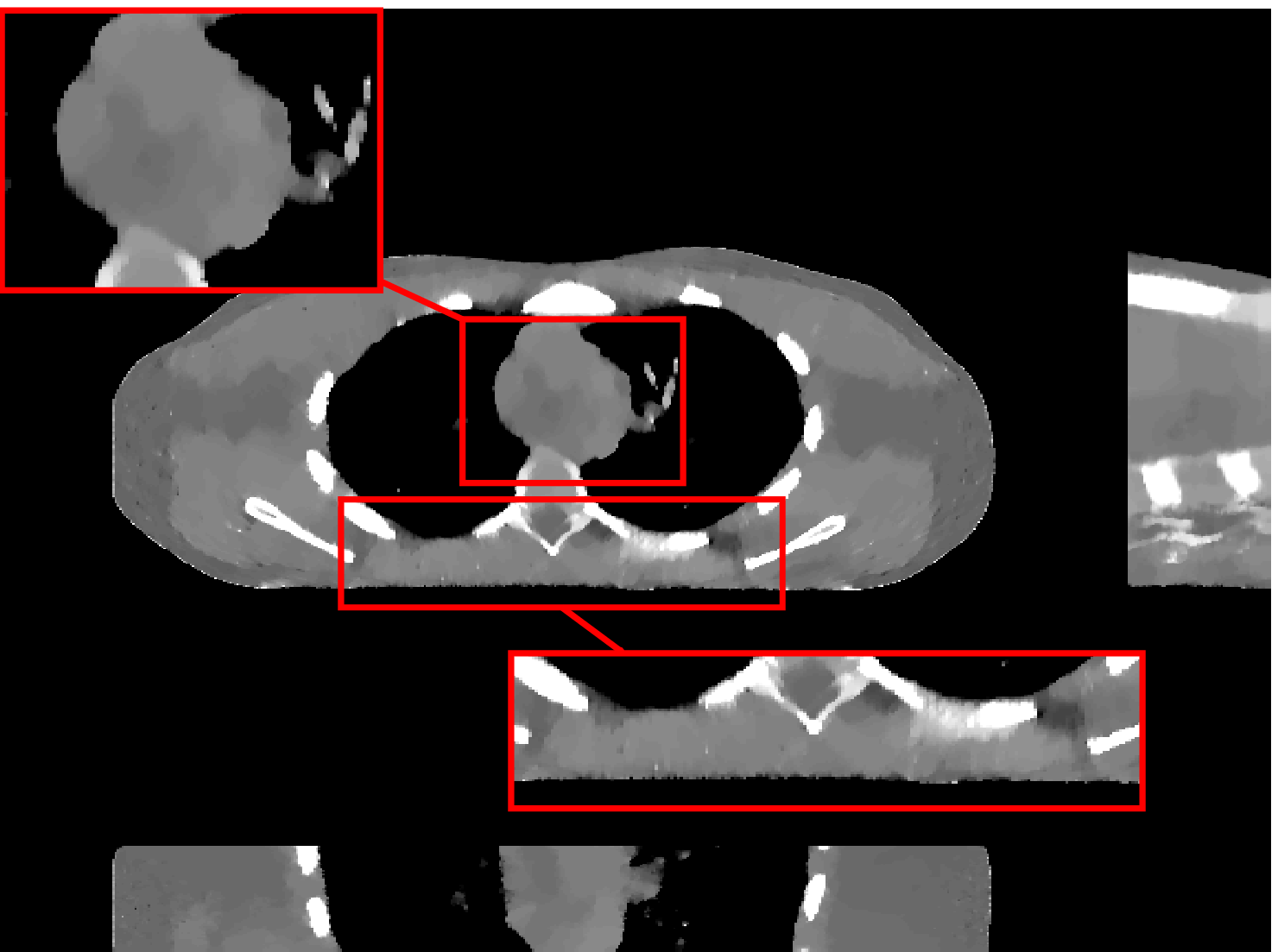}\\
\put(-40,50){$\sigma^2=100^2$}&
\includegraphics[width=.22\linewidth, height=.22\linewidth]{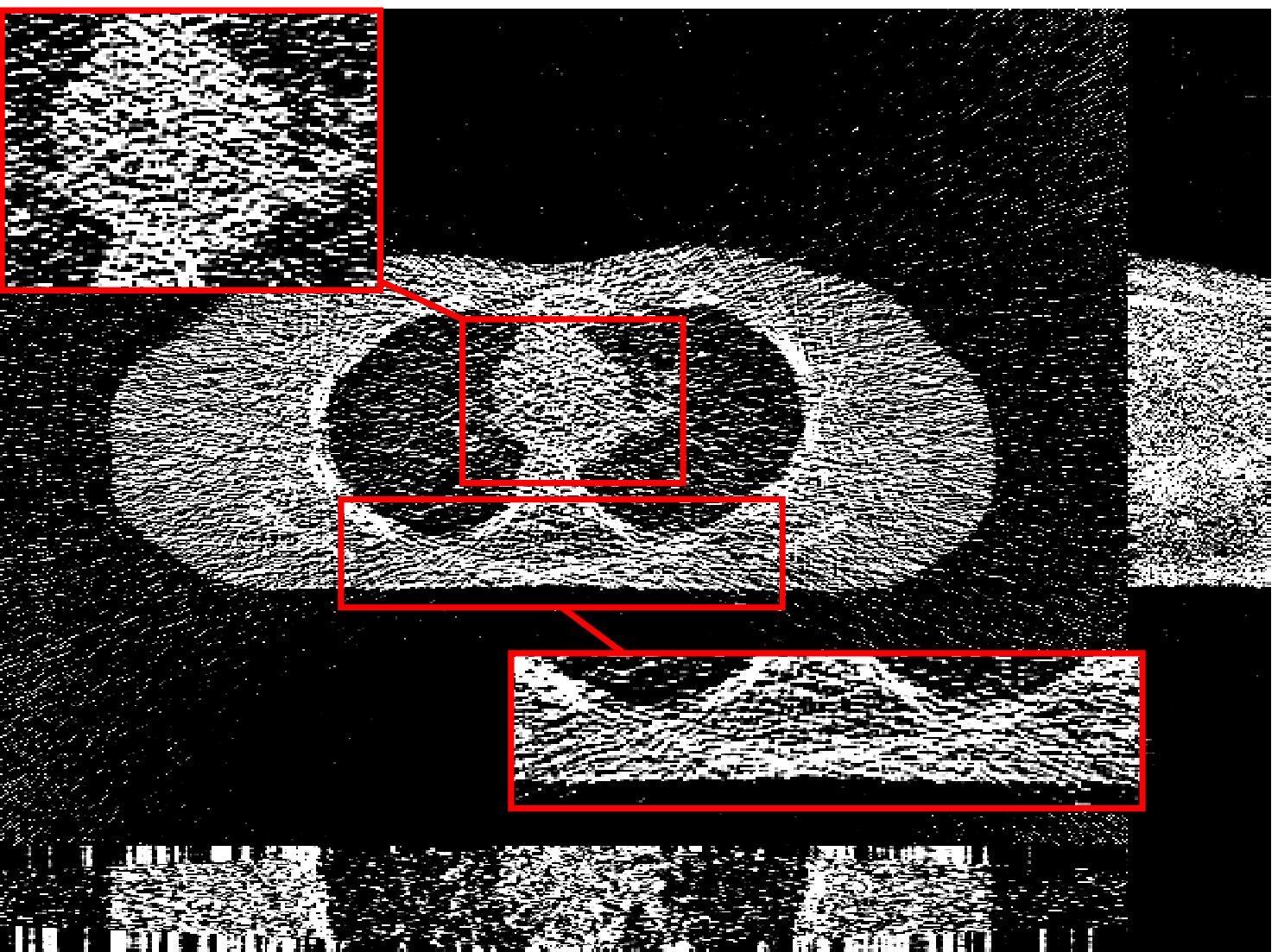}&
\includegraphics[width=.22\linewidth, height=.22\linewidth]{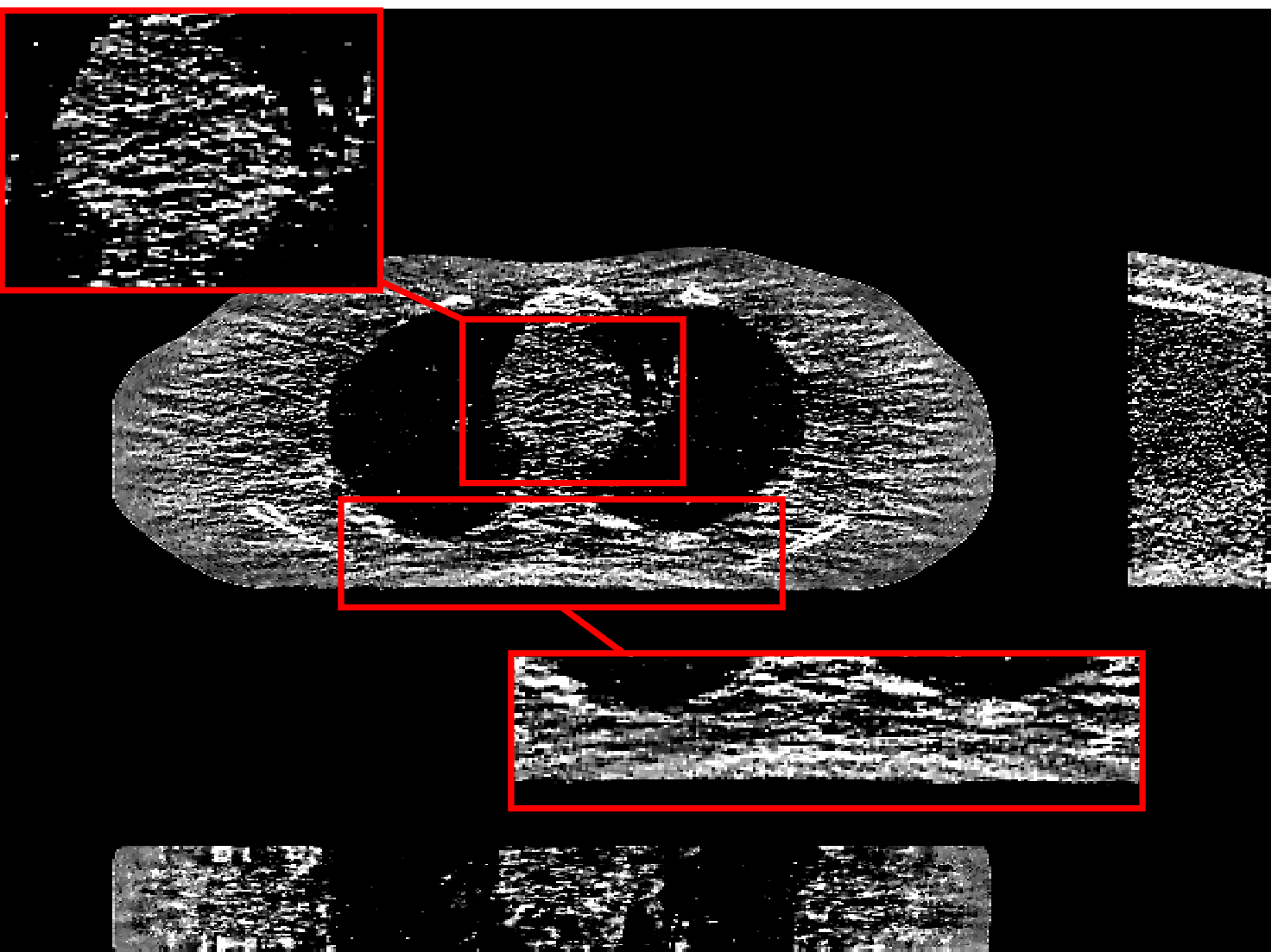}&
\includegraphics[width=.22\linewidth, height=.22\linewidth]{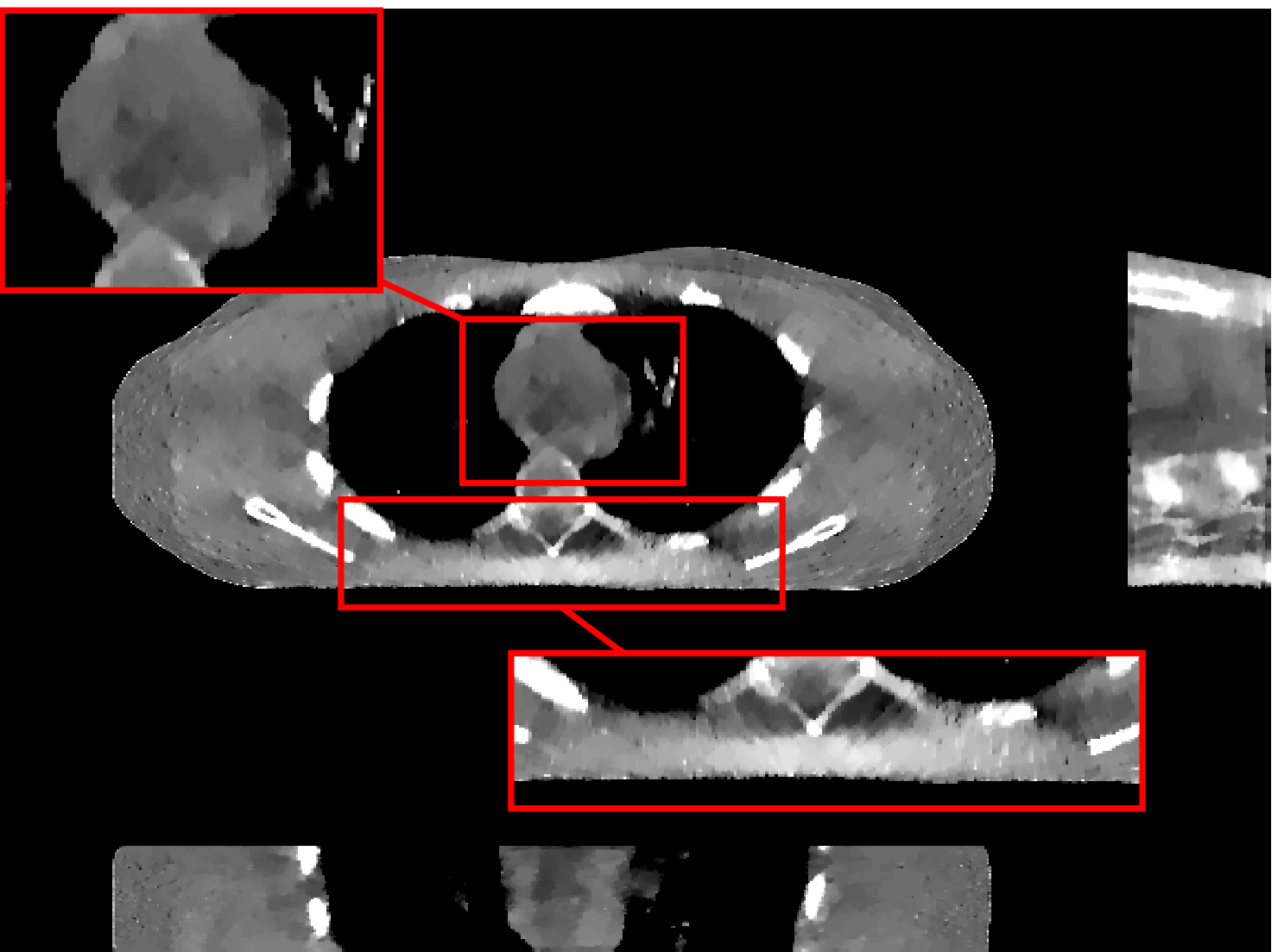}&
\includegraphics[width=.22\linewidth, height=.22\linewidth]{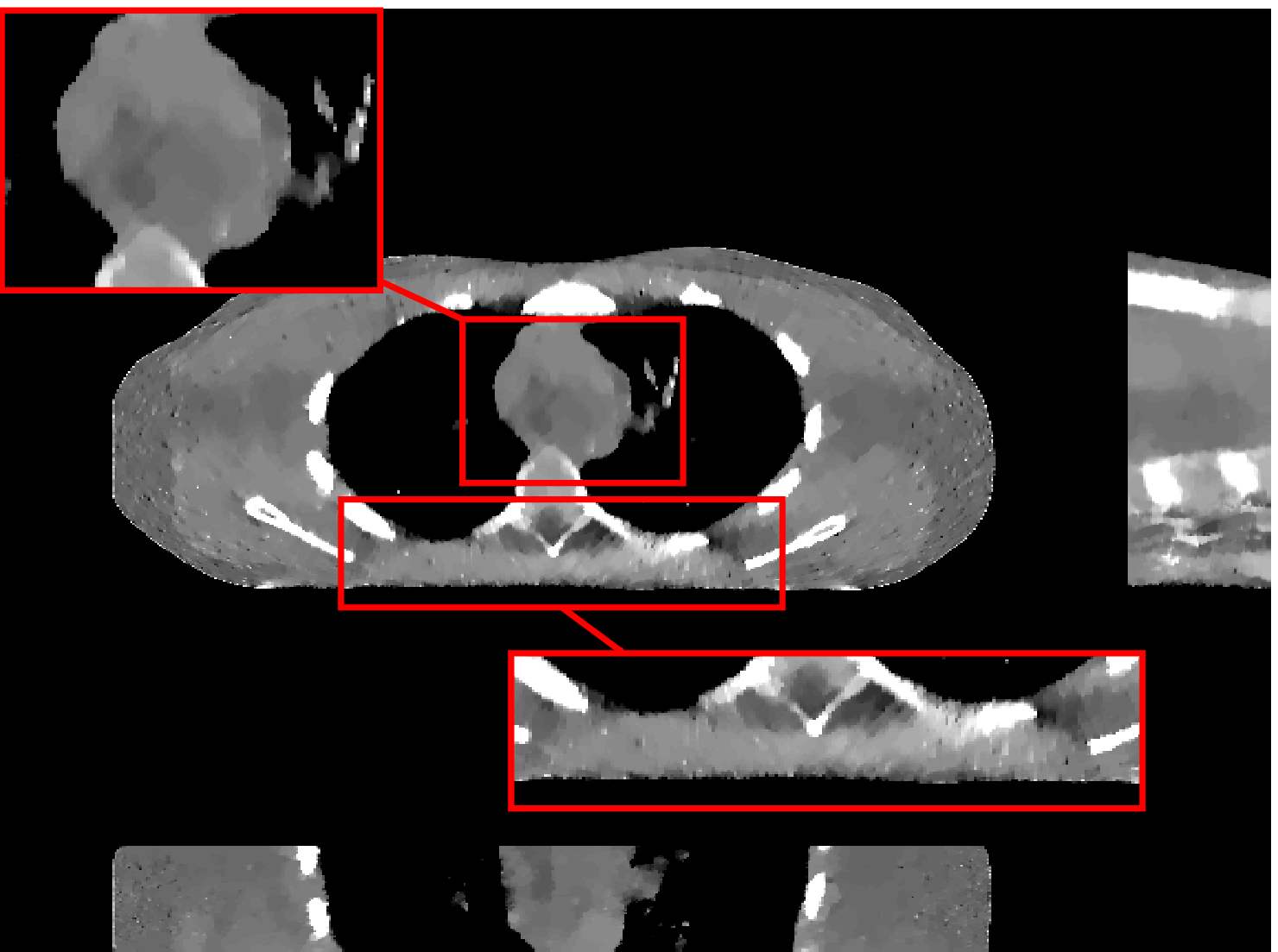}\\
&
{FBP}&
{PWLS}&
{SP}&
{MPG}
\end{tabular}
\caption{ XCAT phantom reconstructed by FBP (first column), PWLS (second column), SP (third column) and the proposed MPG method (forth column) for dose of $I_i = 10^4$ with variance of electronic noise  $\sigma^2 = 50^2$ (first row), $\sigma^2= 60^2$ (second row), $\sigma^2= 70^2$ (third row) and $\sigma^2  = 100^2$ (forth row).
All images are displayed using a window of $[800,1200]$~HU.}
\label{fig:MPG}
\end{center}
\end{figure*}

\begin{figure*}
\begin{center}
\begin{tabular}{c@{\hspace{2pt}}c@{\hspace{2pt}}c@{\hspace{2pt}}c@{\hspace{2pt}}c@{\hspace{2pt}}c}
\put(-20,50){SP}&
\includegraphics[width=.21\linewidth, height=.21\linewidth]{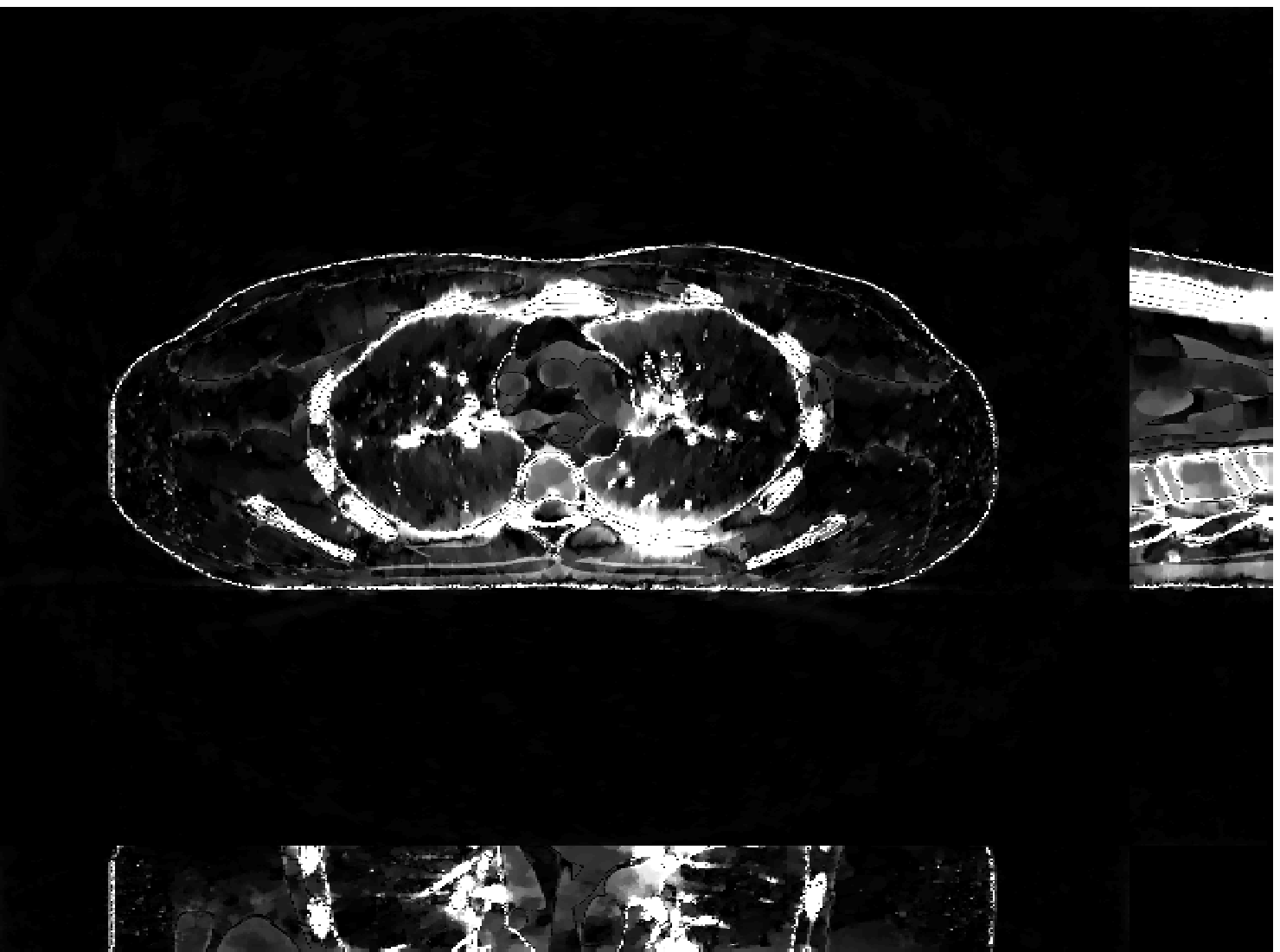}&
\includegraphics[width=.21\linewidth, height=.21\linewidth]{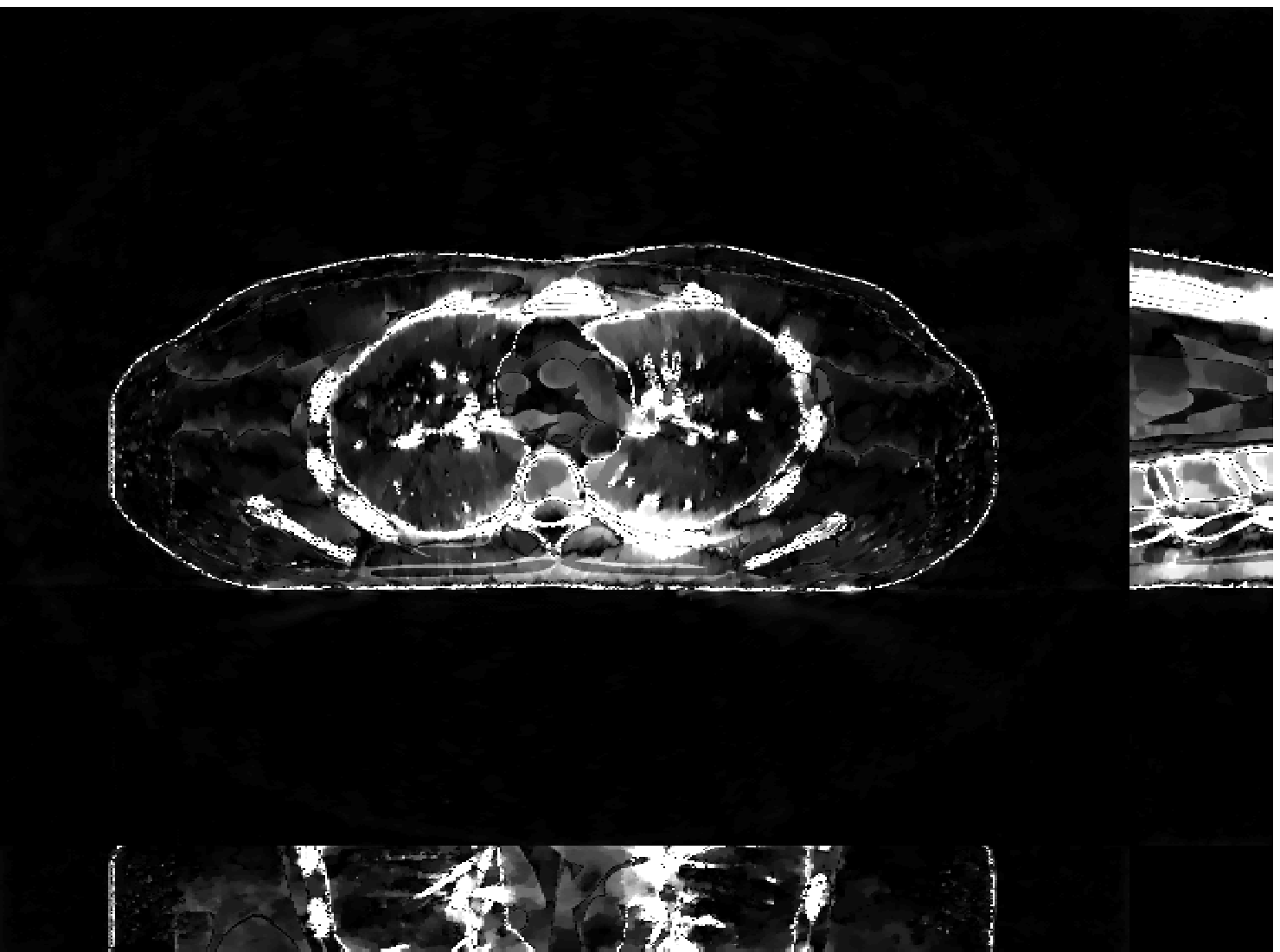}&
\includegraphics[width=.21\linewidth, height=.21\linewidth]{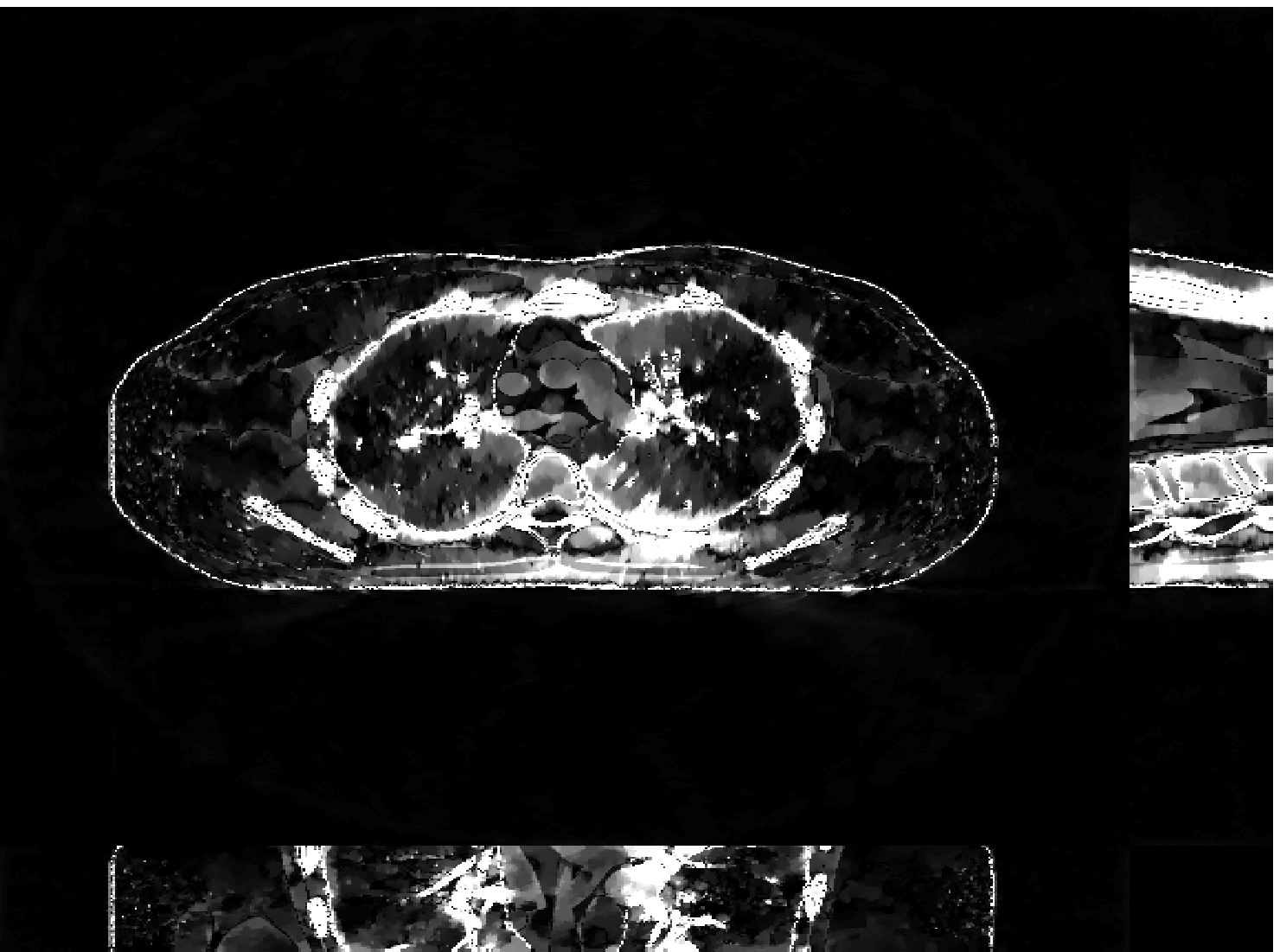}&
\includegraphics[width=.21\linewidth, height=.21\linewidth]{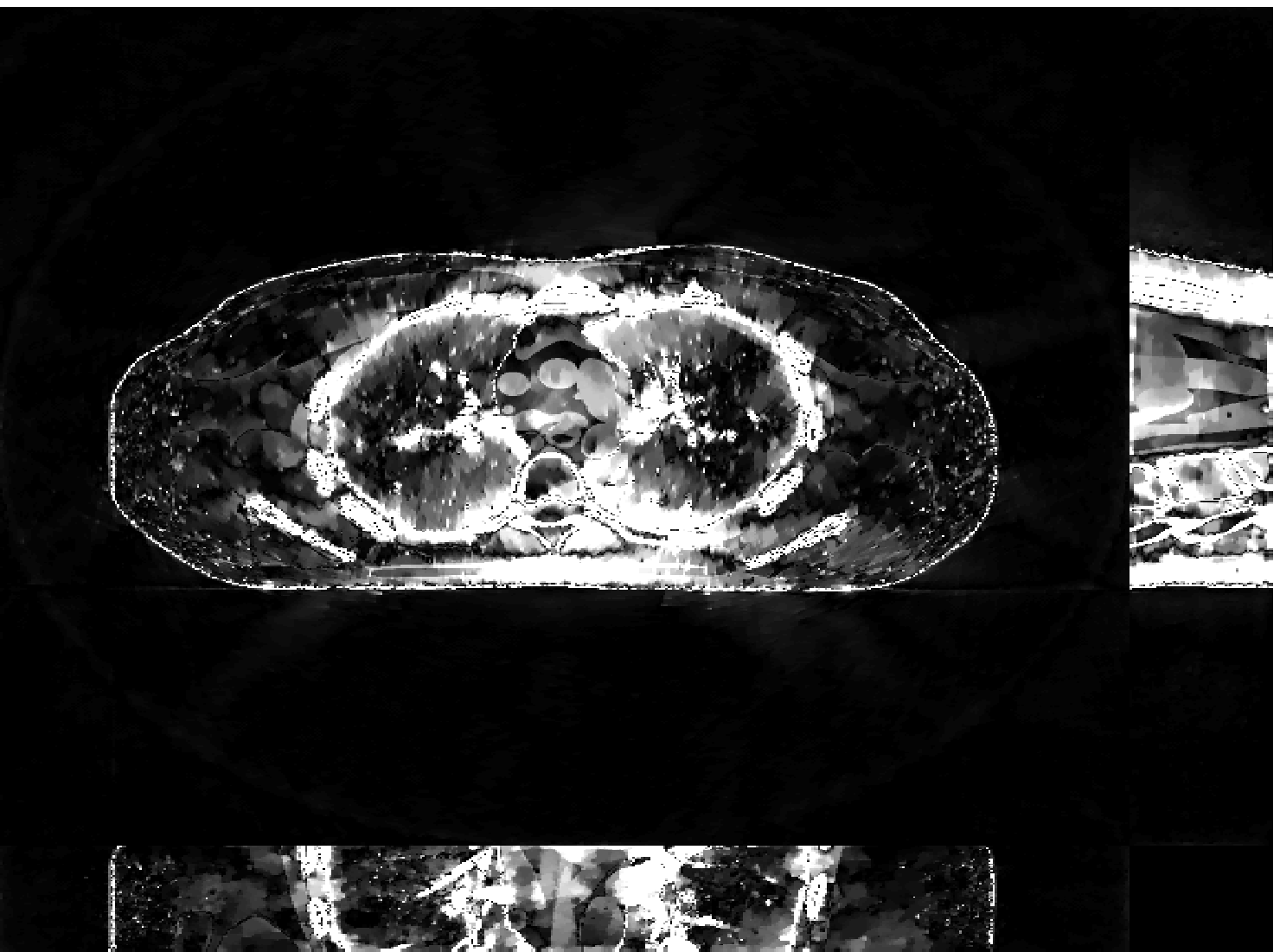}\\
\put(-20,50){MPG}&
\includegraphics[width=.21\linewidth, height=.21\linewidth]{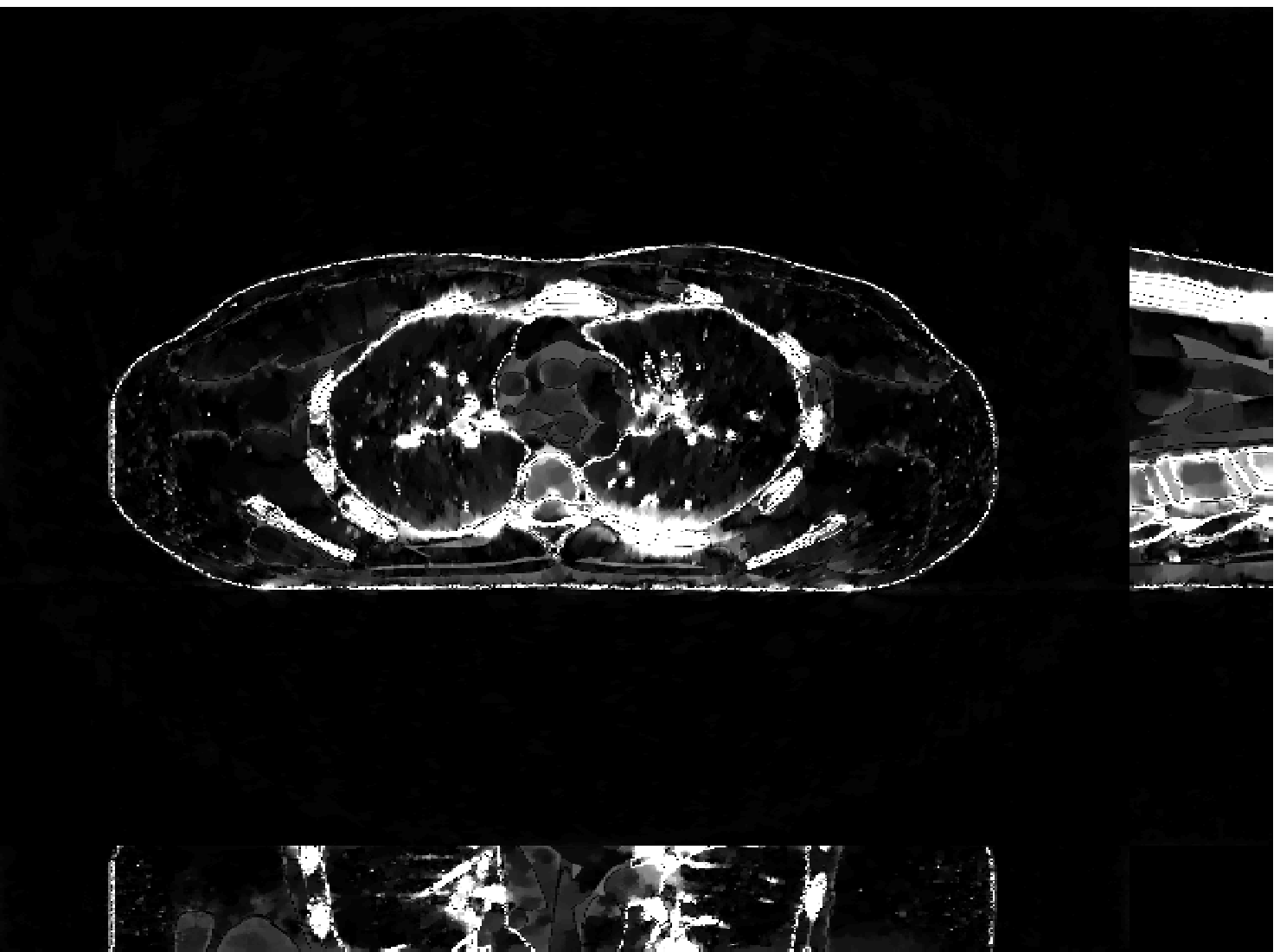}&
\includegraphics[width=.21\linewidth, height=.21\linewidth]{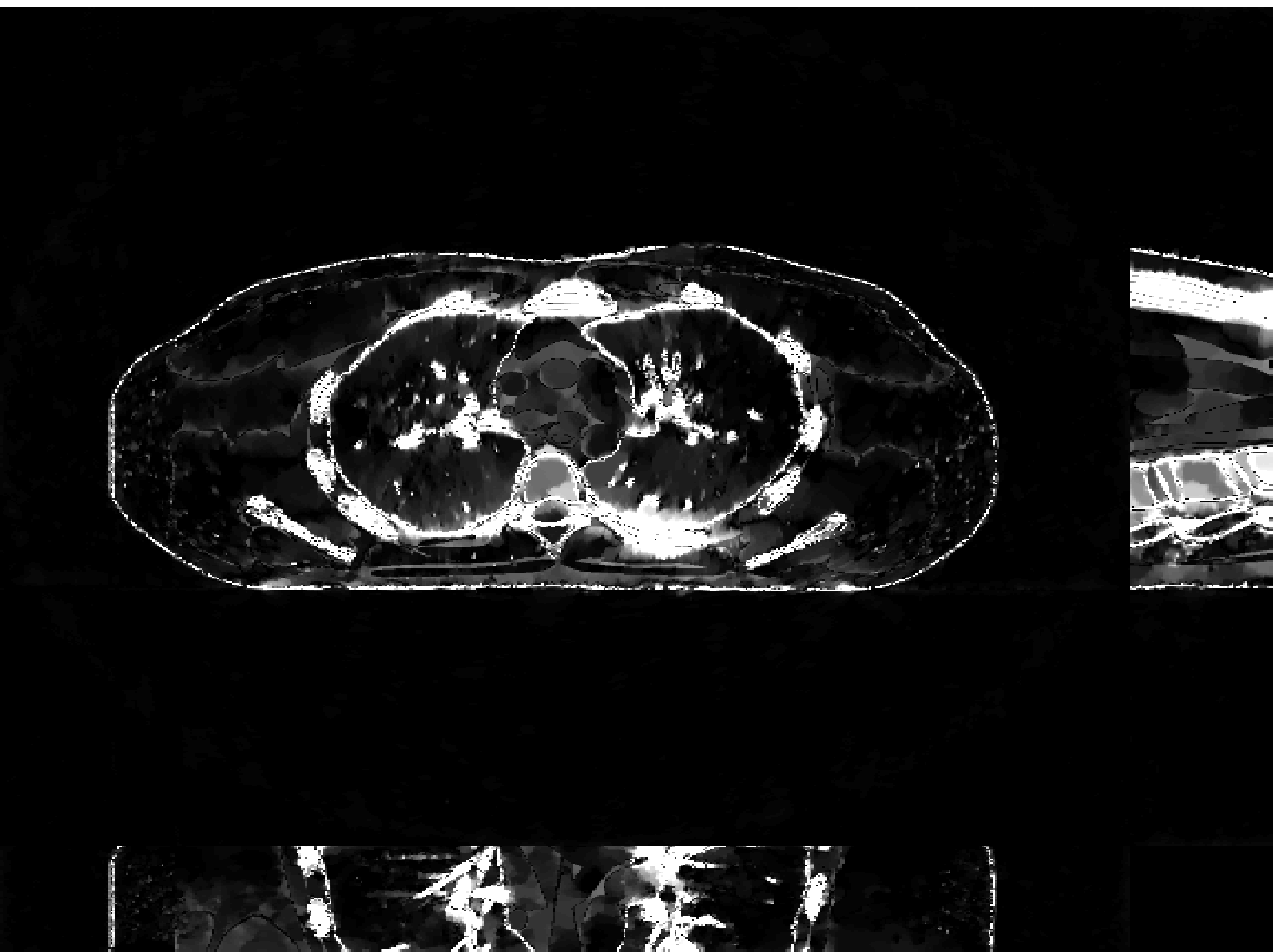}&
\includegraphics[width=.21\linewidth, height=.21\linewidth]{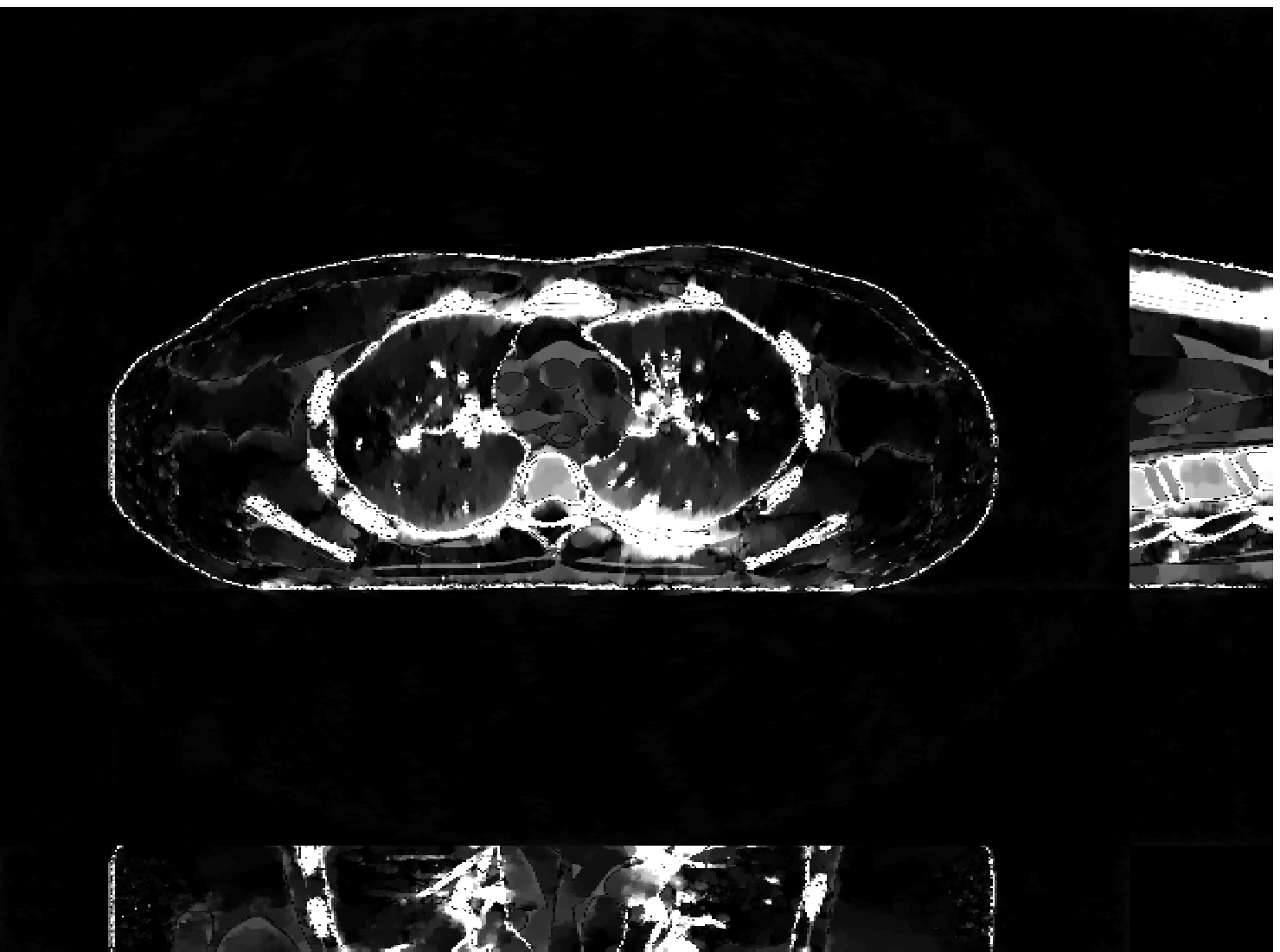}&
\includegraphics[width=.21\linewidth, height=.21\linewidth]{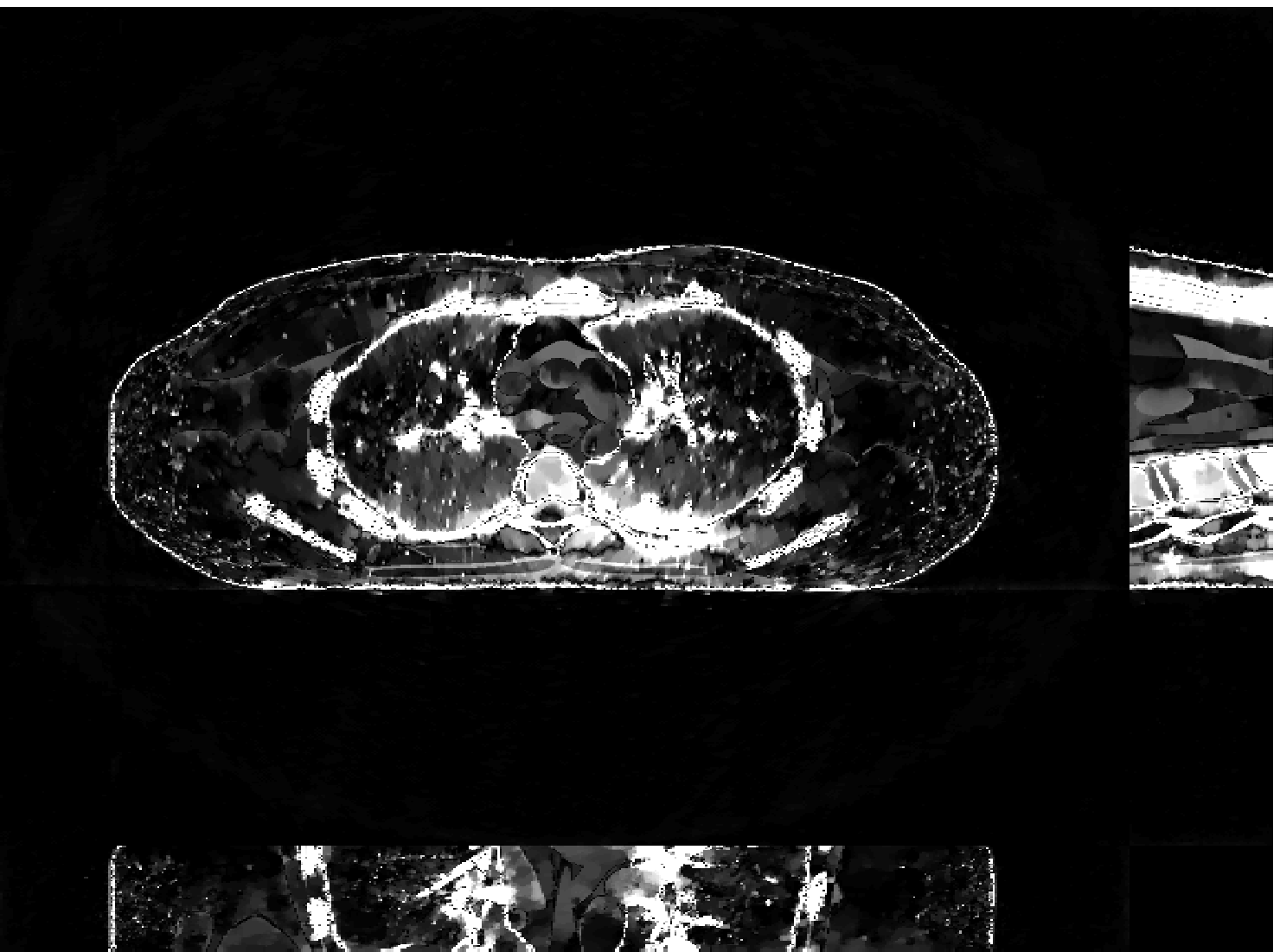}\\
&
$ \sigma^2=50^2$&
$ \sigma^2=60^2$&
$ \sigma^2=70^2$&
$ \sigma^2=100^2$
\end{tabular}
\caption {
Absolute error images of reconstructions by SP (first row) and MPG (second row) for dose of $I_i=10^4$ with variance of electronic noise $\sigma^2 = 50^2$ (first column),  $\sigma^2 = 60^2$ (second column), $\sigma^2 = 70^2$ (third column) and $\sigma^2 = 100^2$ (forth column).
All images are displayed using a window of $[0,100]$~HU.}
\label{fig:error}
\end{center}
\end{figure*}

\begin{figure*}
\begin{center}
\begin{tabular}{c@{\hspace{2pt}}c@{\hspace{2pt}}c@{\hspace{2pt}}c@{\hspace{2pt}}c@{\hspace{2pt}}c@{\hspace{2pt}}c@{\hspace{2pt}}c@{\hspace{2pt}}c}
\put(-40,50){$\sigma^2=50^2$}&
\includegraphics[width=.22\linewidth, height=.22\linewidth]{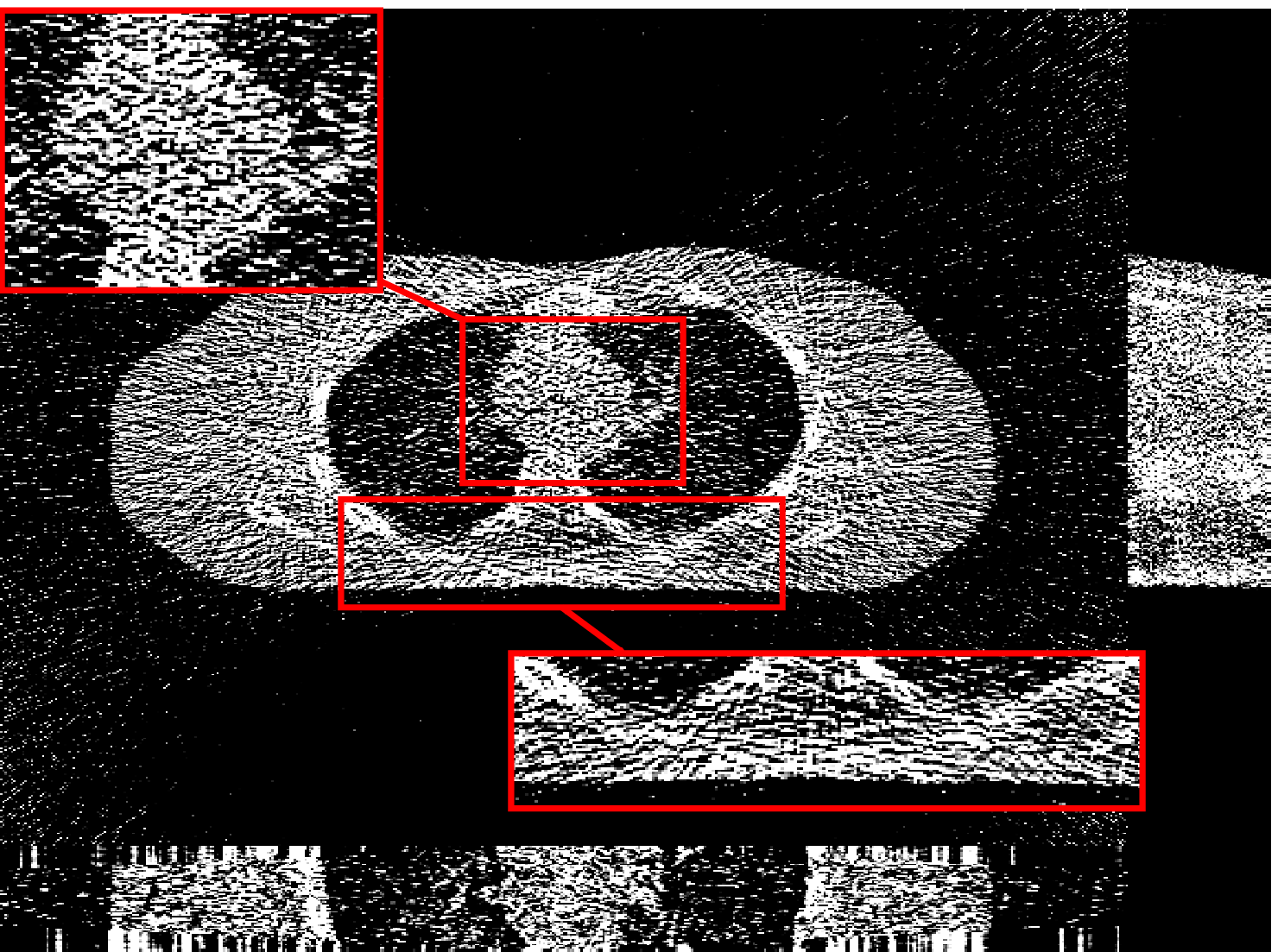}&
\includegraphics[width=.22\linewidth, height=.22\linewidth]{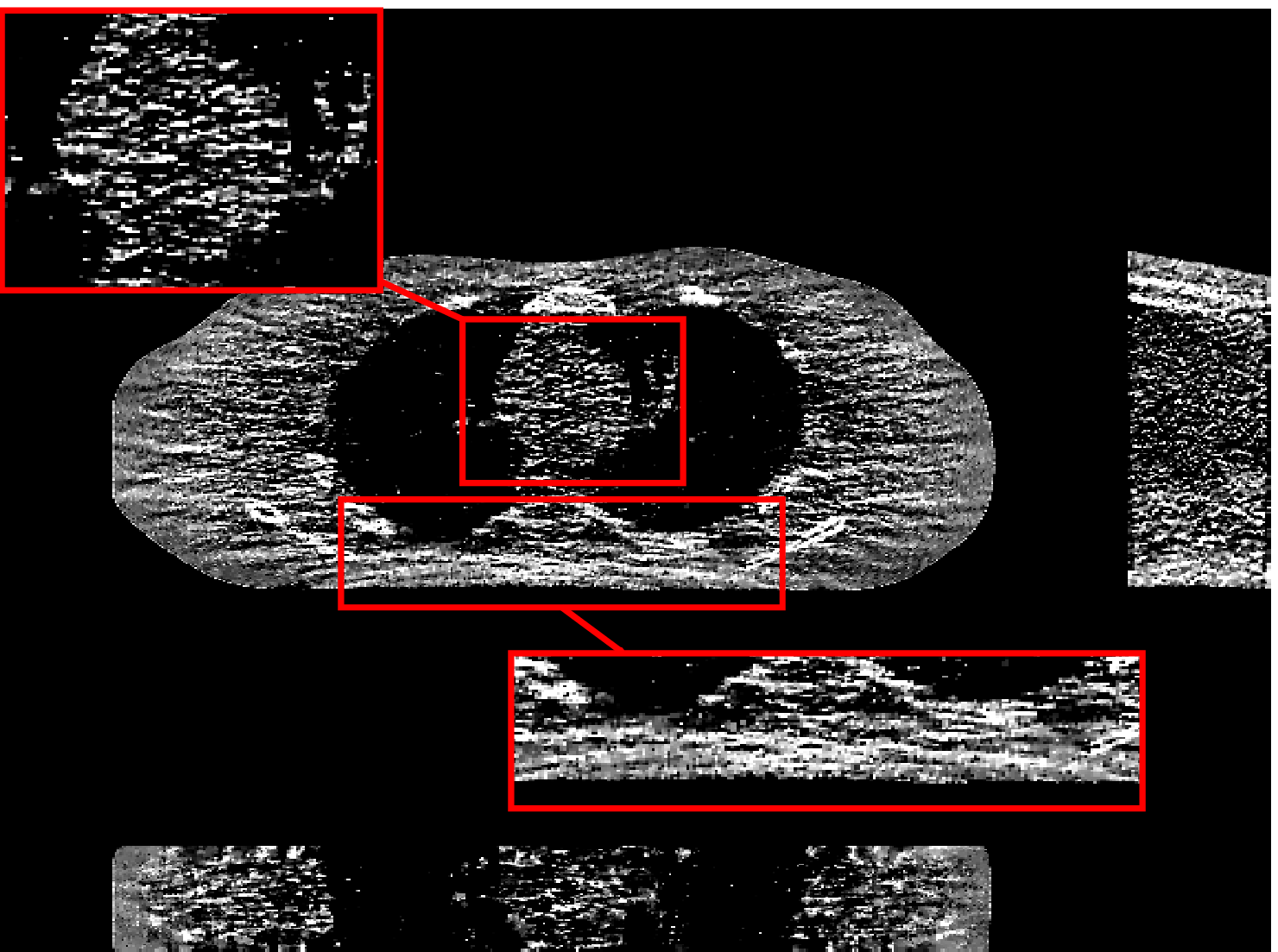}&
\includegraphics[width=.22\linewidth, height=.22\linewidth]{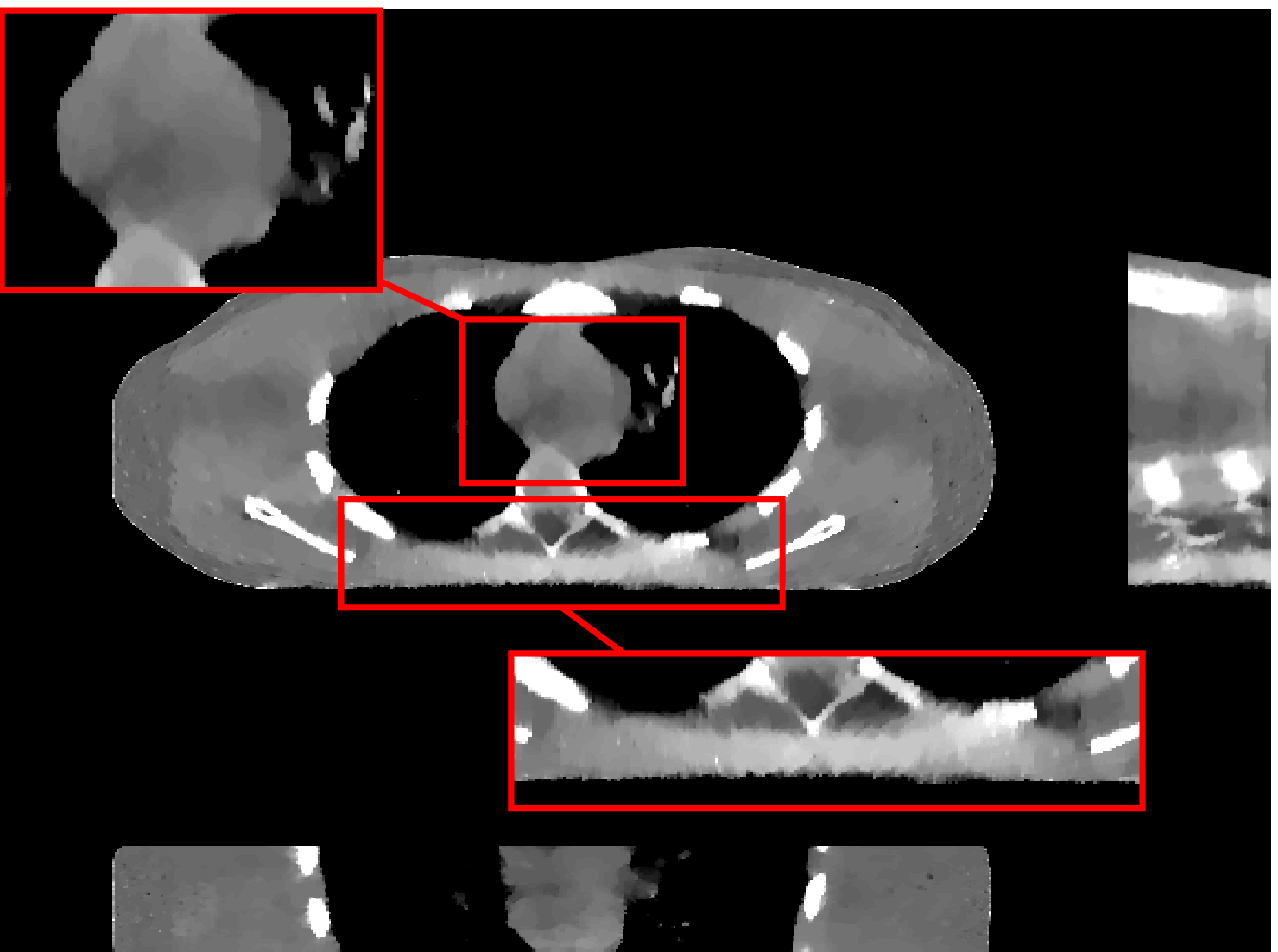}&
\includegraphics[width=.22\linewidth, height=.22\linewidth]{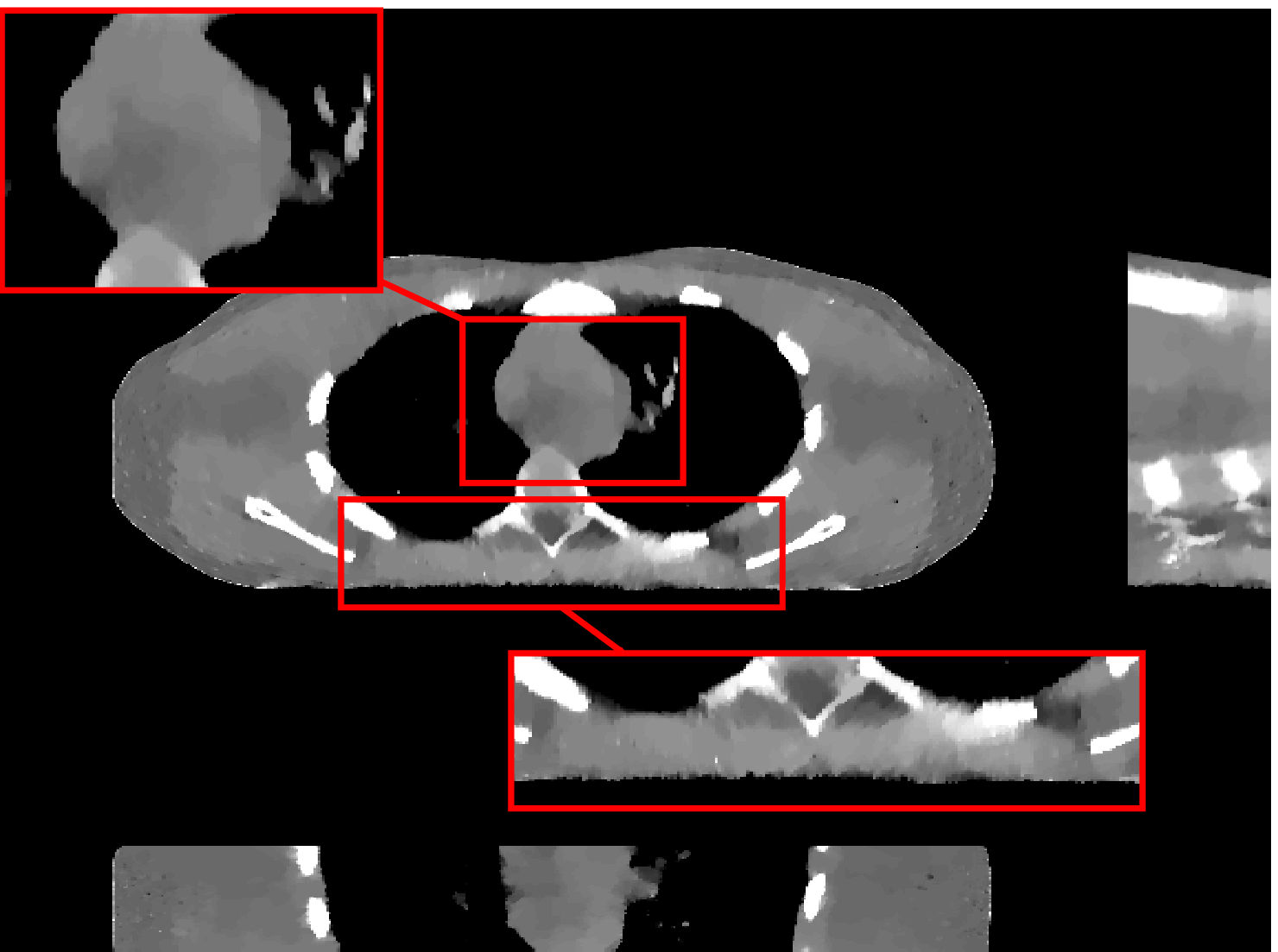}\\
\put(-40,50){$\sigma^2=60^2$}&
\includegraphics[width=.22\linewidth, height=.22\linewidth]{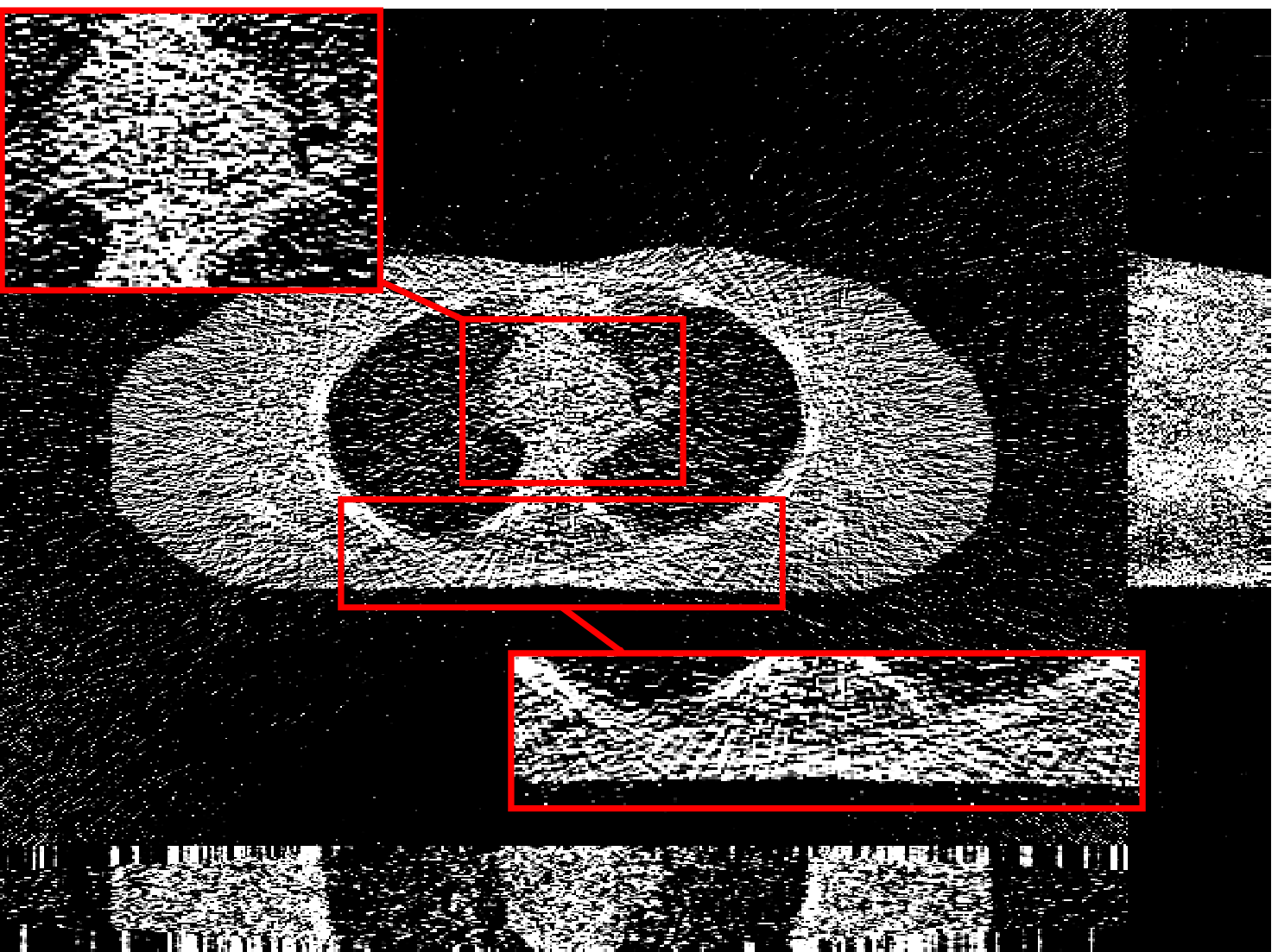}&
\includegraphics[width=.22\linewidth, height=.22\linewidth]{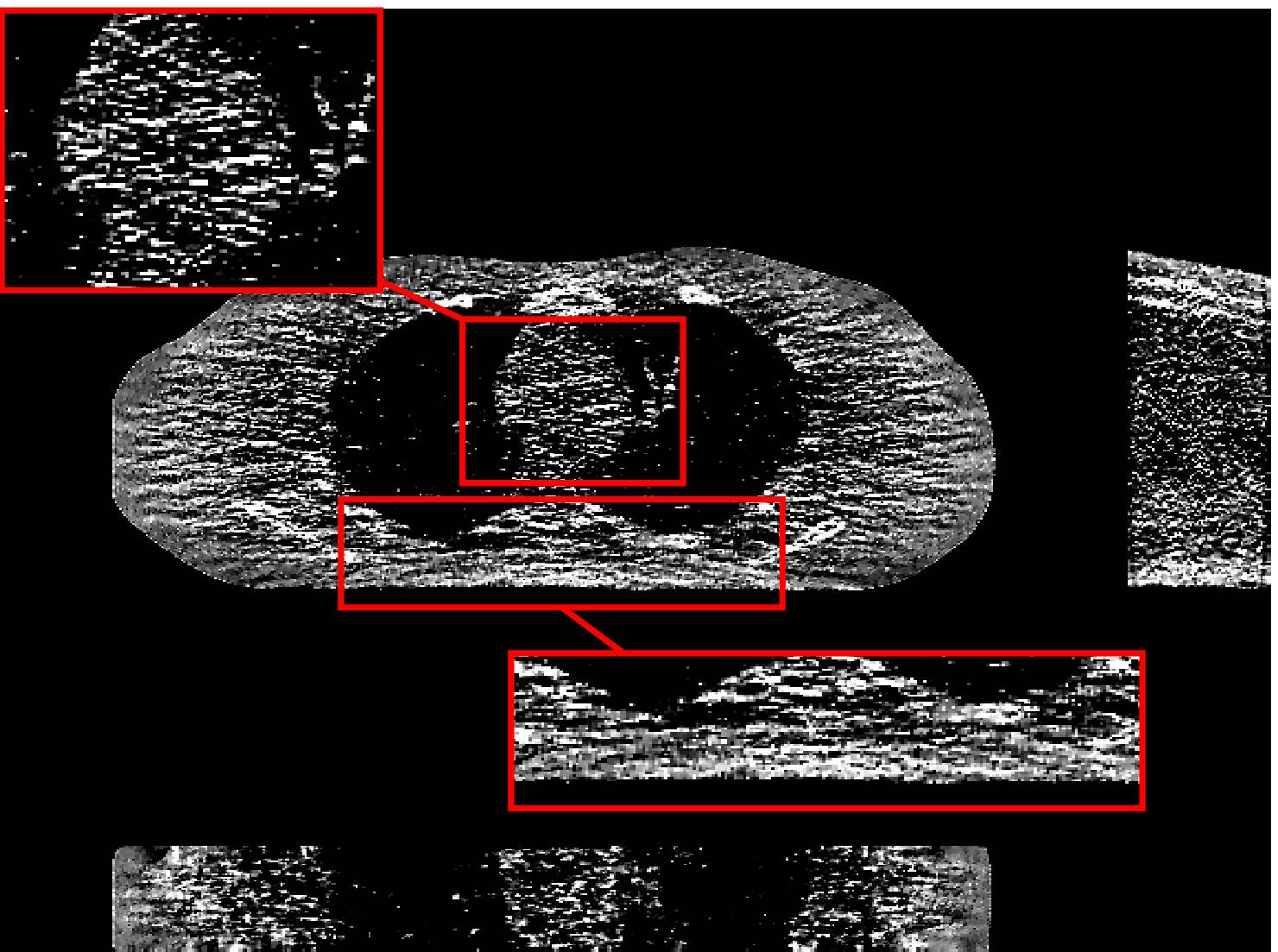}&
\includegraphics[width=.22\linewidth, height=.22\linewidth]{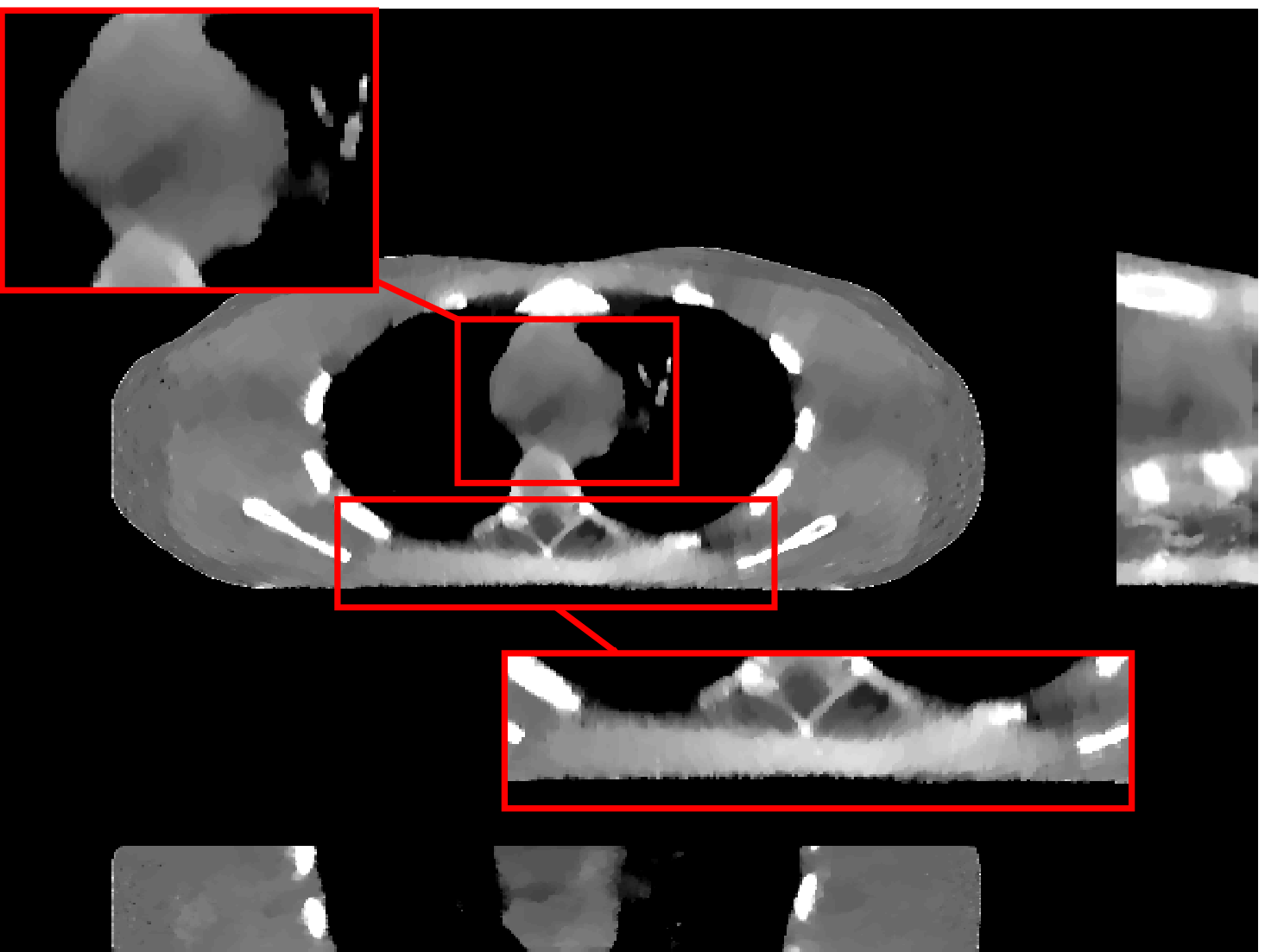}&
\includegraphics[width=.22\linewidth, height=.22\linewidth]{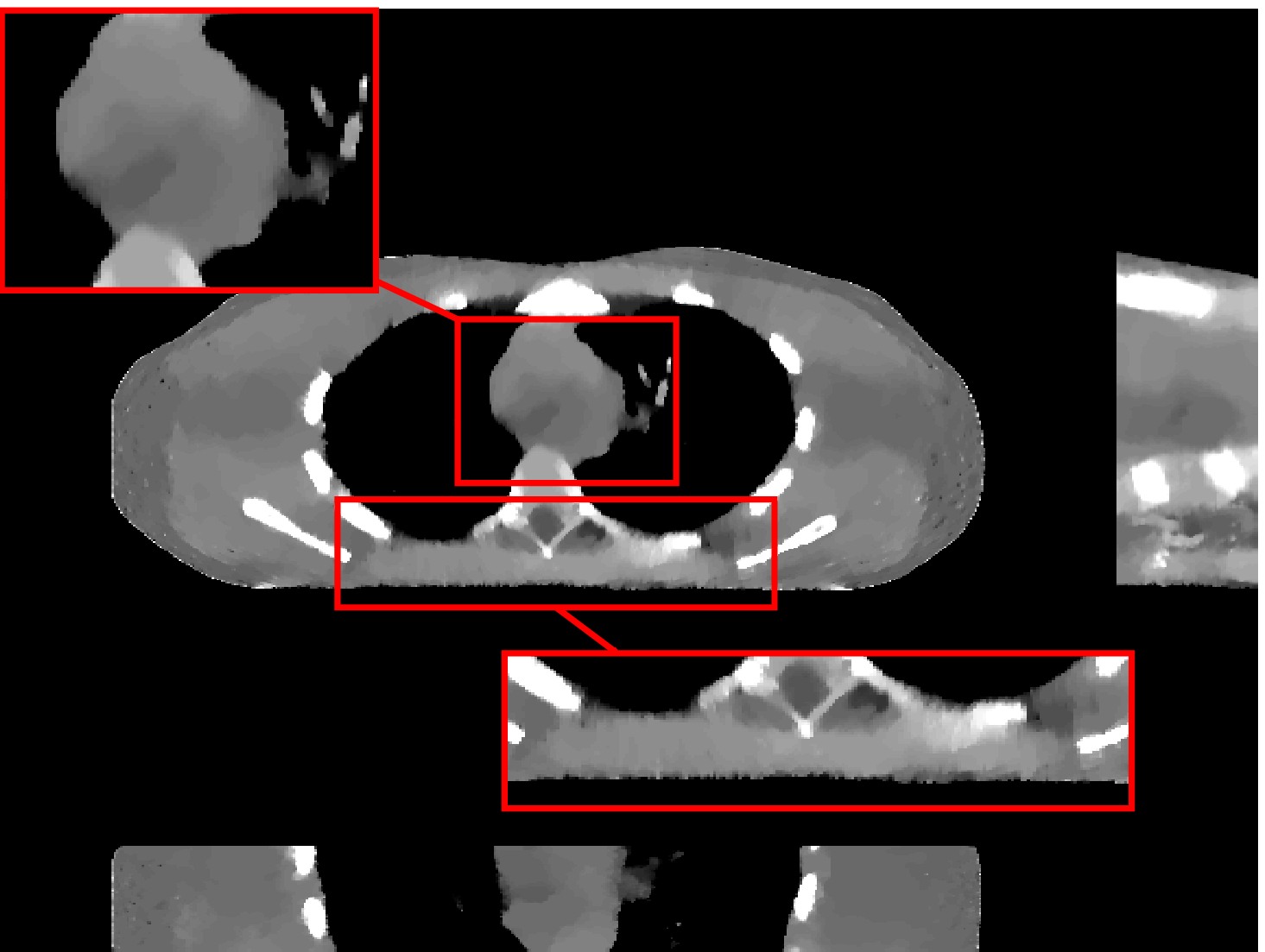}\\
\put(-40,50){$\sigma^2=70^2$}&
\includegraphics[width=.22\linewidth, height=.22\linewidth]{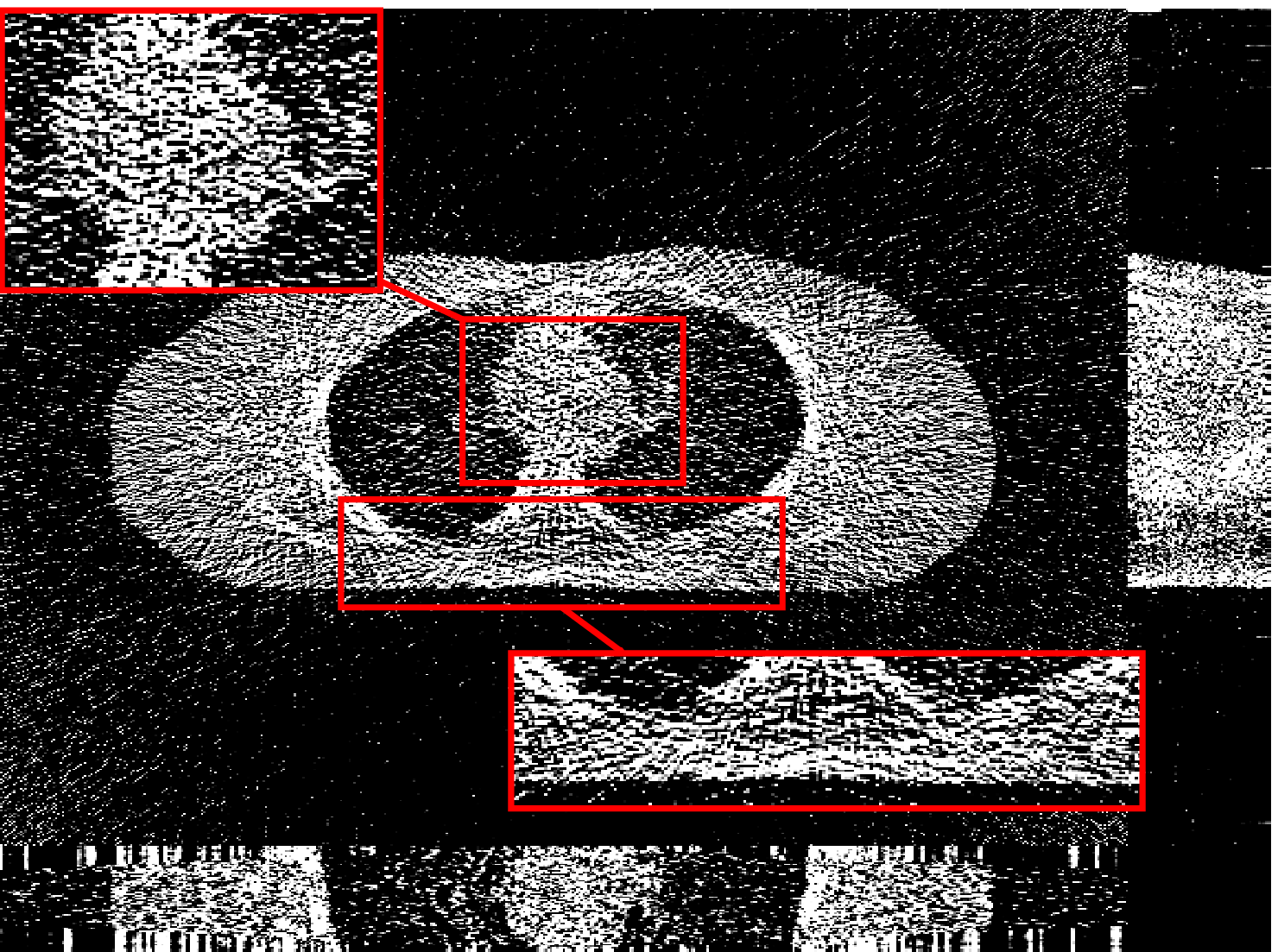}&
\includegraphics[width=.22\linewidth, height=.22\linewidth]{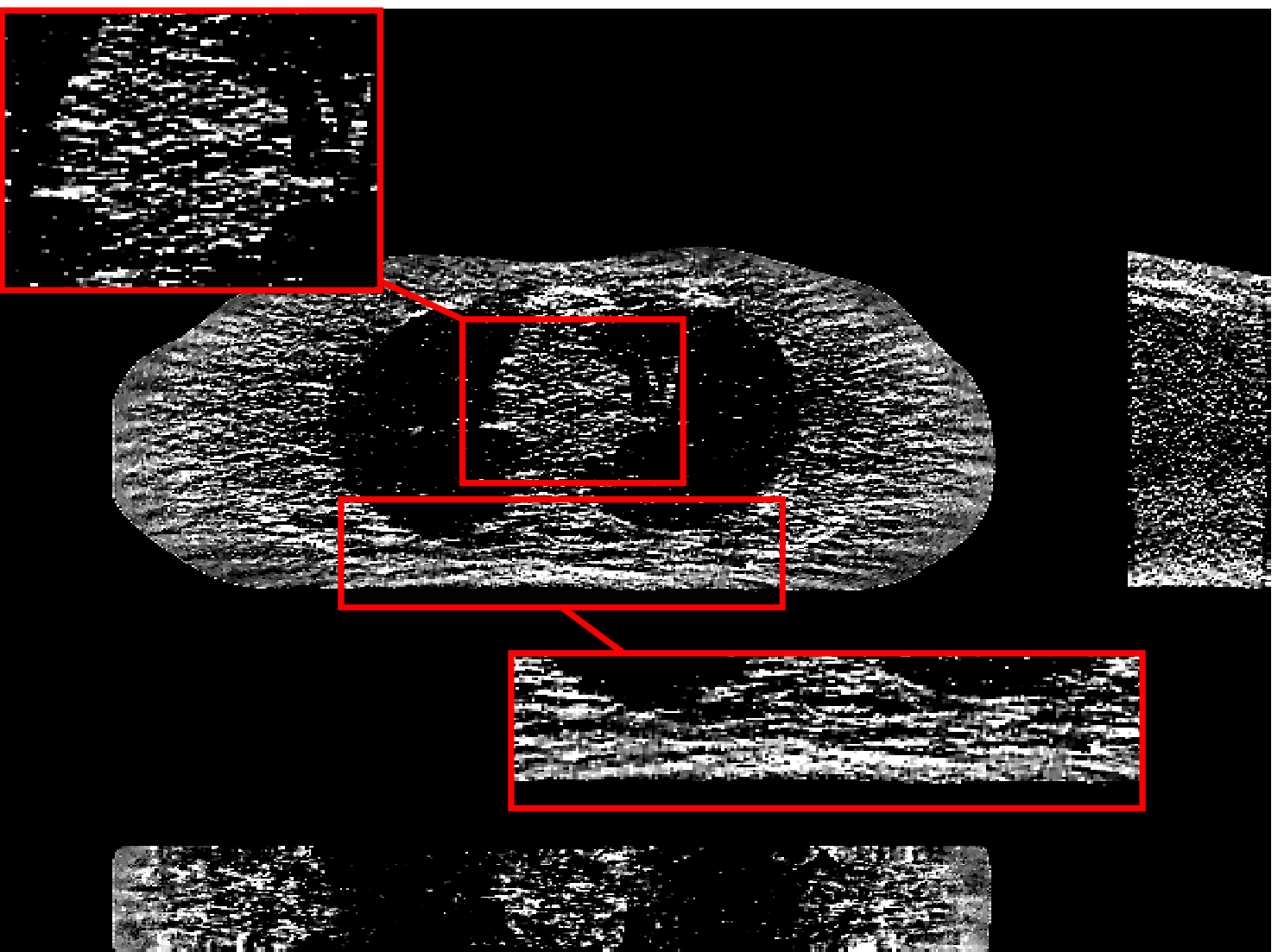}&
\includegraphics[width=.22\linewidth, height=.22\linewidth]{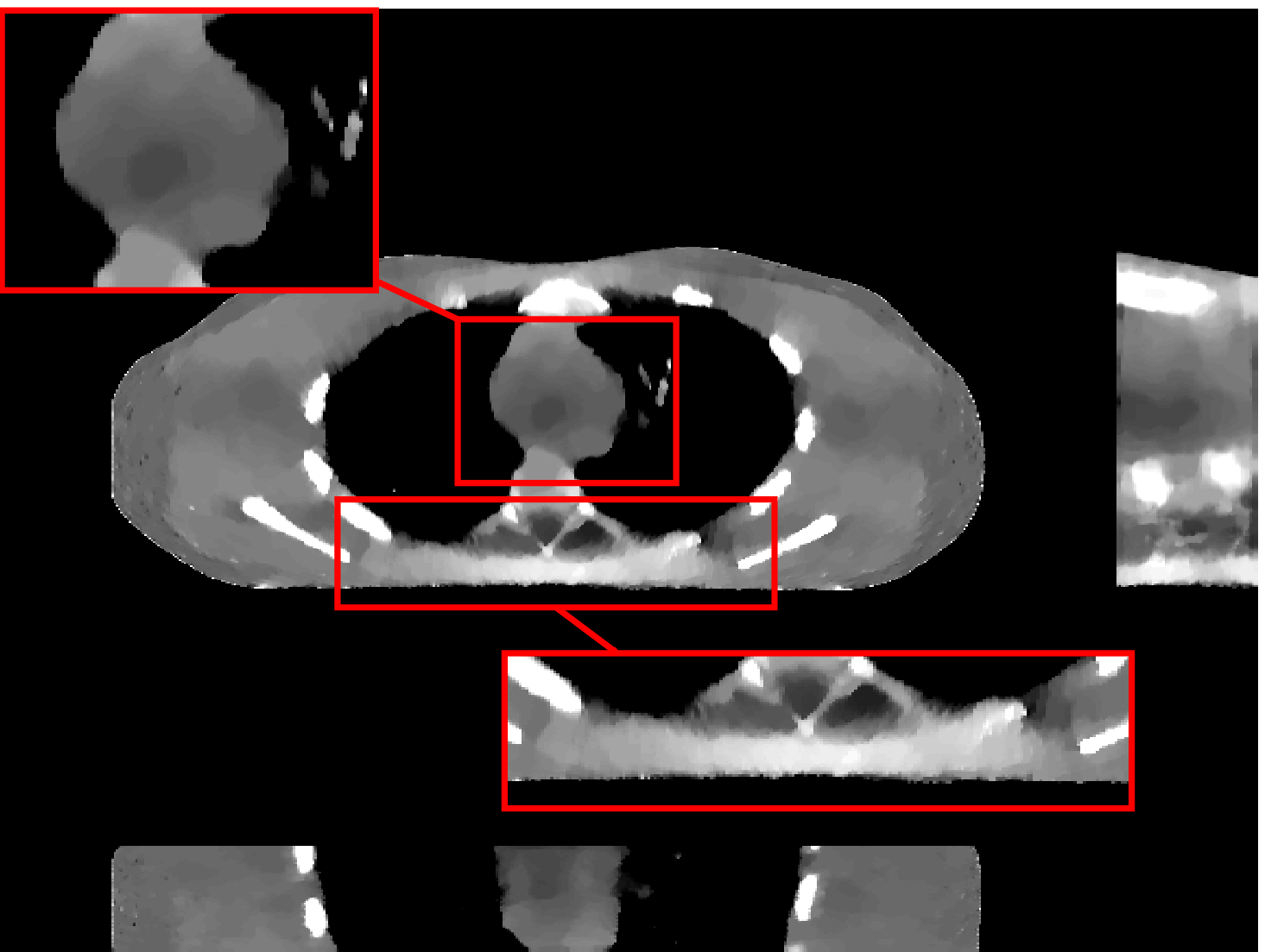}&
\includegraphics[width=.22\linewidth, height=.22\linewidth]{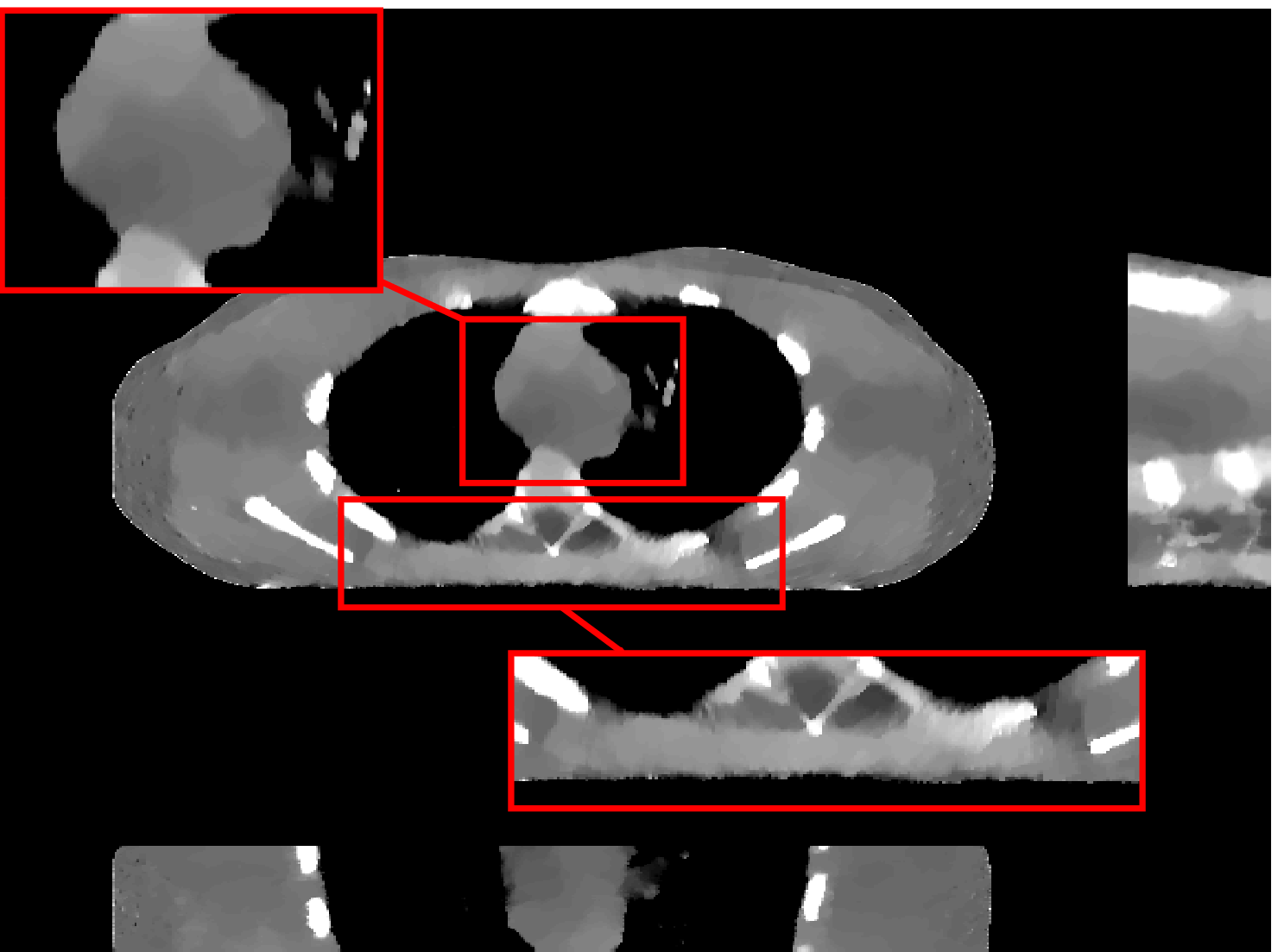}\\
\put(-40,50){$\sigma^2=100^2$}&
\includegraphics[width=.22\linewidth, height=.22\linewidth]{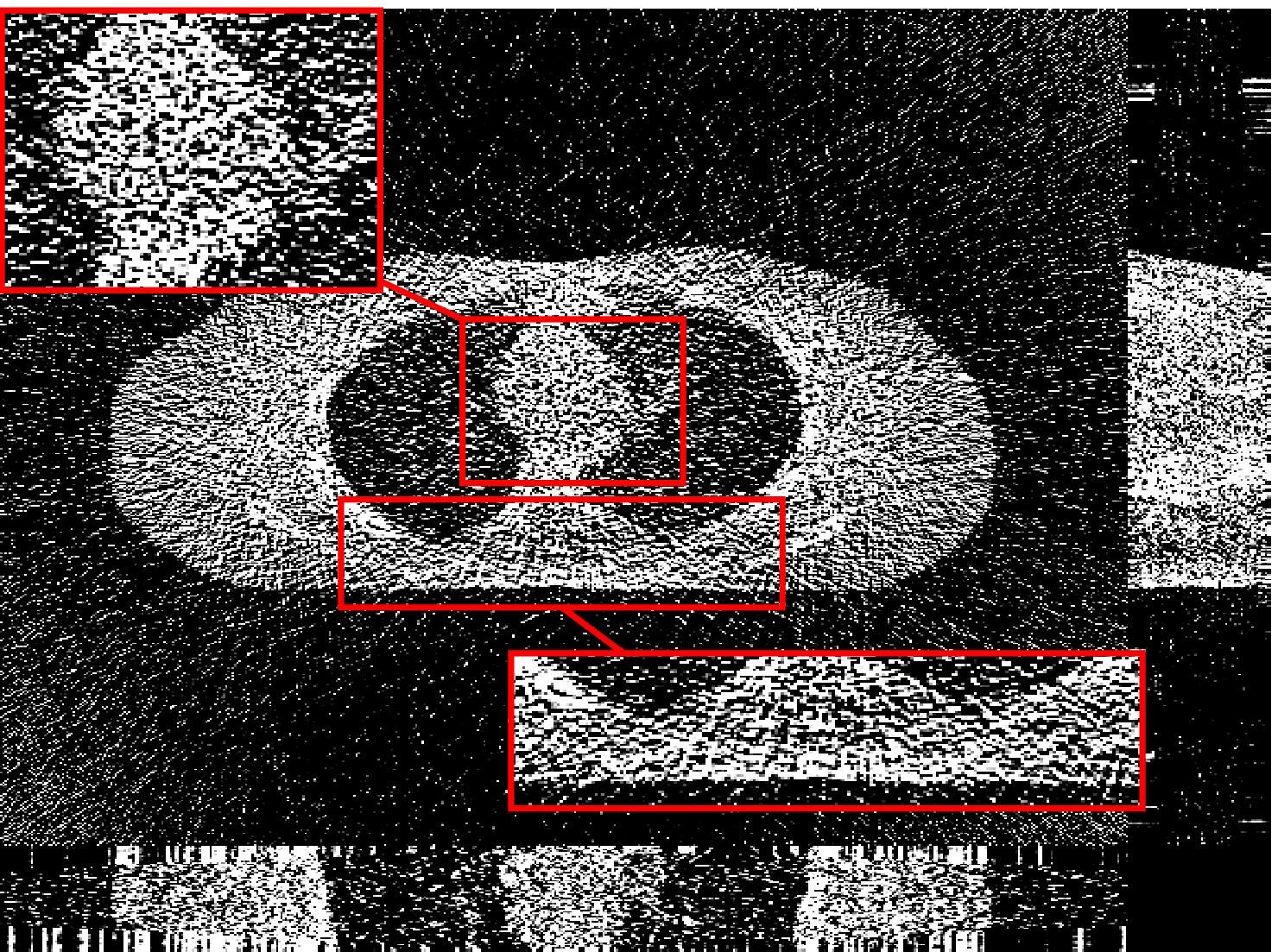}&
\includegraphics[width=.22\linewidth, height=.22\linewidth]{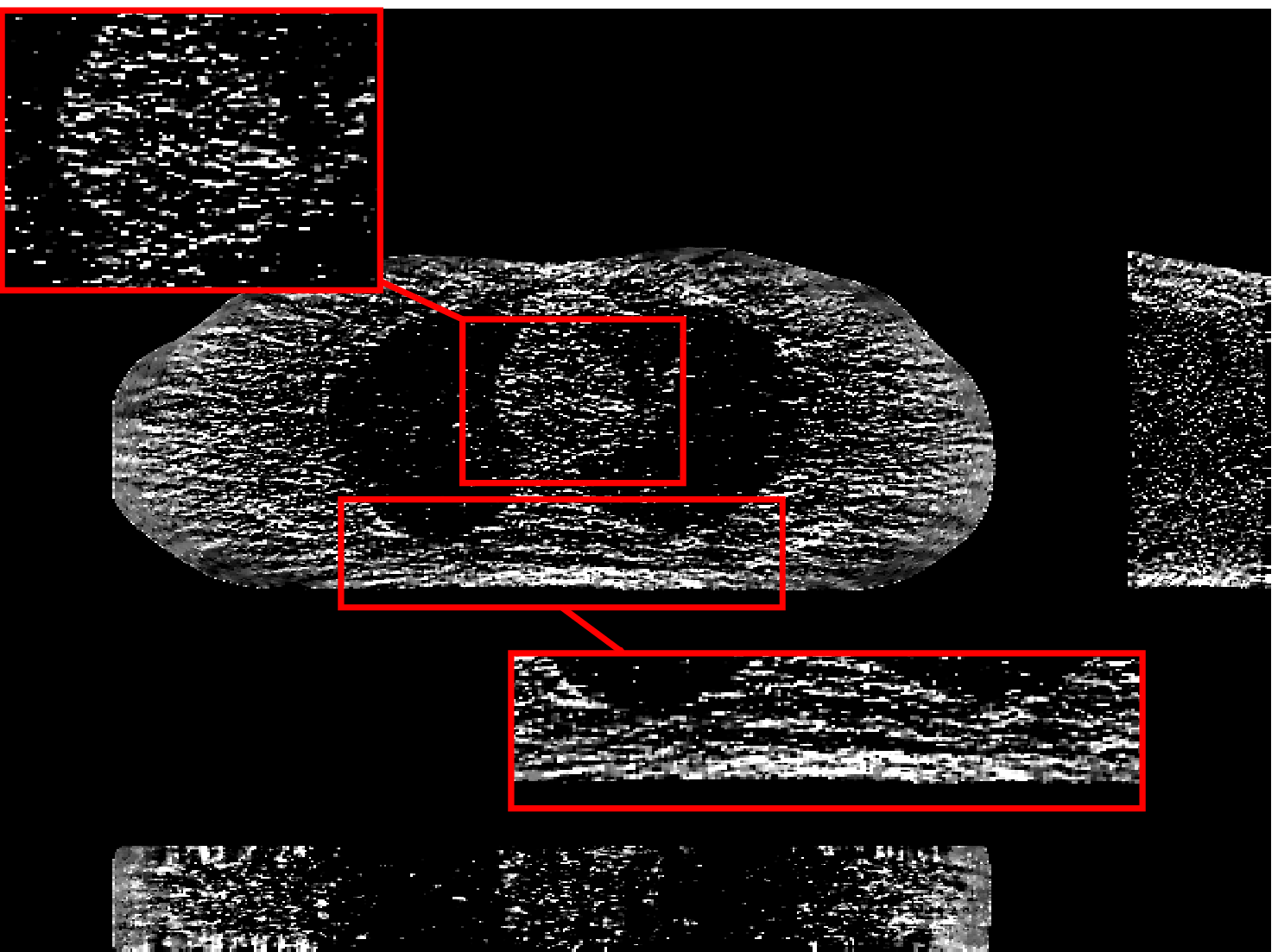}&
\includegraphics[width=.22\linewidth, height=.22\linewidth]{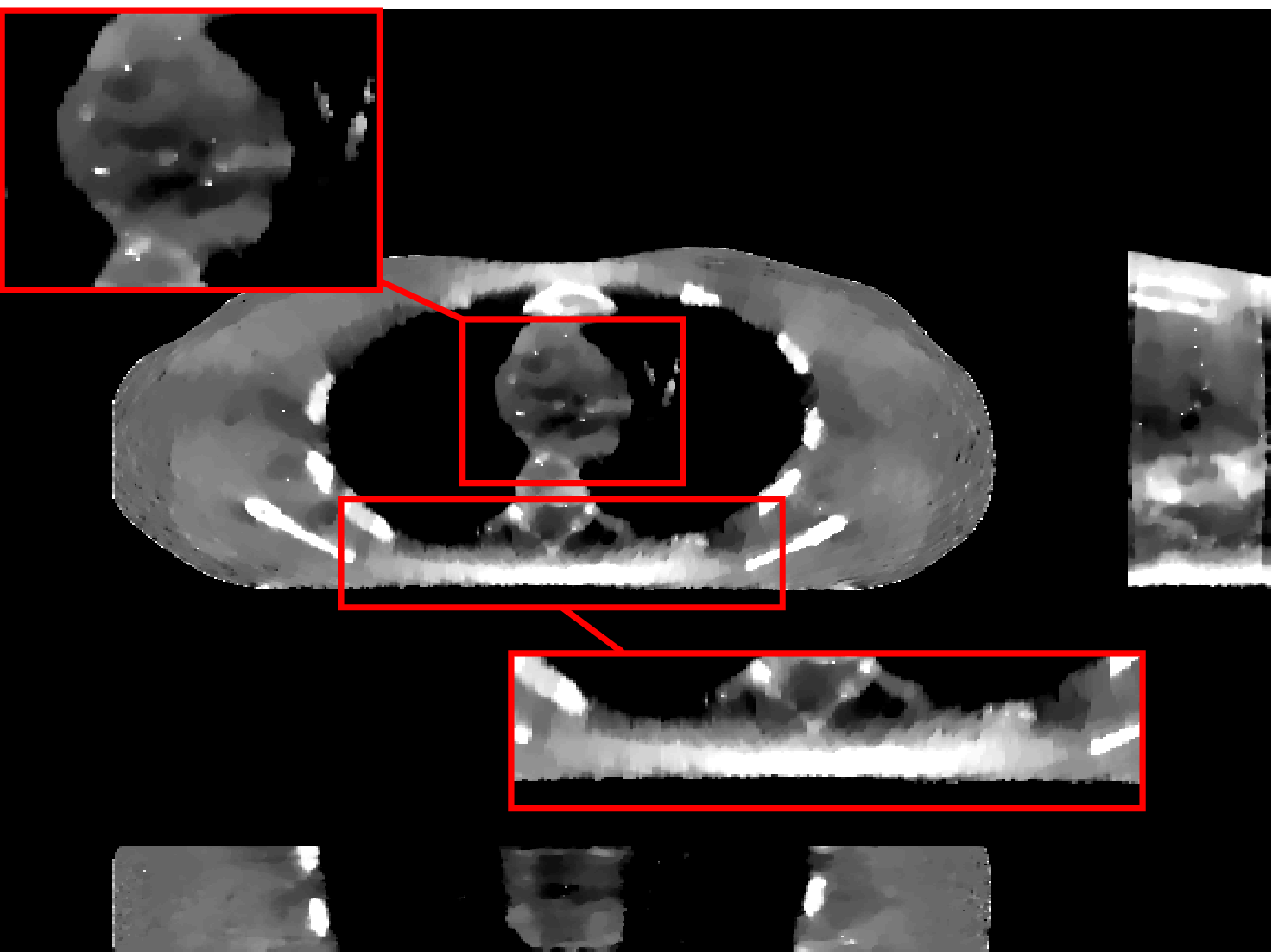}&
\includegraphics[width=.22\linewidth, height=.22\linewidth]{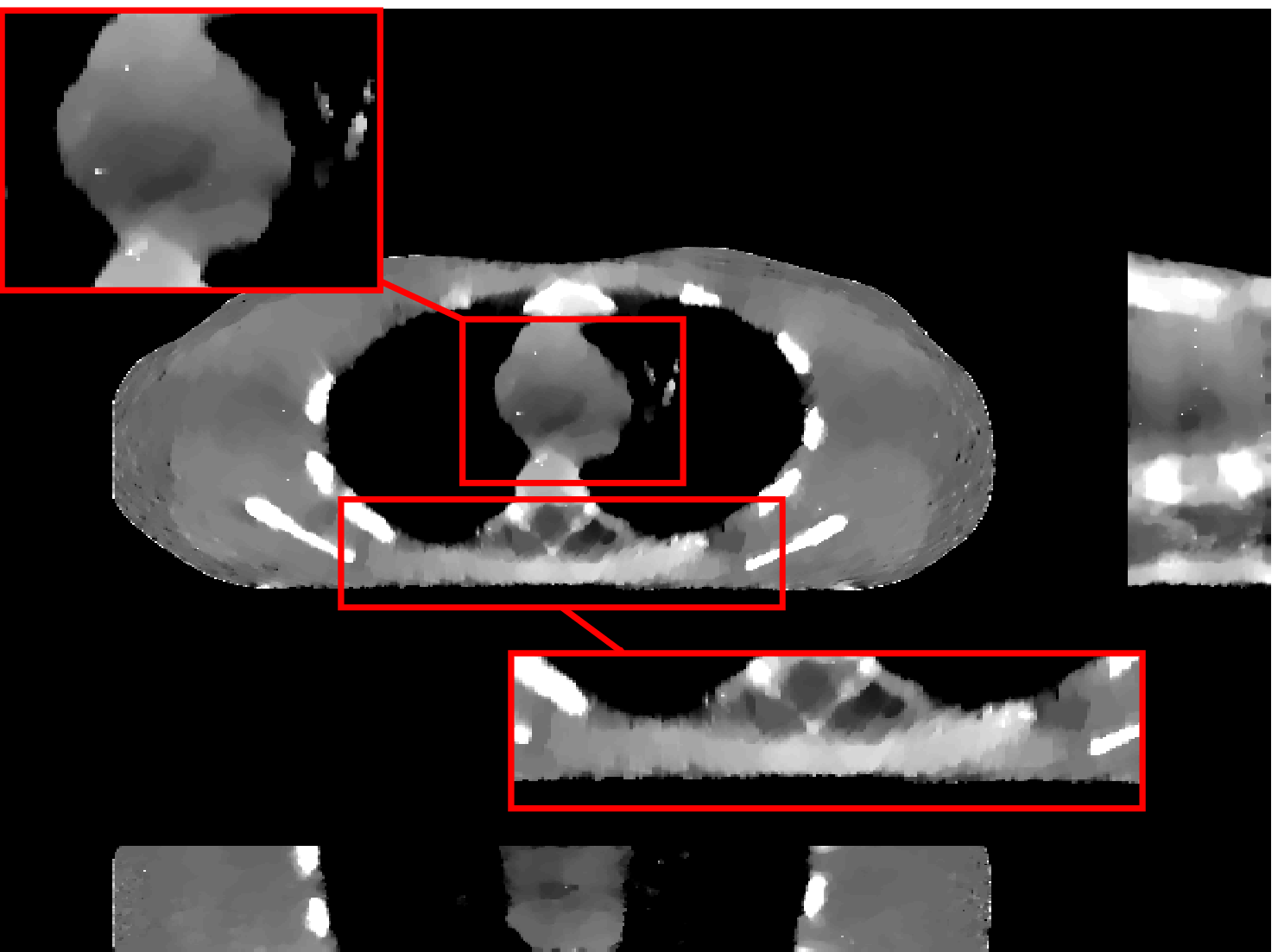}\\
&
{FBP}&
{PWLS}&
{SP}&
{MPG}
\end{tabular}
\caption{XCAT phantom reconstructed by FBP (first column), PWLS (second column), SP (third column) and the proposed MPG method (forth column) for dose of $I = 5 \times 10^3$ with electronic noise variance of $\sigma = 50^2$ (first row), $\sigma = 60^2$ (second row), $\sigma = 70^2$ (third row) and $\sigma = 100^2$ (forth row).
All images are displayed using a window of $[800,1200]$~HU.}
\label{fig:MPGS}
\end{center}
\end{figure*}

\begin{figure*}
\begin{center}
\begin{tabular}{c@{\hspace{2pt}}c@{\hspace{2pt}}c@{\hspace{2pt}}c@{\hspace{2pt}}c@{\hspace{2pt}}c}
\put(-20,50){SP}&
\includegraphics[width=.22\linewidth, height=.22\linewidth]{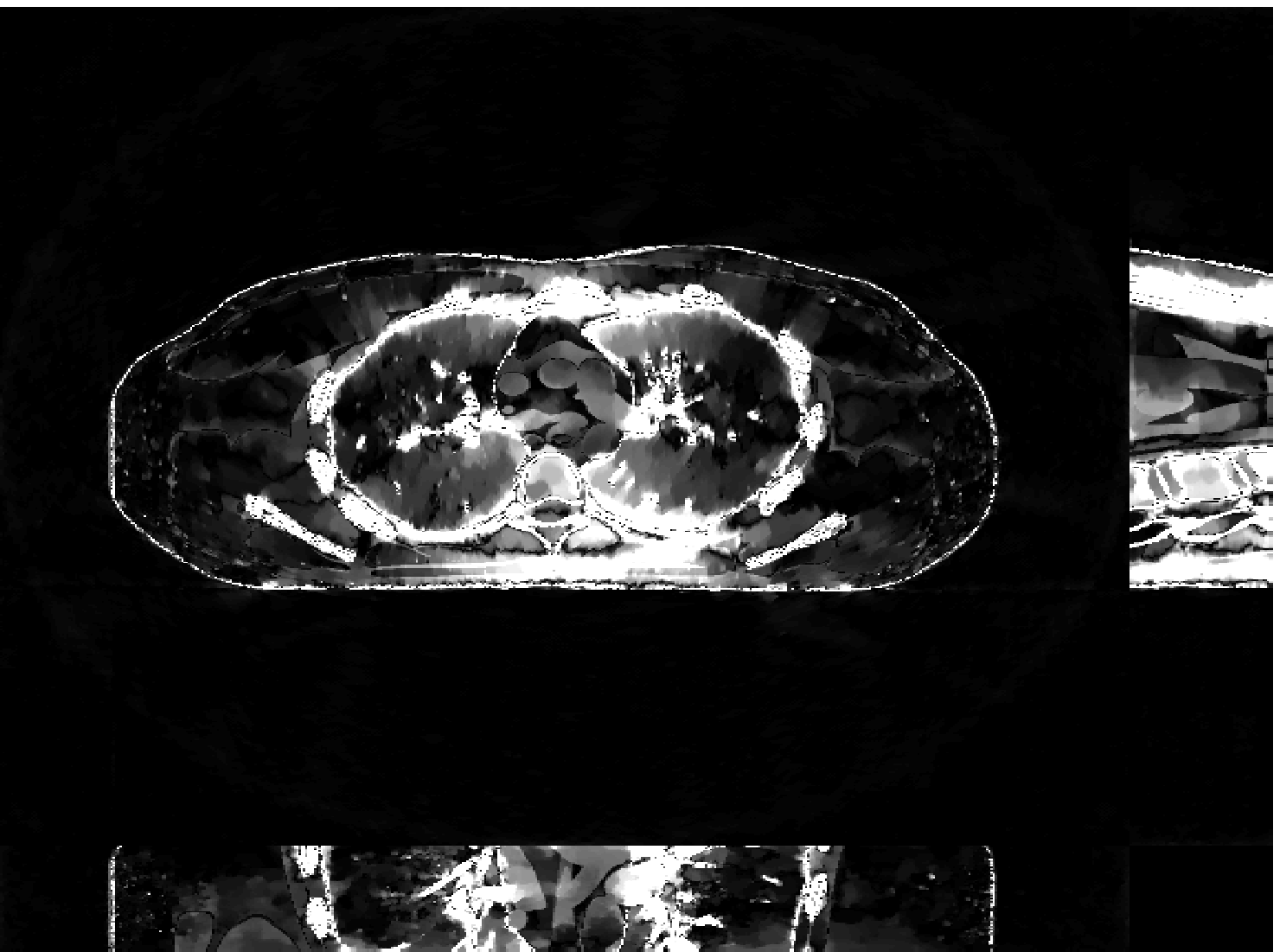}&
\includegraphics[width=.22\linewidth, height=.22\linewidth]{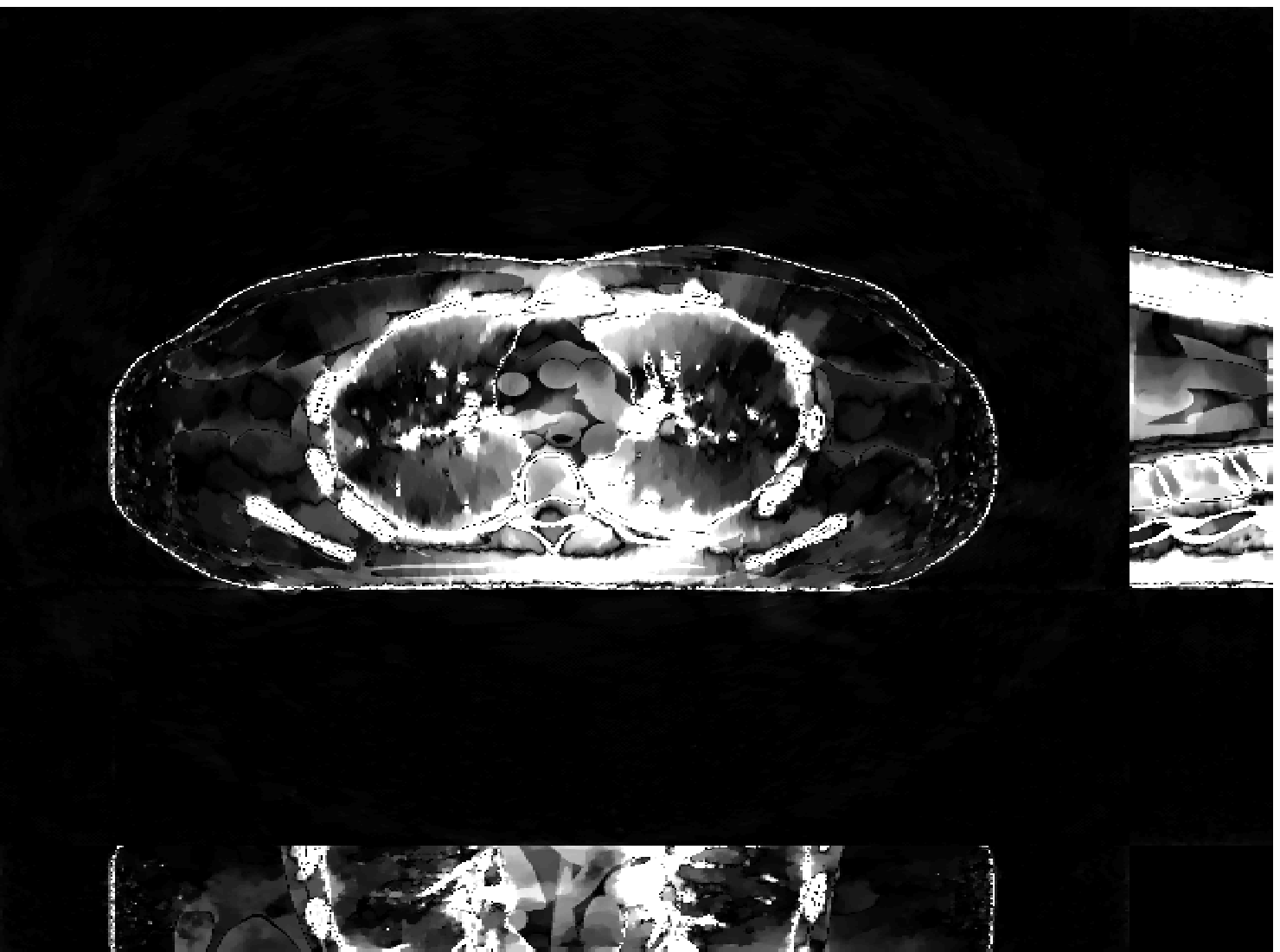}&
\includegraphics[width=.22\linewidth, height=.22\linewidth]{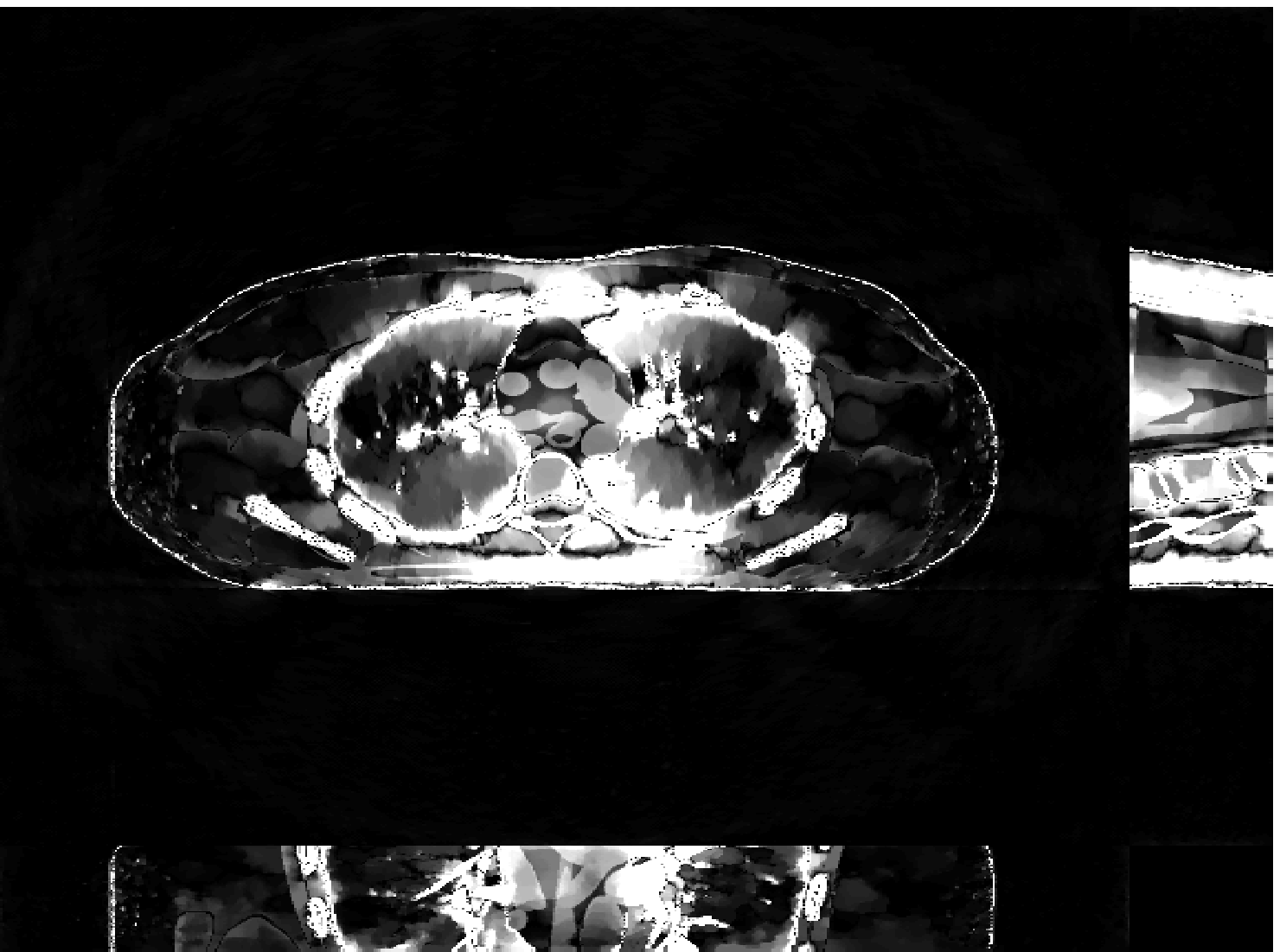}&
\includegraphics[width=.22\linewidth, height=.22\linewidth]{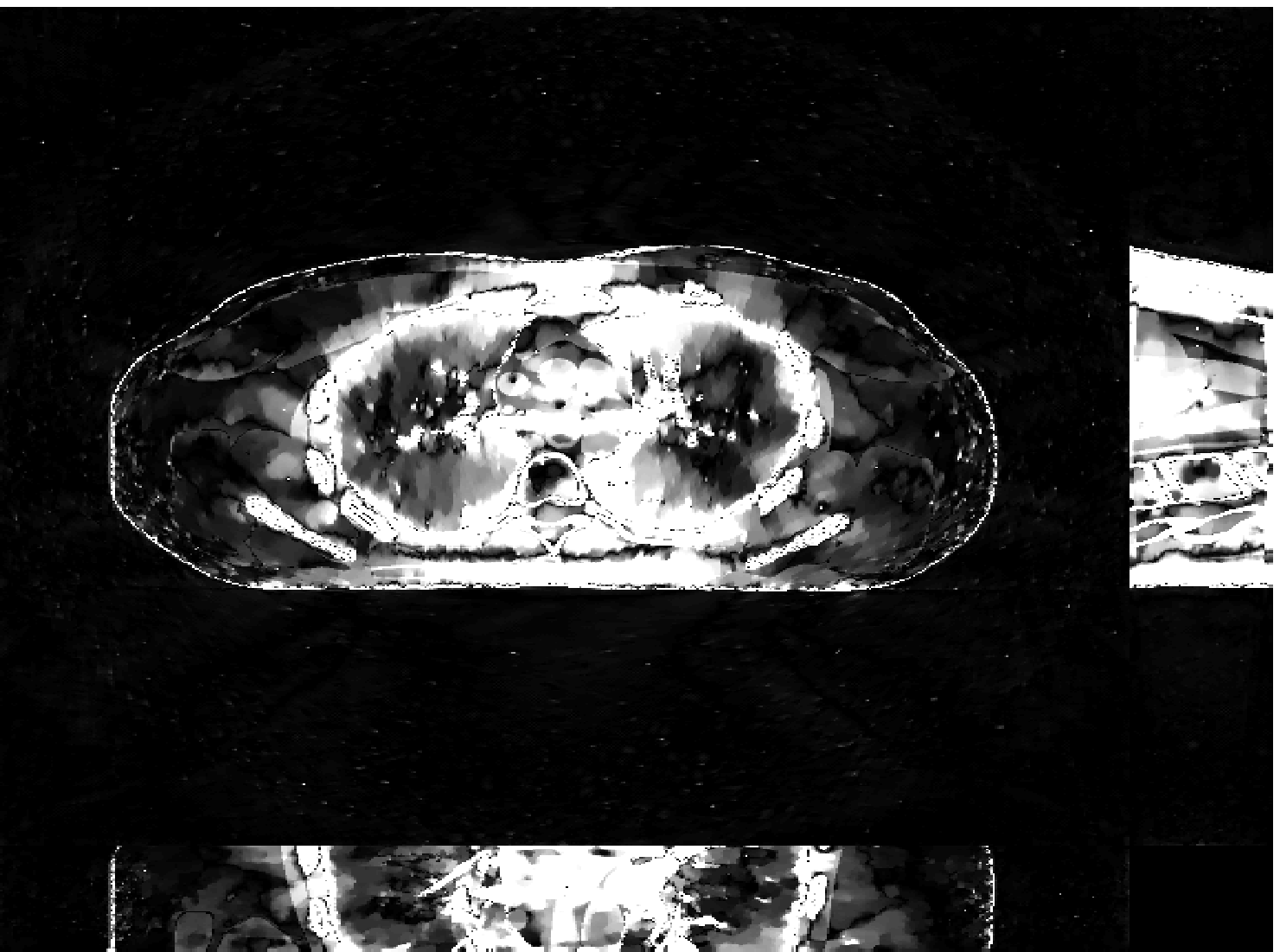}\\
\put(-20,50){MPG}&
\includegraphics[width=.22\linewidth, height=.22\linewidth]{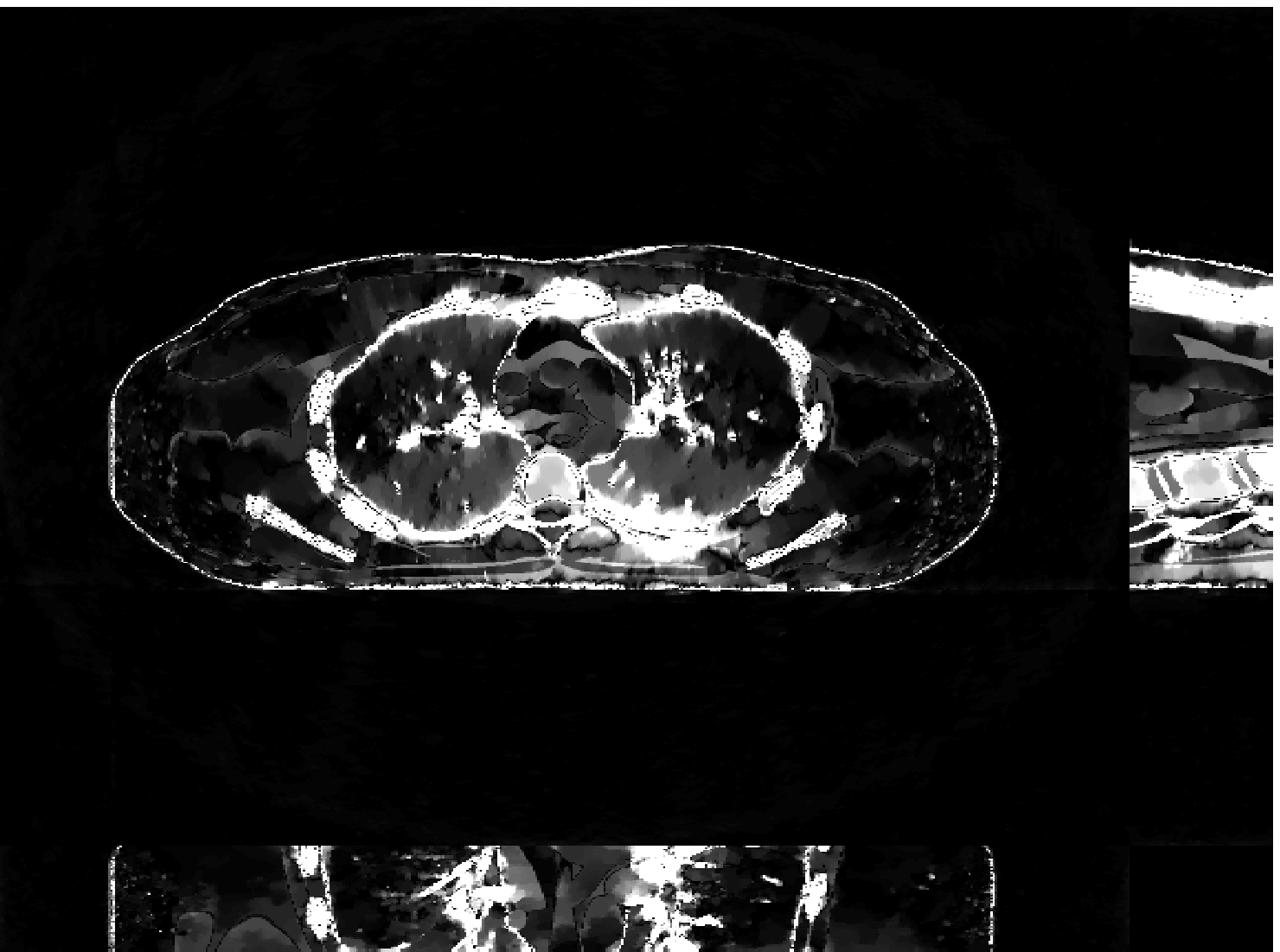}&
\includegraphics[width=.22\linewidth, height=.22\linewidth]{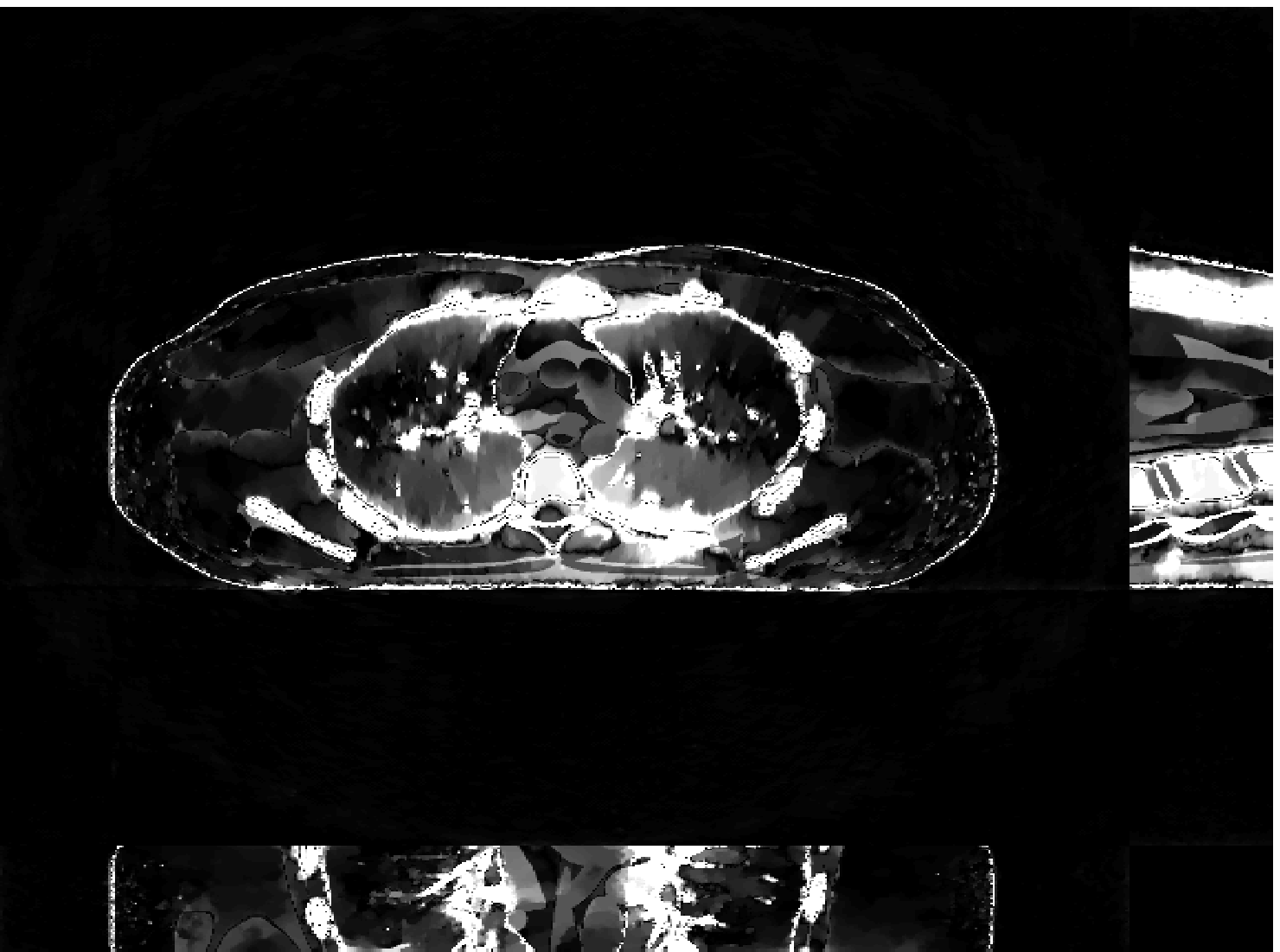}&
\includegraphics[width=.22\linewidth, height=.22\linewidth]{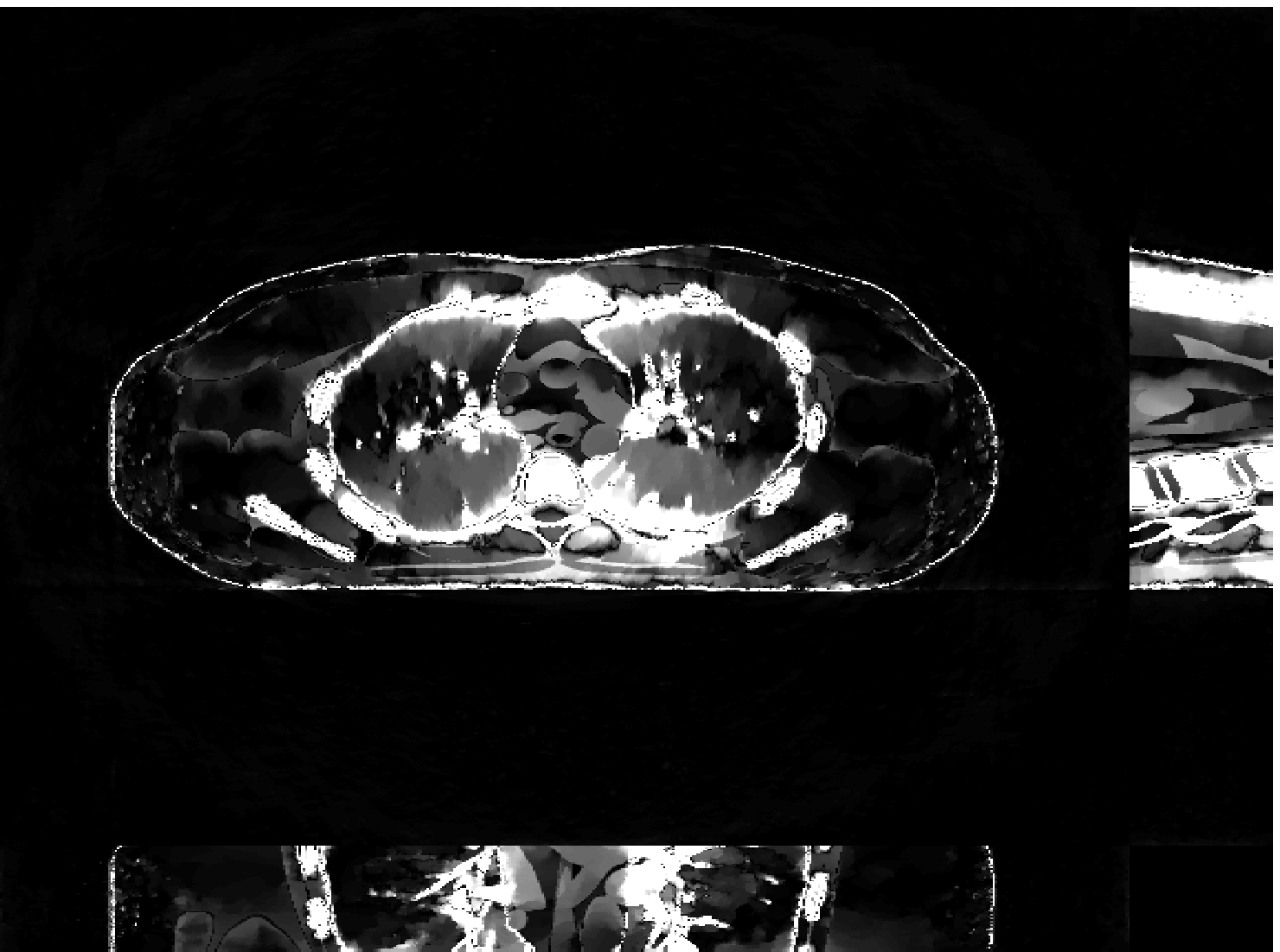}&
\includegraphics[width=.22\linewidth, height=.22\linewidth]{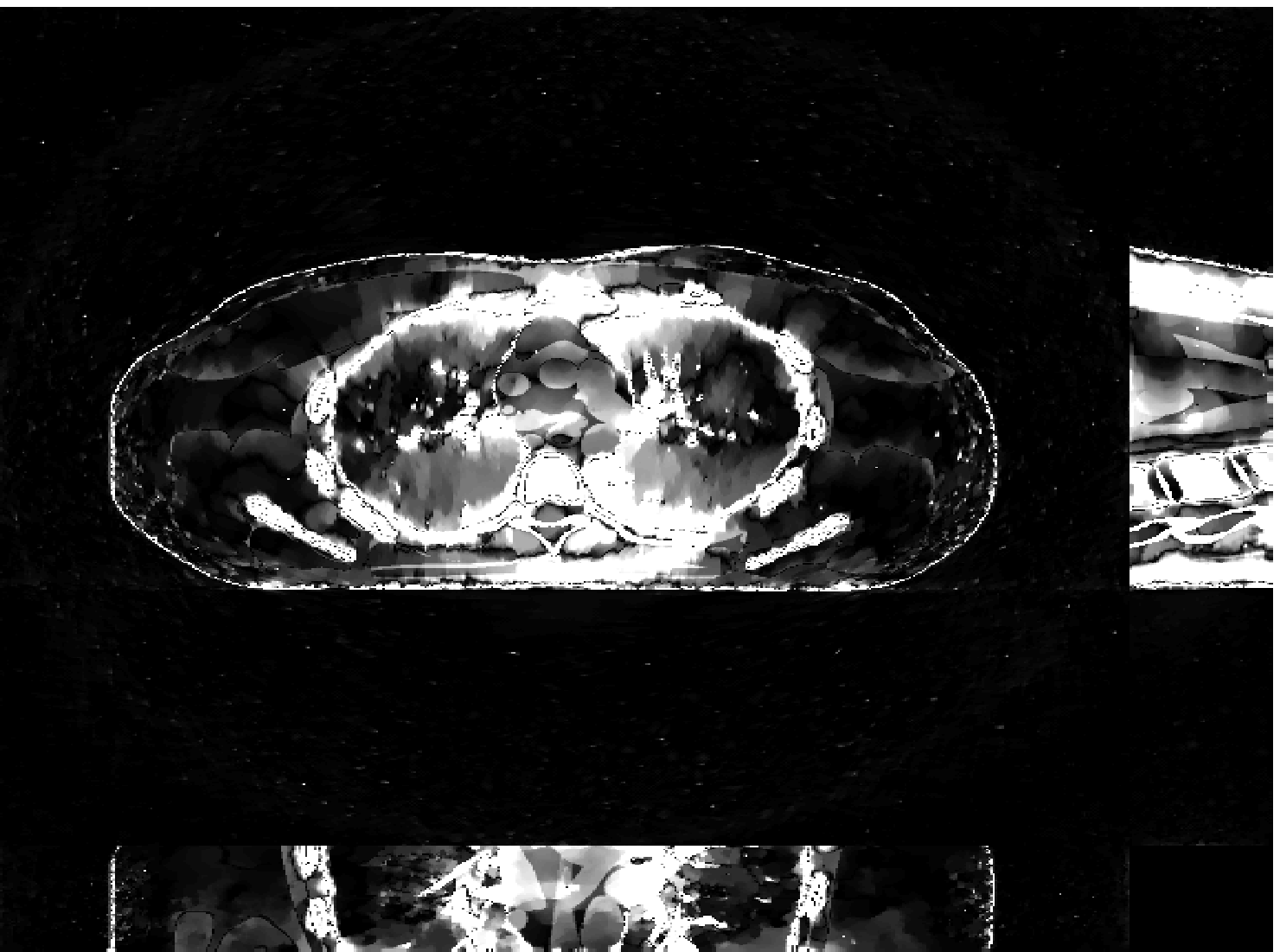}\\
&
$ \sigma^2=50^2$&
$ \sigma^2=60^2$&
$ \sigma^2=70^2$&
$ \sigma^2=100^2$
\end{tabular}
\caption {Absolute error images of reconstructions by SP (first row) and MPG (second row) for dose of $I_i =5 \times 10^3$ with variance of electronic noise  $\sigma^2 = 50^2$ (first column), $\sigma^2 = 60^2$ (second column), $\sigma^2 = 70^2$ (third column) and $\sigma^2 = 100^2$ (forth column).
All images are displayed using a window of $[0,100]$~HU.}
\label{fig:errorS}
\end{center}
\end{figure*}

Table~\ref{table:MPG} and Table~\ref{table:MPGS} show percentages of non-positive measurements, RMSE and SNR of images reconstructed by FBP with ramp filter, PWLS, SP and the proposed MPG method for different electronic noise variances at two dose levels of $I_i=10^4$ and $I_i=5\times10^3$, respectively.
PWLS images have smaller RMSEs and larger SNRs compared to FBP images as expected, while SP and MPG improves RMSE and SNR over PWLS.
MPG further decreases RMSE and increases SNR compared to SP.
Figure~\ref{fig:MPG} and Figure~\ref{fig:MPGS} show images reconstructed by FBP, PWLS, SP and the proposed MPG method for electronic noise variance of $\{50,60,70,100\}$ at two dose levels of $I_i=10^4$ and $I_i=5\times10^3$, respectively.
The PWLS method decreases noise and removes streak artifacts from FBP images, while the SP and MPG method further improve image quality compared to PWLS initializations.
As electronic noise variance $\sigma^2$ becomes larger, the FBP images have increased noise and artifacts; the PWLS method decreases noise and artifacts but introduces bias, especially at the center region; the SP and MPG method significantly improves image quality compared with FBP and PWLS; the MPG method further decreases bias compared with SP.
For small electronic noise variance cases, \emph{i.e.}, $\sigma^2 = \{20,30,40\}$, the SP and MPG images are visually similar.
Figure~\ref{fig:error} and Figure~\ref{fig:errorS} show error images of reconstructions by SP and MPG.
The MPG method better estimates the true image compared to  SP, particularly for large electronic noise variance cases.

\begin{table*}
\centering
\begin{tabular}{|c|c|c|c|c|c|c|c|c|c|c|}
\cline{1-9}
 \multicolumn{2}{|c|}{$I_i=10^4,~\sigma^2$}        &$20^2$        &$30^2$        &$40^2$        &$50^2$        &$60^2$        &$70^2$        &$100^2$   \\\cline{1-9}
\multicolumn{2}{|c|}{Non-positive Percentage ($\%$)}&$0.5$        &$1.0$         &$1.5$         &$2.0$         &$2.5$         &$3.3$         &$4.6$   \\\hline
\multirow{3}*{RMSE       } &FBP                    &$160.1$       &$215.8$       &$263.1$       &$303.7$       &$340.3$       &$372.9$       &$458.9$   \\\cline{2-9}
                       &PWLS                       &$52.3$        &$55.7$        &$60.6$        &$65.8$        &$70.2$        &$76.3$        &$133.3$   \\\cline{2-9}
                       &SP                         &$52.0$        &$53.3$        &$55.5$        &$58.0$        &$60.3$        &$62.1$        &$69.9$    \\\cline{2-9}
                       &MPG                        &$\bold{51.8}$ &$\bold{53.0}$ &$\bold{55.0}$ &$\bold{57.0}$ &$\bold{58.8}$ &$\bold{60.5}$ &$\bold{64.1}$     \\\hline
\multirow{3}*{SNR}     &FBP                        &$8.3$         &$5.7$         &$4.0$         &$2.7$         &$1.7$         &$1.0$         &$-0.9$   \\\cline{2-9}
                       &PWLS                       &$18.0$        &$17.5$        &$16.7$       &$16.0$         &$15.4$       &$14.7$         &$9.9 $    \\\cline{2-9}
                       &SP                         &$18.1$        &$17.8$        &$17.5$       &$17.1$         &$16.8$       &$16.5$         &$15.5$   \\\cline{2-9}
                       &MPG                        &$\bold{18.1}$ &$\bold{17.9}$ &$\bold{17.6}$&$\bold{17.3}$&$\bold{17.0}$&$\bold{16.7}$&$\bold{16.2}$   \\\hline
\end{tabular}
\caption{Percentages of non-positive values in measurements, RMSE and SNR of images reconstructed by FBP, PWLS, SP and MPG with different levels of electronic noise for dose of $I_i=10^4$.}
\label{table:MPG}
\end{table*}
\begin{table}
\begin{tabular}{|c|c|c|c|c|c|c|c|c|c|c|}
\cline{1-6}
 \multicolumn{2}{|c|}{$I_i=5\times10^3,~\sigma^2$} &$50^2$         &$60^2$          &$70^2$          &$100^2$        \\\cline{1-6}
\multicolumn{2}{|c|}{Non-positive Percentage ($\%$)}&$4.6$         &$5.5$           &$6.3$           &$8.6$        \\\hline
\multirow{3}*{RMSE    }&FBP                        &$408.3$        &$453.5$         &$493.4$         &$592.2$      \\\cline{2-6}
                       &PWLS                       &$126.2$        &$145.1$         &$162.3$         &$186.9$      \\\cline{2-6}
                       &SP                         &$66.8$         &$70.7$          &$73.7$          &$83.0$       \\\cline{2-6}
                       &MPG                        &$\bold{64.4}$  &$\bold{67.0}$   &$\bold{69.3}$   &$\bold{75.5}$       \\\hline
\multirow{3}*{SNR}     &FBP                        &$0.2$          &$-0.8$          &$-1.5$          &$-3.1$          \\\cline{2-6}
                       &PWLS                       &$10.3$         &$9.1$           &$8.2$           &$6.9$        \\\cline{2-6}
                       &SP                         &$15.9$         &$15.4$          &$15.0$          &$14.0$        \\\cline{2-6}
                       &MPG                        &$\bold{16.2}$  &$\bold{15.9}$   &$\bold{15.6}$   &$\bold{14.8}$       \\\hline
\end{tabular}
\caption{Percentages of non-positive values in measurements, RMSE and SNR of images reconstructed by FBP, PWLS, SP and MPG with different levels of electronic noise for dose of $I_i=5\times10^3$.}
\label{table:MPGS}
\end{table}
\subsection{Synthetic Clinical Data Results}

\begin{figure*}
\begin{center}
\begin{tabular}{c@{\hspace{2pt}}c@{\hspace{2pt}}c@{\hspace{2pt}}c@{\hspace{2pt}}c@{\hspace{2pt}}c@{\hspace{2pt}}c@{\hspace{2pt}}c@{\hspace{2pt}}c@{\hspace{2pt}}c}
\put(-40,50){$\sigma^2=20^2$}&
\includegraphics[width=.22\linewidth, height=.22\linewidth]{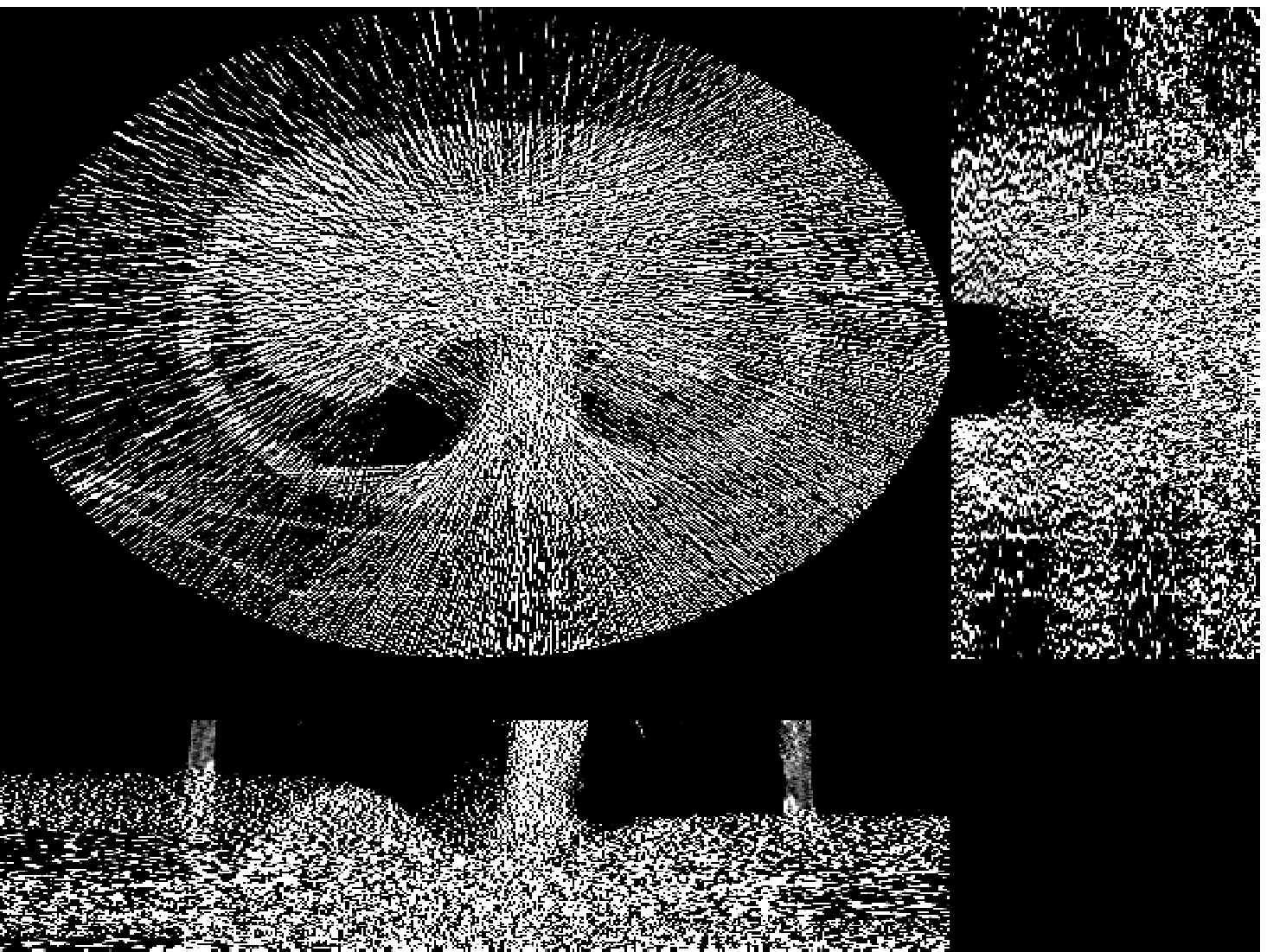}&
\includegraphics[width=.22\linewidth, height=.22\linewidth]{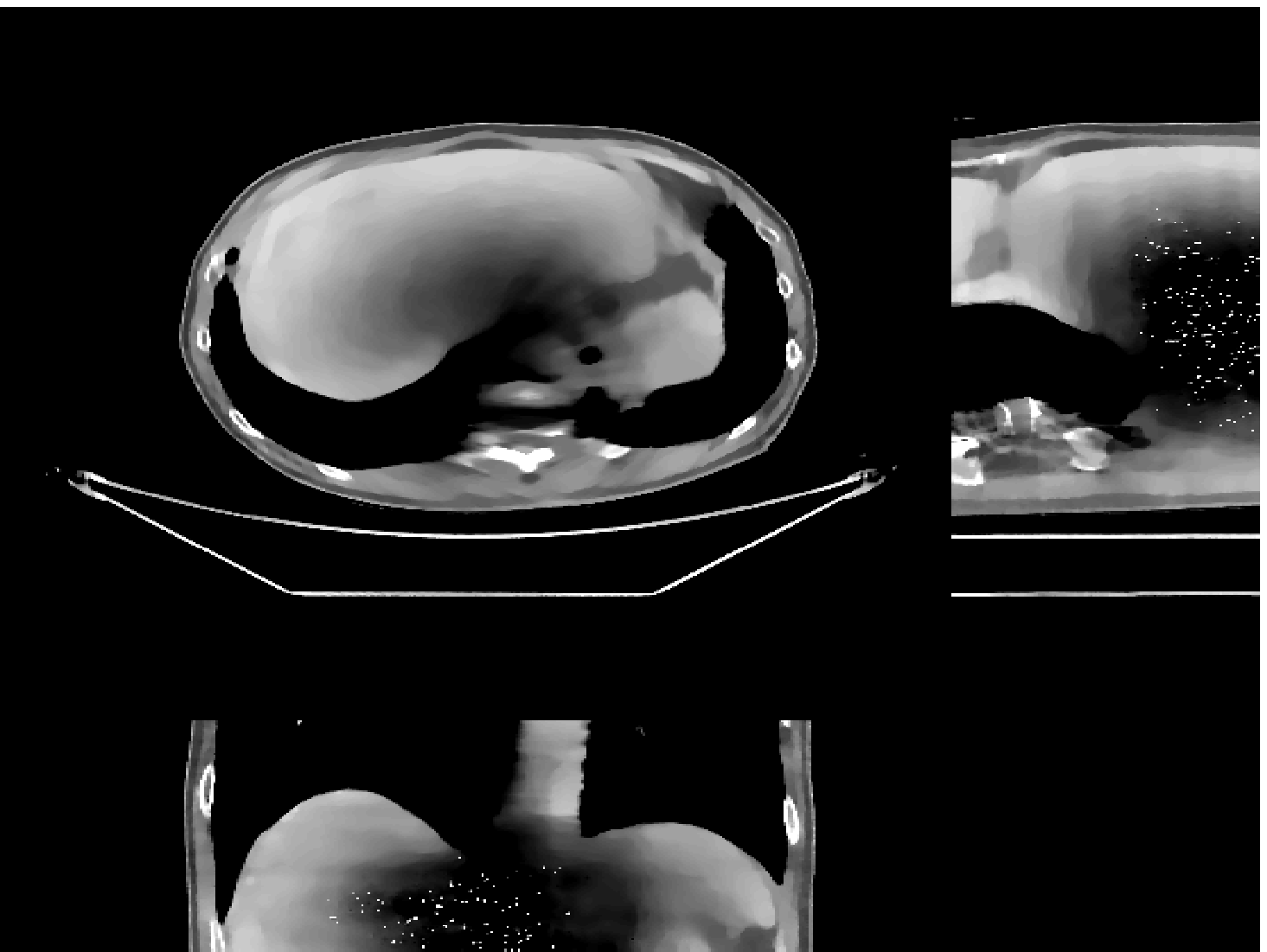}&
\includegraphics[width=.22\linewidth, height=.22\linewidth]{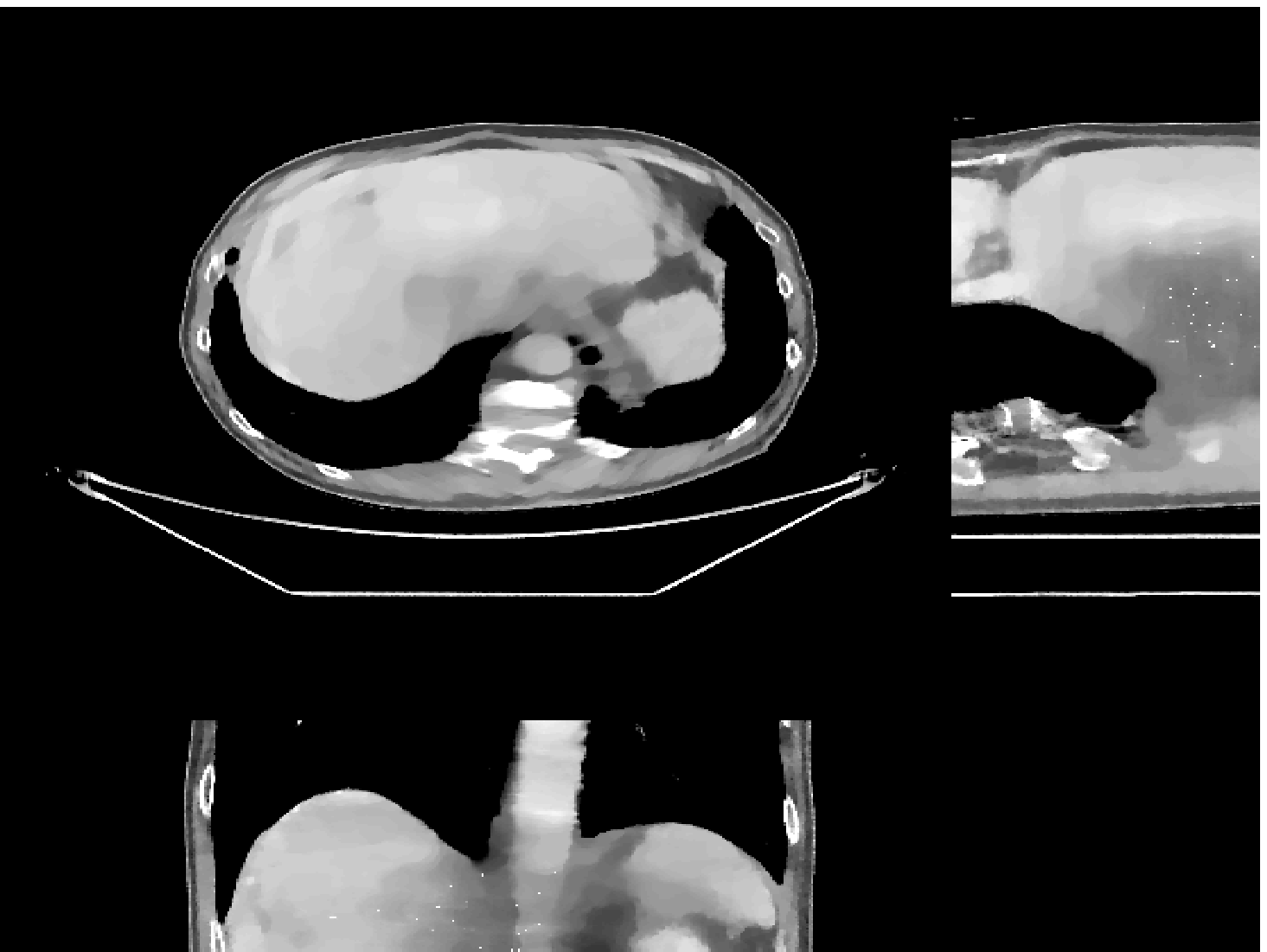}&
\includegraphics[width=.22\linewidth, height=.22\linewidth]{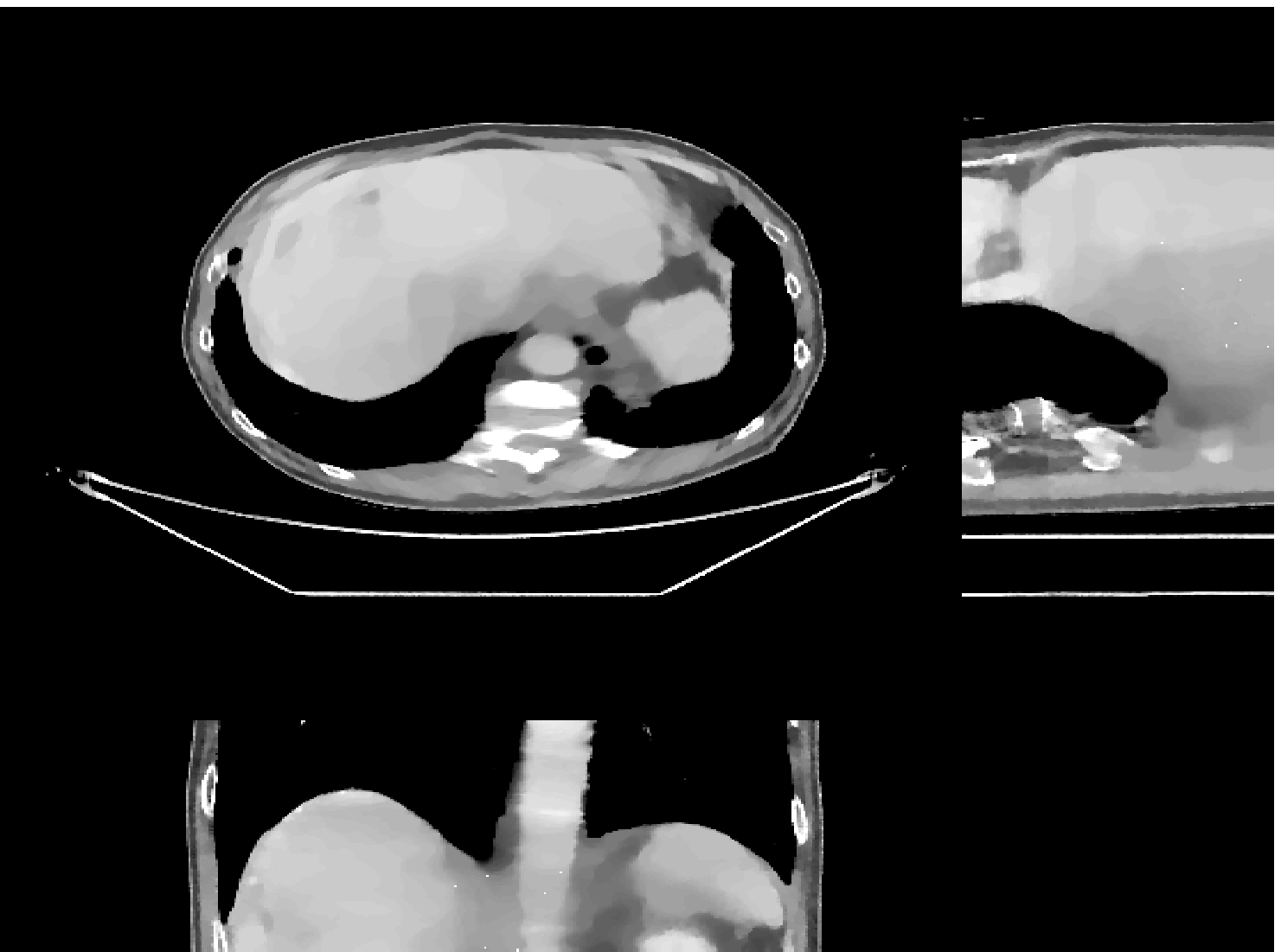}\\
\put(-40,50){$\sigma^2=30^2$}&
\includegraphics[width=.22\linewidth, height=.22\linewidth]{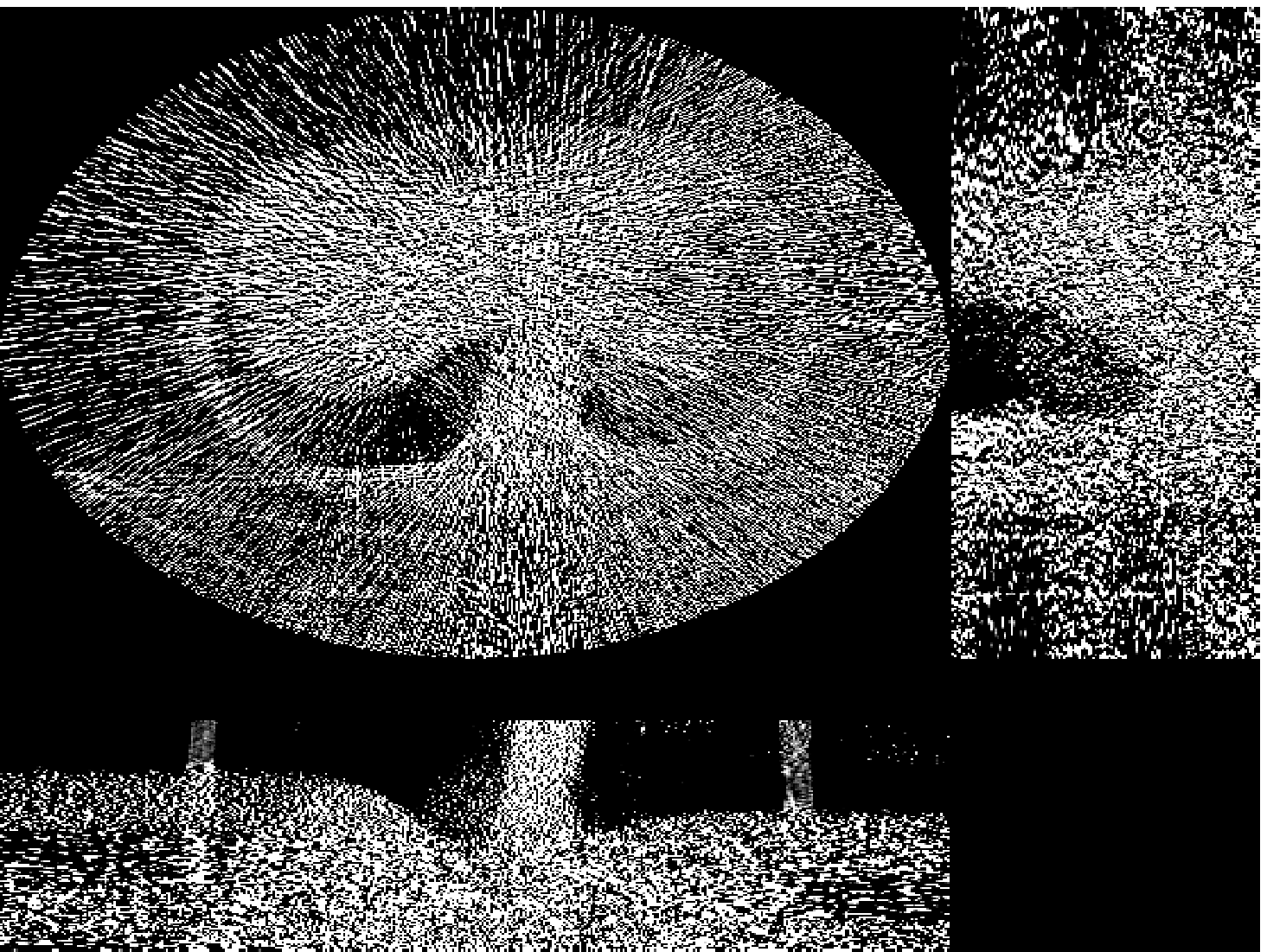}&
\includegraphics[width=.22\linewidth, height=.22\linewidth]{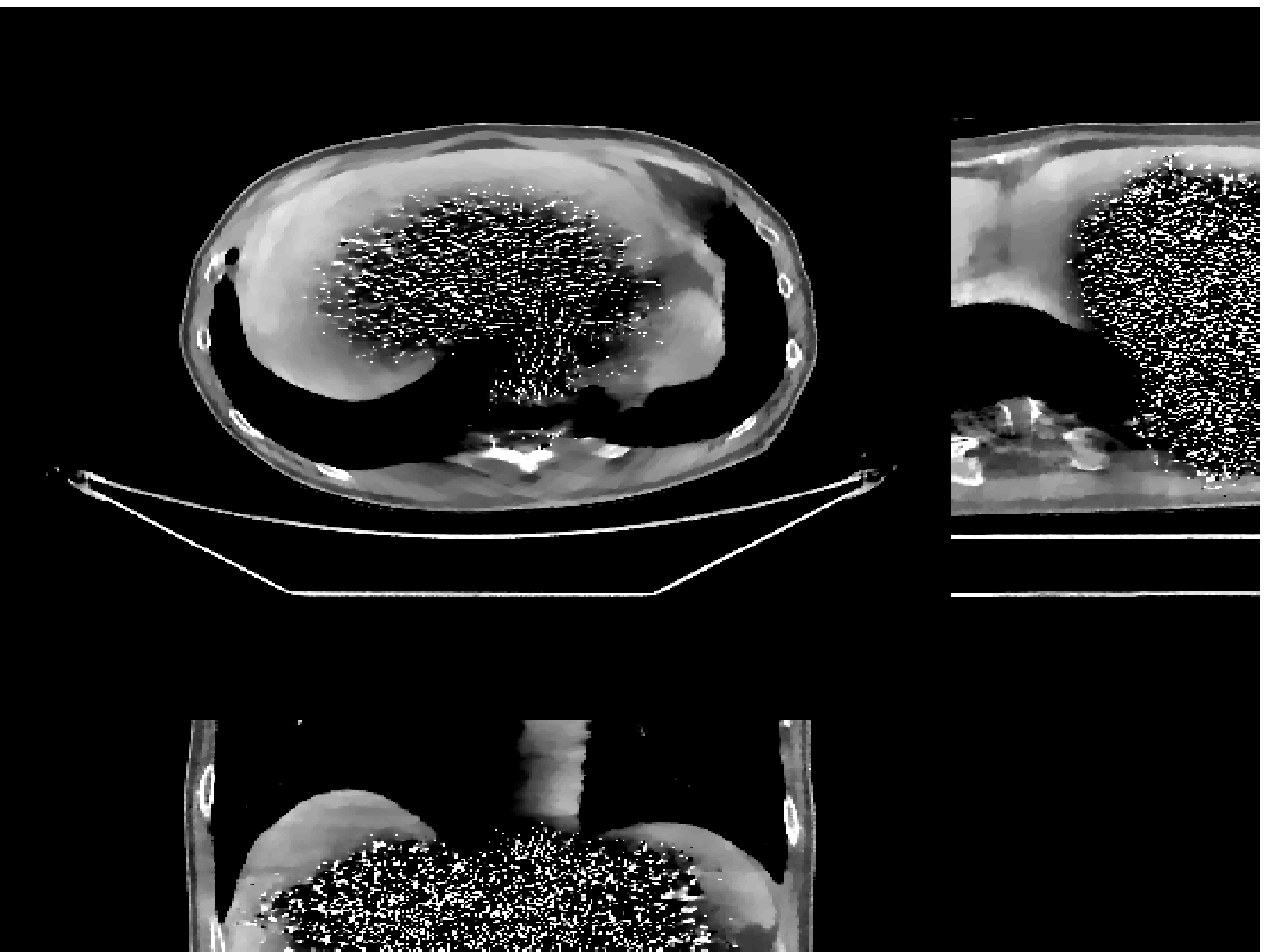}&
\includegraphics[width=.22\linewidth, height=.22\linewidth]{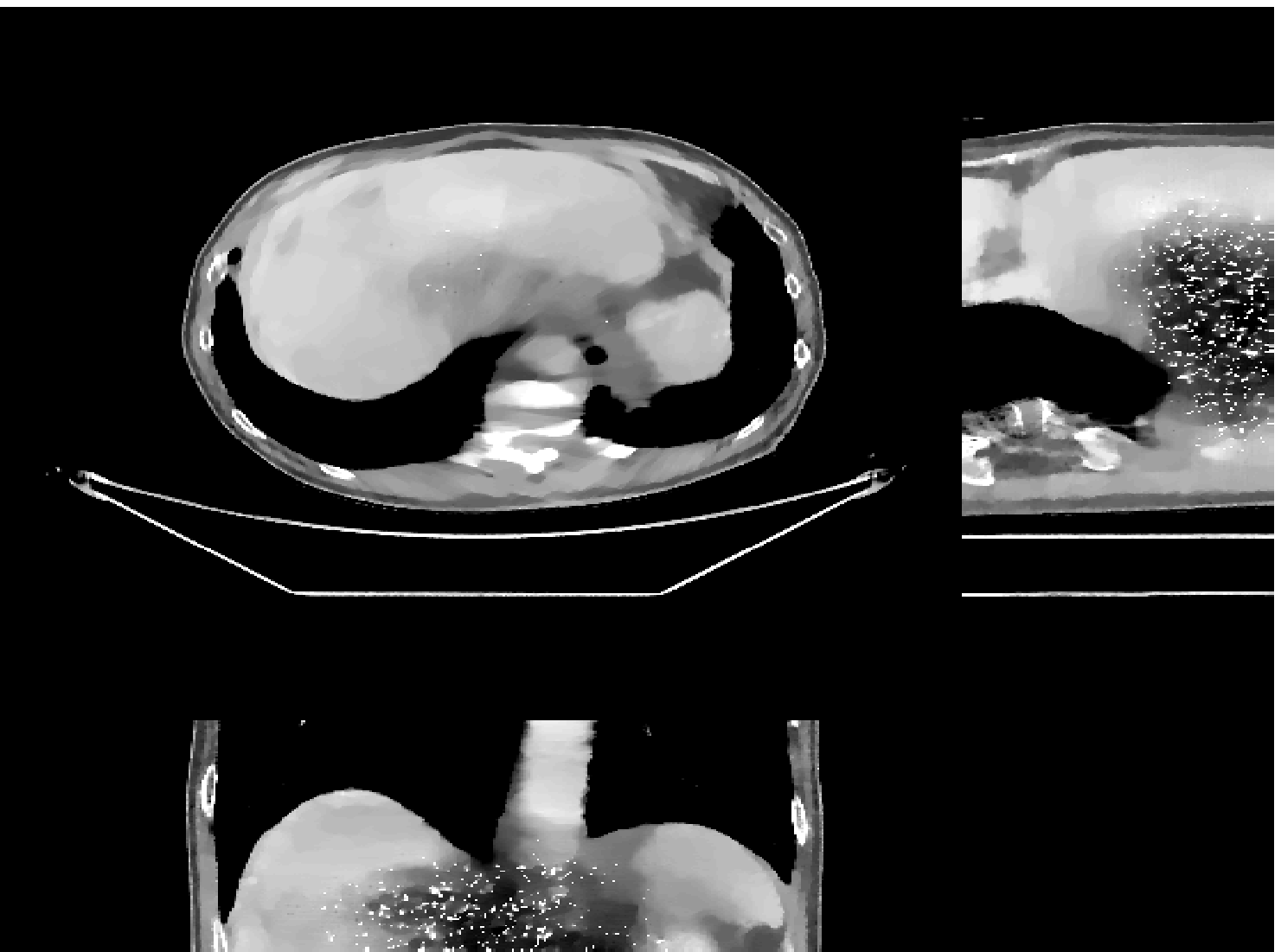}&
\includegraphics[width=.22\linewidth, height=.22\linewidth]{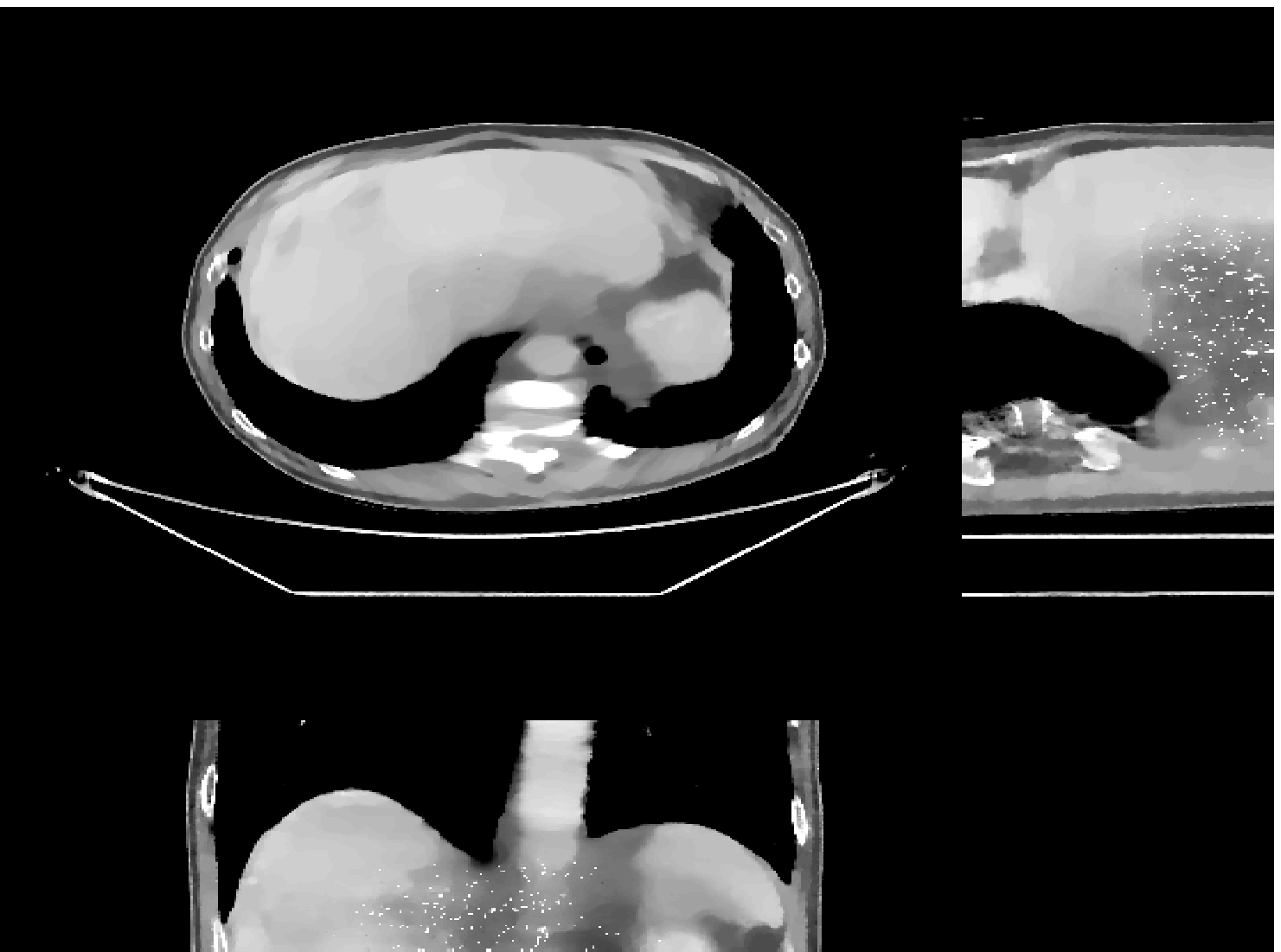}\\
\put(-40,50){$\sigma^2=40^2$}&
\includegraphics[width=.22\linewidth, height=.22\linewidth]{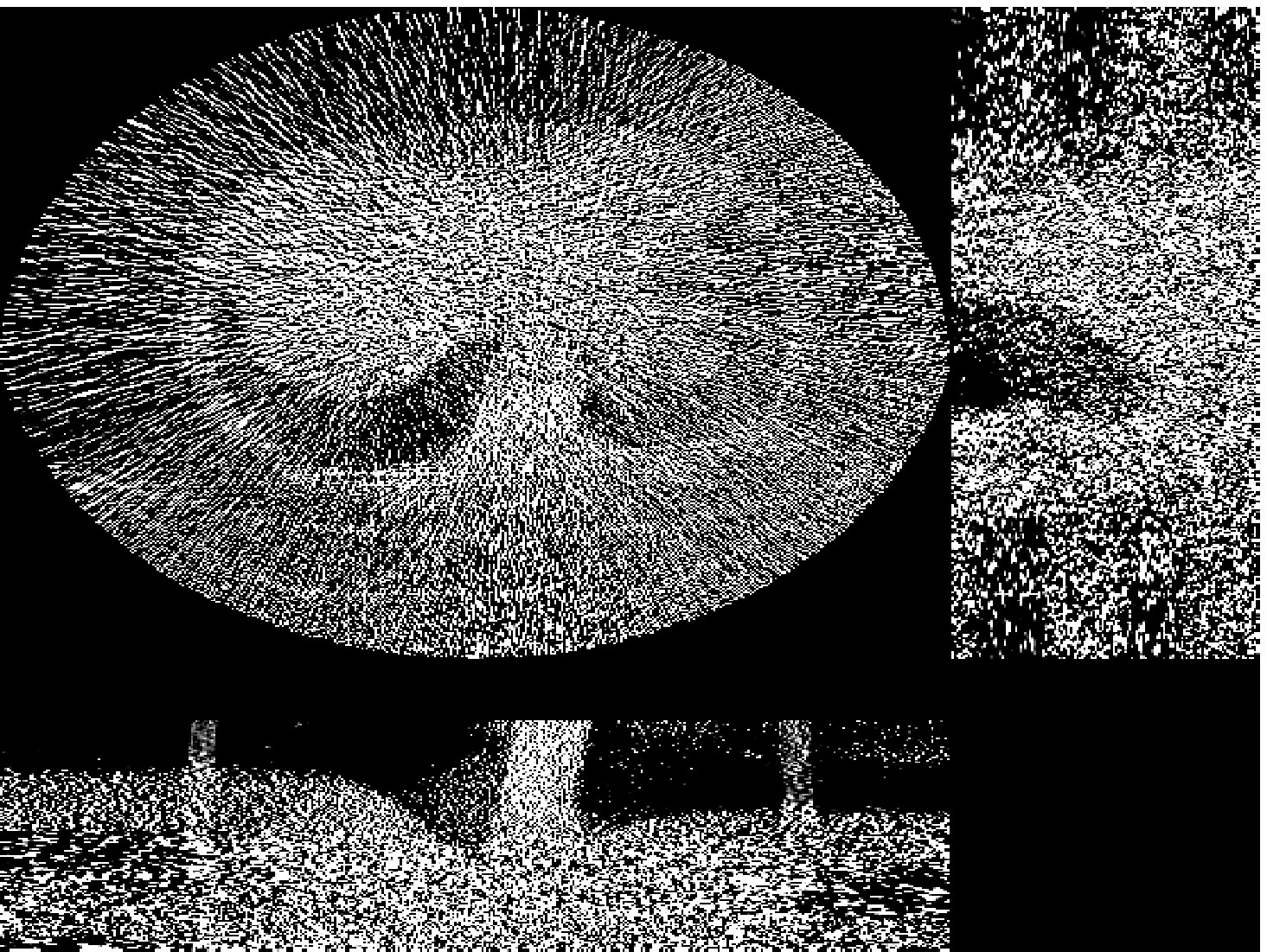}&
\includegraphics[width=.22\linewidth, height=.22\linewidth]{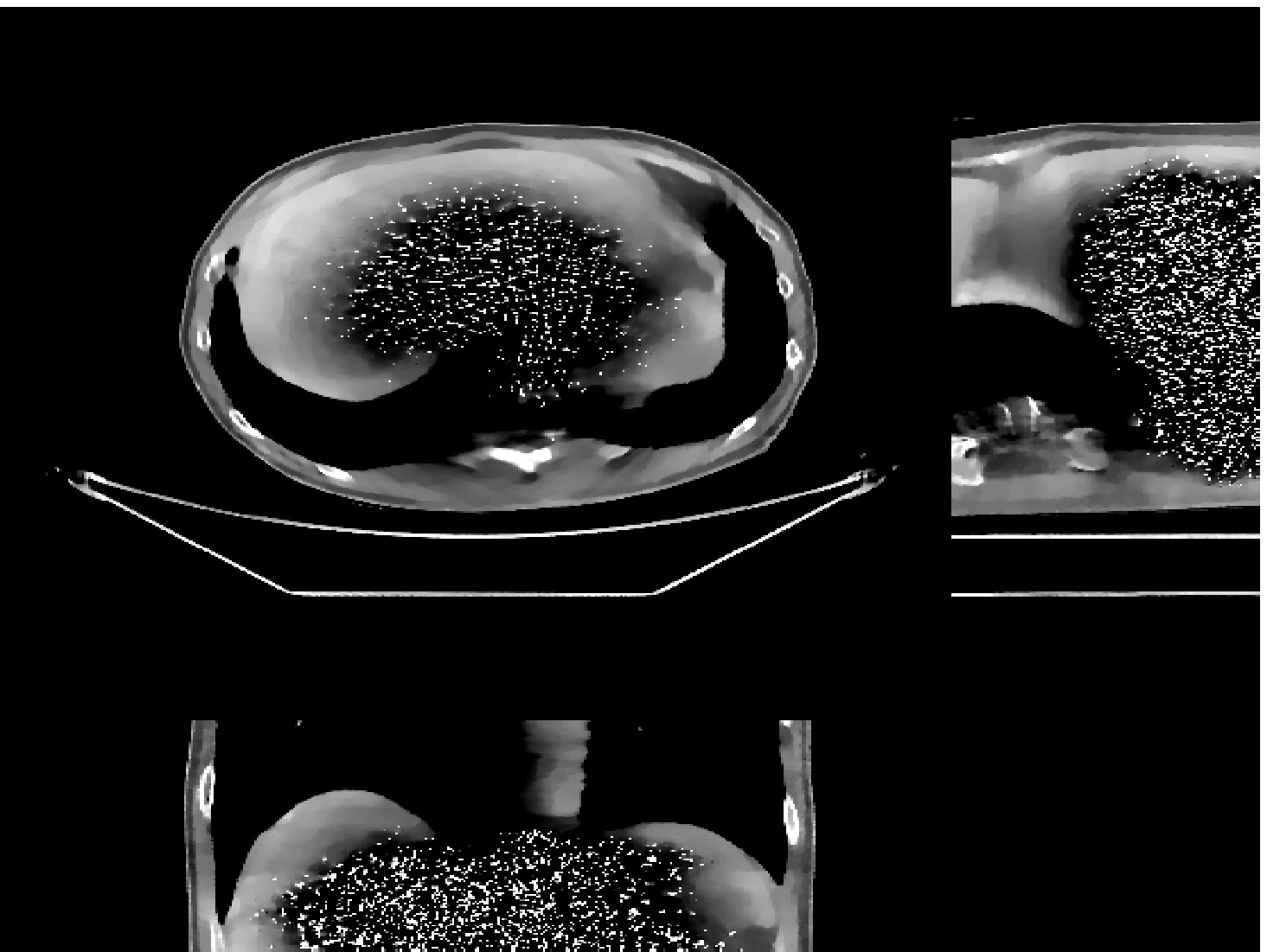}&
\includegraphics[width=.22\linewidth, height=.22\linewidth]{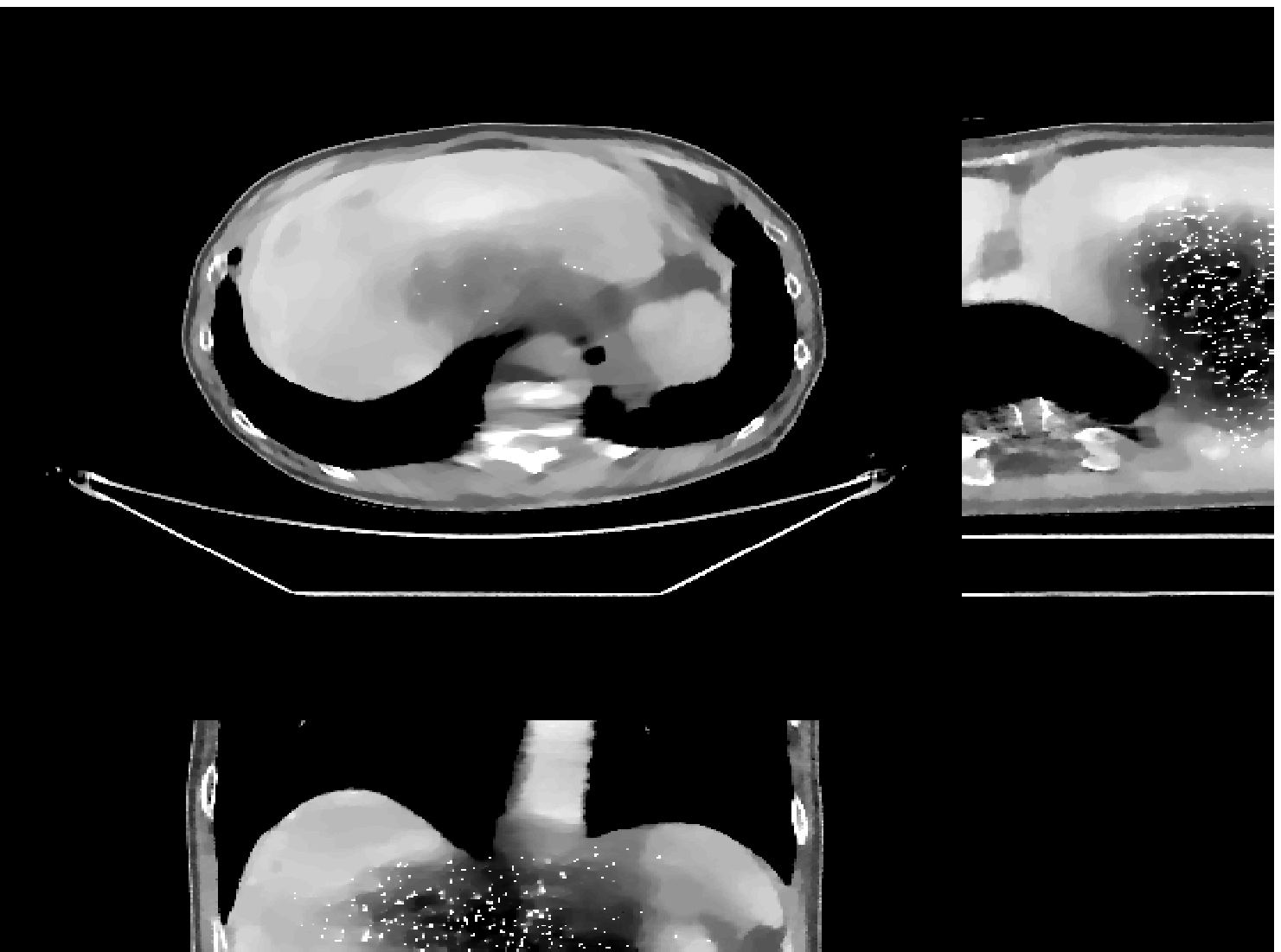}&
\includegraphics[width=.22\linewidth, height=.22\linewidth]{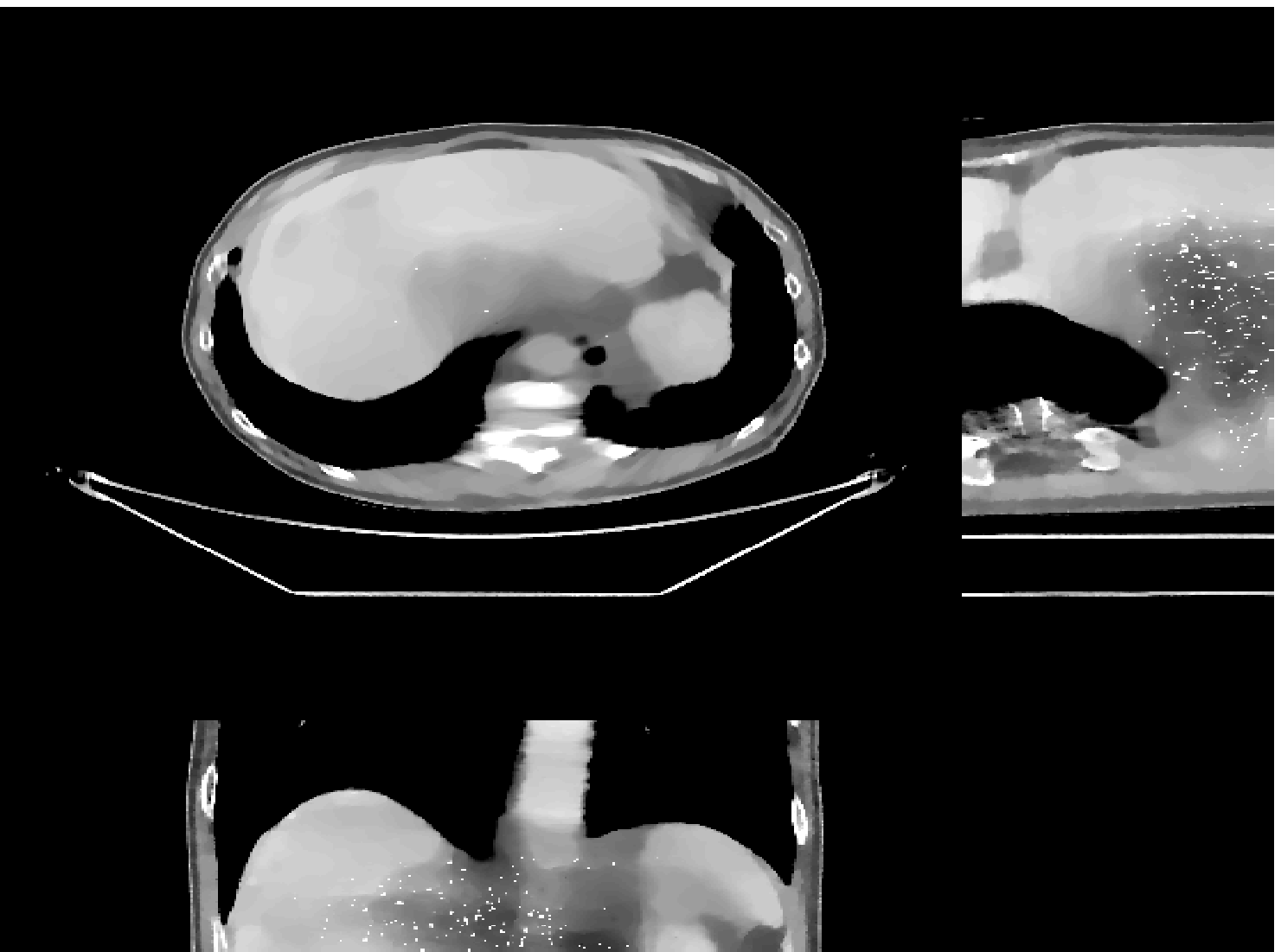}\\
\put(-40,50){$\sigma^2=50^2$}&
\includegraphics[width=.22\linewidth, height=.22\linewidth]{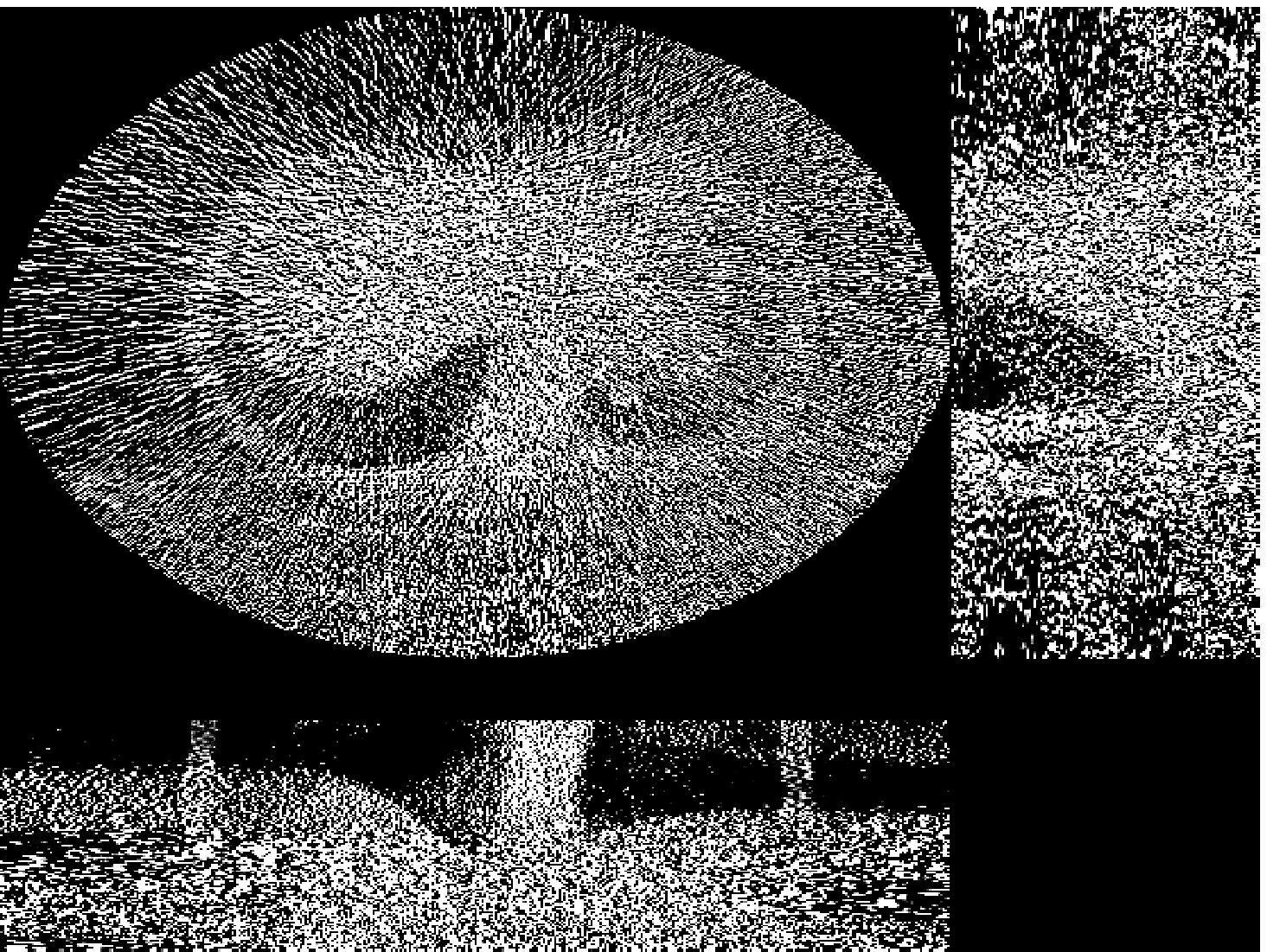}&
\includegraphics[width=.22\linewidth, height=.22\linewidth]{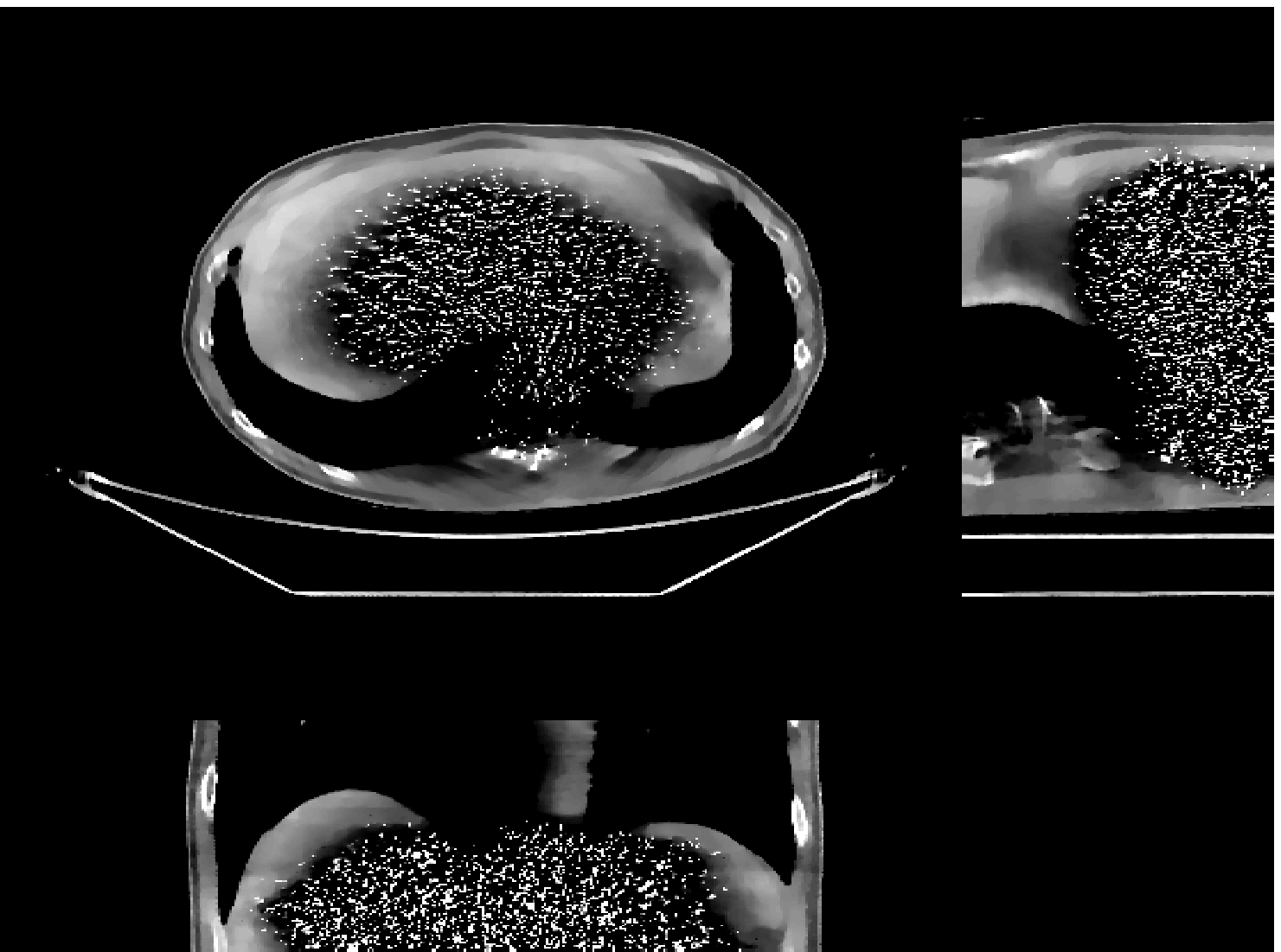}&
\includegraphics[width=.22\linewidth, height=.22\linewidth]{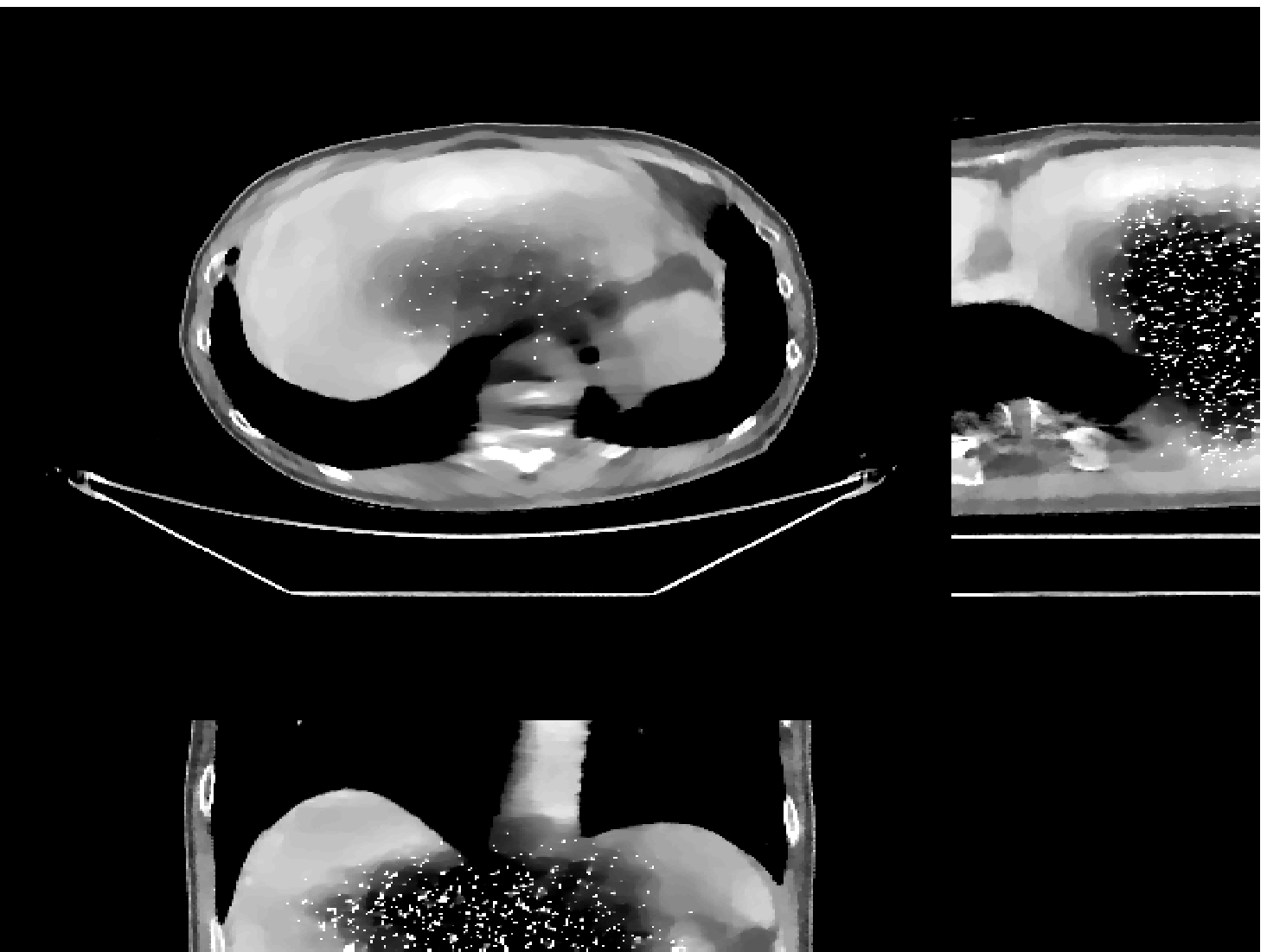}&
\includegraphics[width=.22\linewidth, height=.22\linewidth]{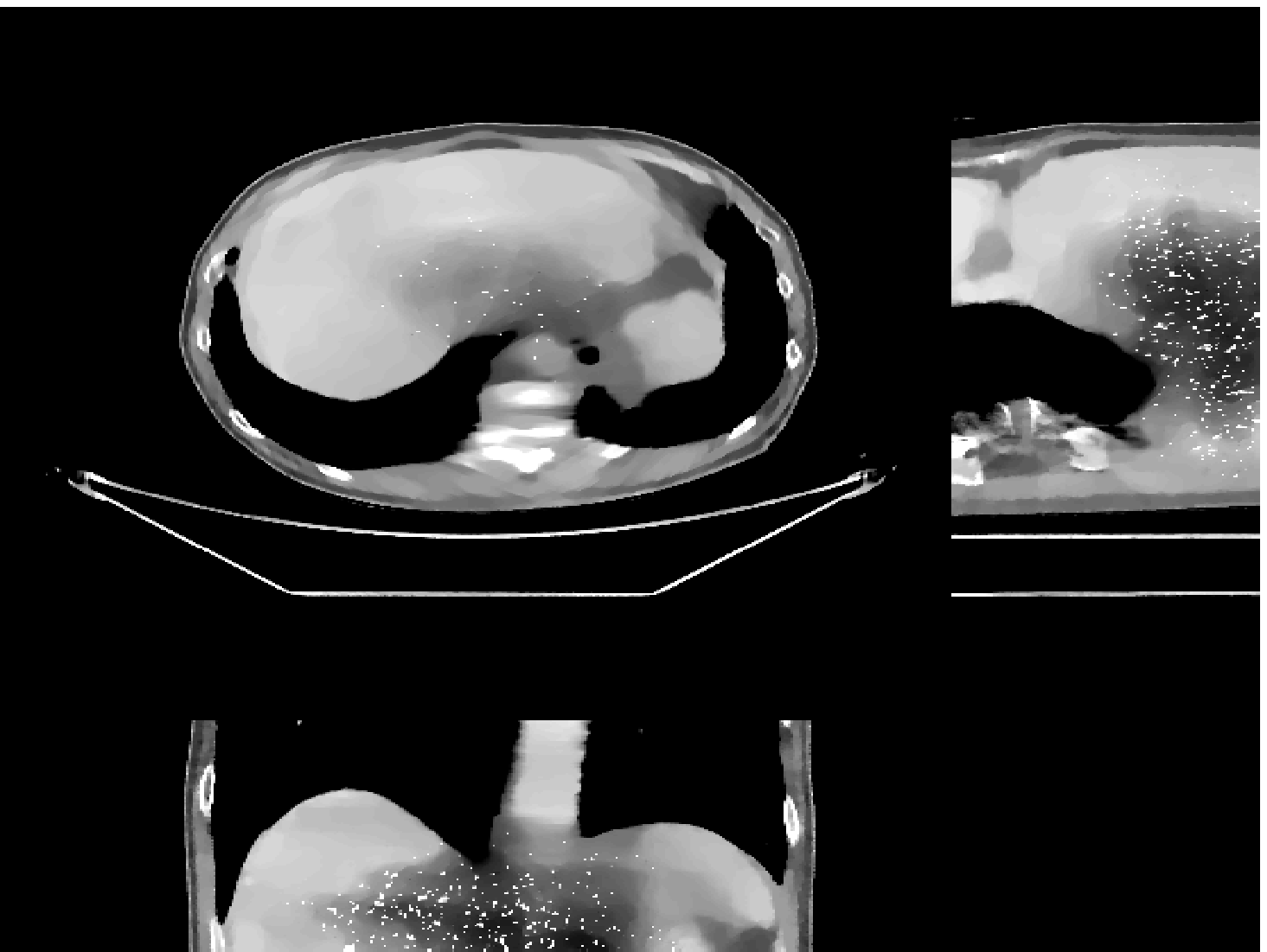}\\
\put(-40,50){$\sigma^2=60^2$}&
\includegraphics[width=.22\linewidth, height=.22\linewidth]{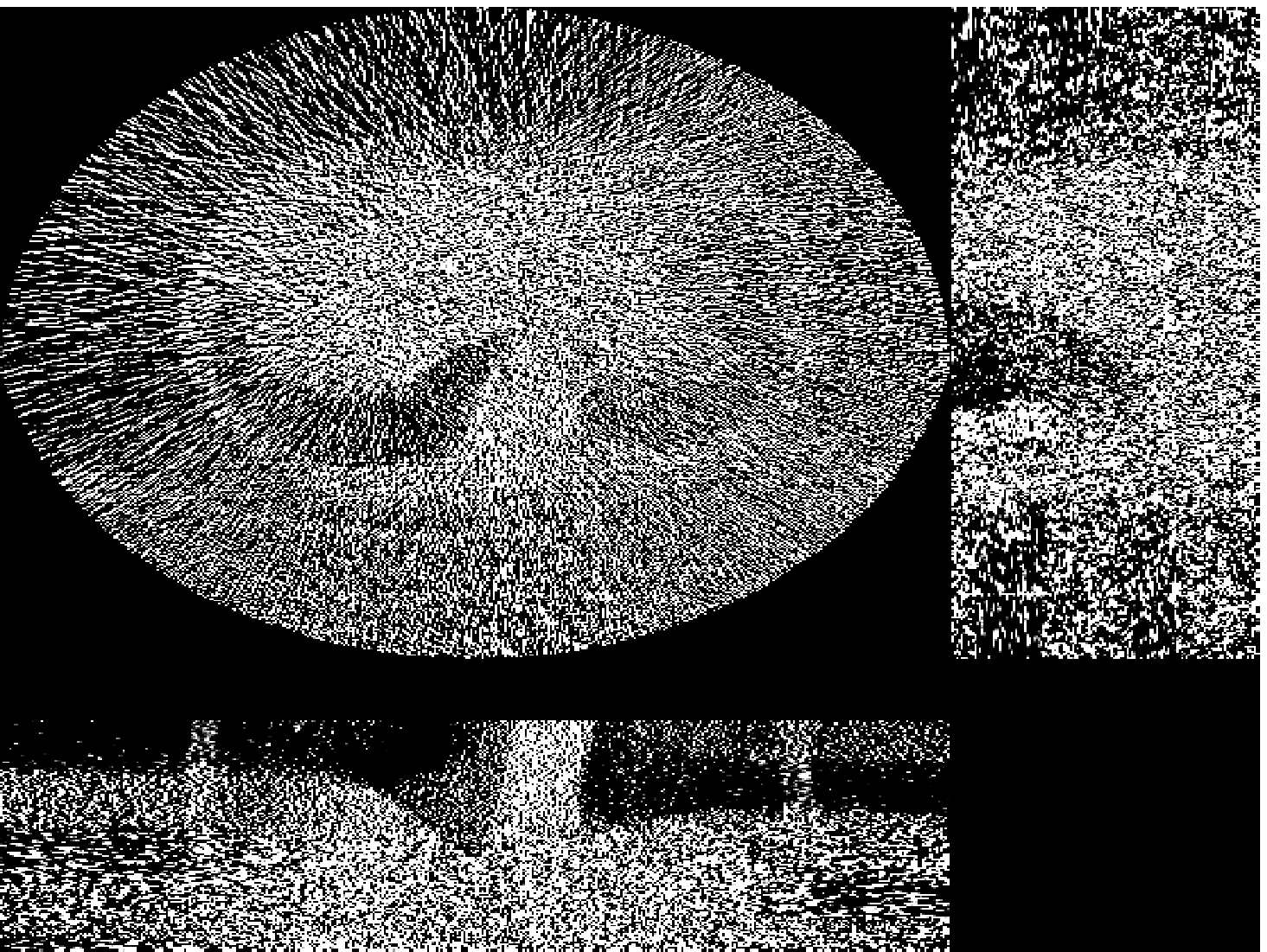}&
\includegraphics[width=.22\linewidth, height=.22\linewidth]{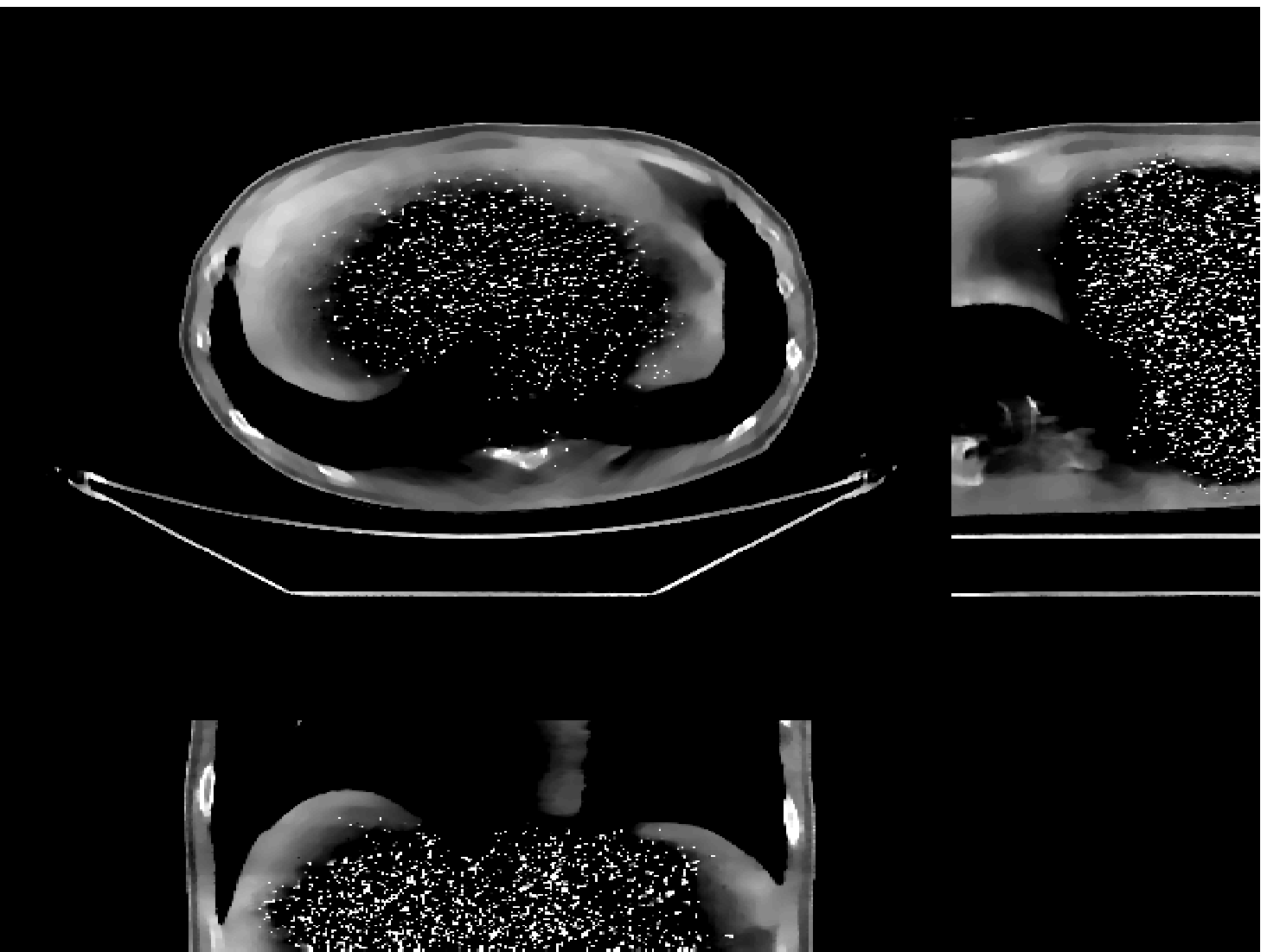}&
\includegraphics[width=.22\linewidth, height=.22\linewidth]{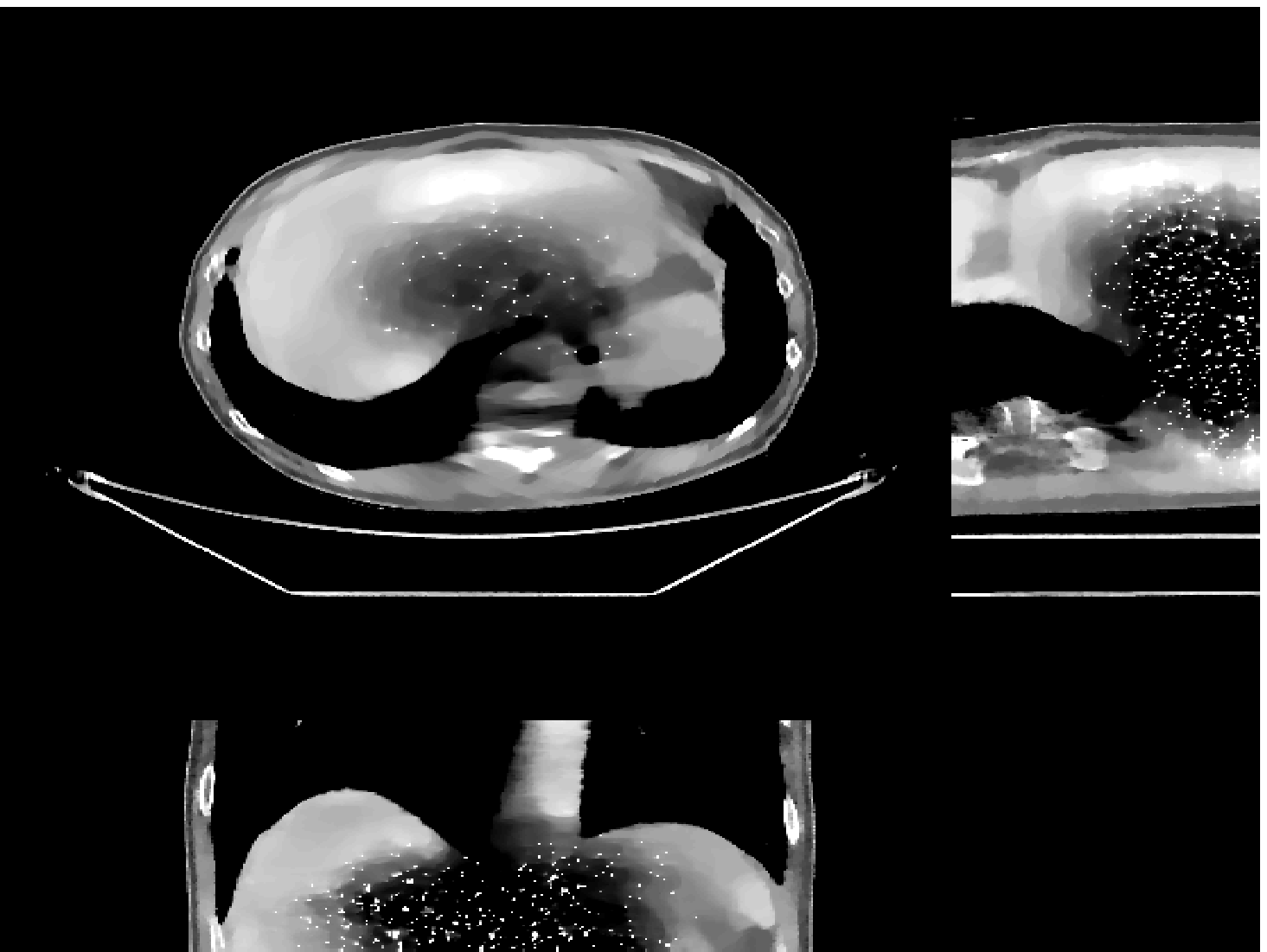}&
\includegraphics[width=.22\linewidth, height=.22\linewidth]{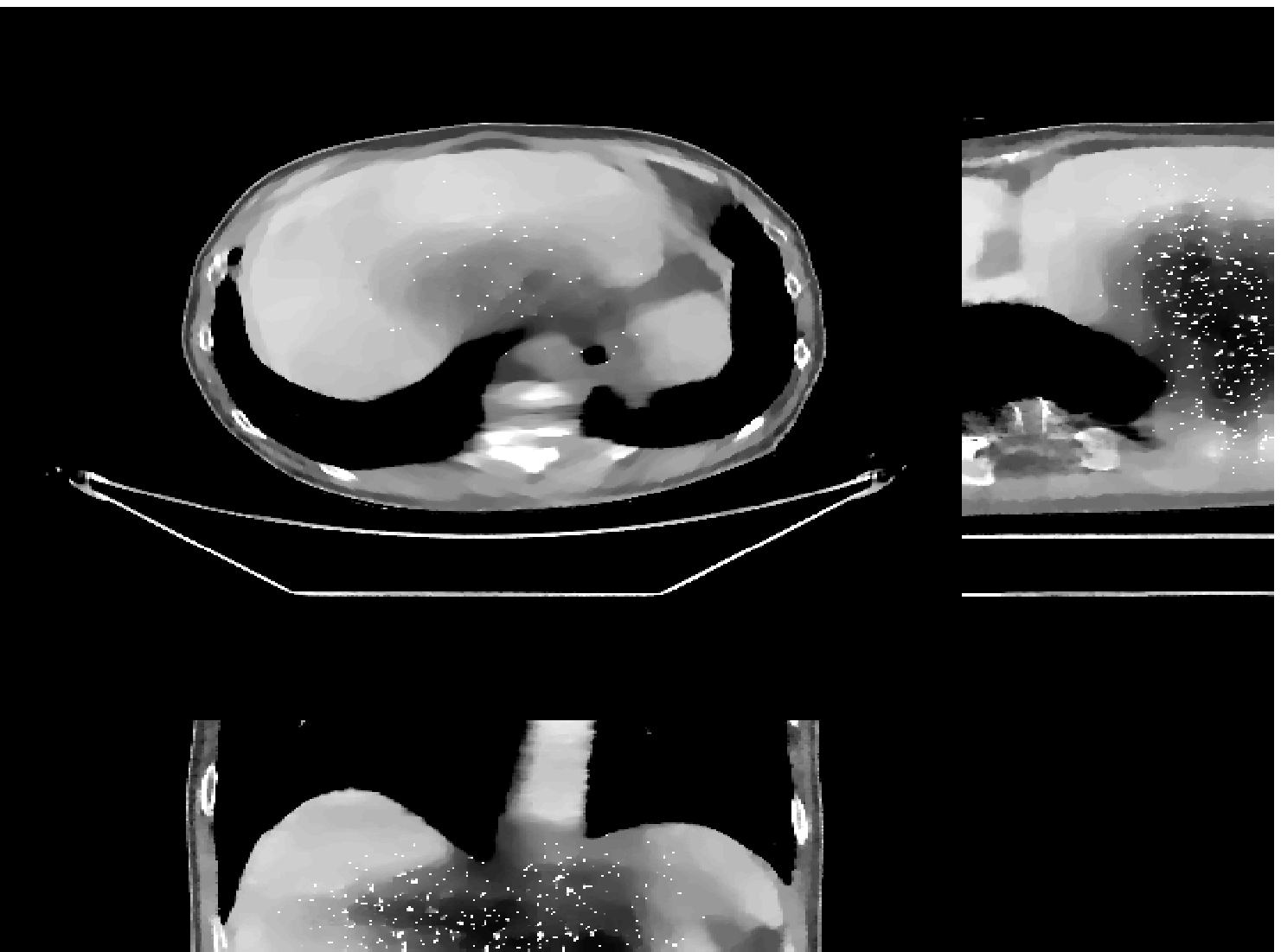}\\
&
{FBP}&
{PWLS}&
{SP}&
{MPG}
\end{tabular}
\caption{
Clinical CT volumes reconstructed by FBP (first column), PWLS (second column), SP (third column) and the proposed MPG method (forth column) for dose of $I_i = 10^4$  with variance of electronic noise $\sigma^2 = 20^2$ (first row),  $\sigma^2 = 30^2$ (second row), $\sigma^2 = 40^2$ (third  row), $\sigma^2 = 50^2$ (forth  row) and $\sigma^2 = 60^2$ (fifth row).
All images are displayed using a window of $[800,1200]$~HU.
}
\label{fig:MPGClinical}
\end{center}
\end{figure*}

We reconstructed a $420\times420\times222$ image volume with $\Delta_x=\Delta_y=1.1667$~mm and $\Delta_z=0.625$~mm using PWLS with edge-preserving regularizer
from a chest region helical CT scan. The size of the sinogram was $888\times64\times3611$ and pitch was $1.0$ (about $3.7$ rotations with rotation time $0.4$ seconds).
The tube current and tube voltage of the X-ray source were $750$~mA and $120$~kVp, respectively.
Figure~\ref{TrueB} shows the reconstructed clinical volume in axial, coronal and sagittal view.
Using this reconstructed clinical volume, we generated a synthetic $888\times64\times3611$ helical CT scan with mono-energetic source of $I_i= 10^4$ incident photons per ray.
We added electronic noise at different levels, \emph{i.e.}, $\sigma=\{20, 30,40,50,60\}$, to the generated synthetic pre-log data.
Table~\ref{table:MPGClinical} shows percents of non-positive measurements for difference electronic noise levels.
Figure~\ref{fig:MPGClinical} shows images reconstructed by the FBP, PWLS, SP and MPG method.
The FBP images are full of artifacts and noise, especially when electronic noise becomes large.
The PWLS method initialized with FBP images improves image quality compared to FBP images, but produces bias, especially for large electronic noise cases.
The SP method initialized with PWLS images reconstruct images with less noise and smaller bias, but bias still exists as electronic noise becomes larger.
The MPG method initialized with PWLS images produces images with smaller bias and less noise compared with SP reconstructions.
The MPG method reconstruct images with the best quality compared with FBP, PWLS and SP.
\begin{table}
\centering
\begin{tabular}{|c|c|c|c|c|c|c|c|c|c|c|}
\cline{1-7}
 \multicolumn{2}{|c|}{$I_i=10^4,~\sigma^2$}            &$20^2$    &$30^2$    &$40^2$    &$50^2$   &$60^2$     \\\cline{1-7}
\multicolumn{2}{|c|}{Non-positive Percentage ($\%$)}    &$2.6$    &$4.0$     &$5.2$     &$6.1$    &$6.9$    \\\hline
\end{tabular}
\caption{Percentage of non-positive
values of the measurements with different electronic noise level when $I_i=10^4$ for the synthetic clinical data.}
\label{table:MPGClinical}
\end{table}

\section{Discussion and Conclusion}
\label{conclusion}
We proposed a novel SIR method, called MPG (mixed Poisson-Gaussian) for ULDCT imaging. MPG method models the noisy measurements using mixed Poisson-Gaussian distribution which accounts for both quantum noise  and electronic noise that dominates when the X-ray dose is at an ultra-low level.
We used the reweighted least square method to develop a tractable likelihood function that can be incorporated into SIR reconstruction framework.
The proposed MPG method can accommodate  edge-preserving regularizers  that preserve edges and can be useful for under-sampled data by reducing the number of views for further dose reduction.
We minimize the MPG cost function using ADMM which divides the original optimization problem into several sub-problems that are easier to solve.
The proposed MPG method is able to directly use negative and zero values  in the raw data without any pre-processing.
Experimental results on  simulated 3D cone-beam data and synthetic helical scans that generated from clinical data  indicate that the proposed MPG method outperforms the PWLS and SP method.
We were not able to test the proposed MPG method on pre-log clinical data because this kind of un-processed ULDCT data is proprietary to CT vendors. The exact value of electronic noise variance depends on CT scanners, and is propriety to CT vendors too. We tested the proposed MPG method for different electronic noise variances to demonstrate robustness of the MPG method. In future work we will investigate optimization methods to accelerate MPG reconstruction.	
\bibliographystyle{Plain}

\begin{thebibliography}{10}
	\bibitem{brenner2007computed}
	D.~J. Brenner and E.~J. Hall, ``Computed tomography-an increasing source of
	radiation exposure,'' {\em New England Journal of Medicine}, vol.~357,
	no.~22, pp.~2277--2284, 2007.
	
	\bibitem{thibault2007three}
	J.-B. Thibault, K.~D. Sauer, C.~A. Bouman, and J.~Hsieh, ``A three-dimensional
	statistical approach to improved image quality for multislice helical {CT},''
	{\em Medical physics}, vol.~34, no.~11, pp.~4526--4544, 2007.
	
	\bibitem{long13aou}
	Y.~Long, L.~Cheng, X.~Rui, B.~De~Man, A.~Alessio, E.~Asma, and P.~E. Kinahan,
	``Analysis of ultra-low dose {CT} acquisition protocol and reconstruction
	algorithm combinations for {PET} attenuation correction,'' in {\em Proc.
		Intl. Mtg. on Fully 3D Image Recon. in Rad. and Nuc. Med}, pp.~400--403,
	2013.
	
	\bibitem{whiting06pop}
	B.~R. Whiting, P.~Massoumzadeh, O.~A. Earl, J.~A. O’Sullivan, D.~L. Snyder,
	and J.~F. Williamson, ``Properties of preprocessed sinogram data in {X}-ray
	computed tomography,'' {\em Medical physics}, vol.~33, no.~9, pp.~3290--3303,
	2006.
	
	\bibitem{nuyts13mtp}
	J.~Nuyts, B.~De~Man, J.~A. Fessler, W.~Zbijewski, and F.~J. Beekman,
	``Modelling the physics in the iterative reconstruction for transmission
	computed tomography,'' {\em Physics in medicine and biology}, vol.~58,
	no.~12, p.~R63, 2013.
	
	\bibitem{thibault2006recursive}
	J.-B. Thibault, C.~A. Bouman, K.~D. Sauer, and J.~Hsieh, ``A recursive filter
	for noise reduction in statistical iterative tomographic imaging,'' in {\em
		Proceedings of SPIE}, vol.~6065, pp.~264--273, International Society for
	Optics and Photonics, 2006.
	
	\bibitem{wang2006penalized}
	J.~Wang, T.~Li, H.~Lu, and Z.~Liang, ``Penalized weighted least-squares
	approach to sinogram noise reduction and image reconstruction for low-dose
	{X}-ray computed tomography,'' {\em Medical Imaging, IEEE Transactions on},
	vol.~25, no.~10, pp.~1272--1283, 2006.
	
	\bibitem{hsieh1998adaptive}
	J.~Hsieh, ``Adaptive streak artifact reduction in computed tomography resulting
	from excessive {X}-ray photon noise,'' {\em Medical Physics}, vol.~25,
	no.~11, pp.~2139--2147, 1998.
	
	\bibitem{la2006monotonic}
	P.~J. La~Rivi{\`e}re, ``Monotonic iterative reconstruction algorithms for
	targeted reconstruction in emission and transmission computed tomography,''
	in {\em Nuclear Science Symposium Conference Record, 2006. IEEE}, vol.~5,
	pp.~2924--2928, IEEE, 2006.
	
	\bibitem{la2006penalized}
	P.~J. La~Rivi{\`e}re, J.~Bian, and P.~A. Vargas, ``Penalized-likelihood
	sinogram restoration for computed tomography,'' {\em Medical Imaging, IEEE
		Transactions on}, vol.~25, no.~8, pp.~1022--1036, 2006.
	
	\bibitem{elbakri2003efficient}
	I.~A. Elbakri and J.~A. Fessler, ``Efficient and accurate likelihood for
	iterative image reconstruction in {X}-ray computed tomography,'' in {\em
		Medical Imaging 2003}, pp.~1839--1850, International Society for Optics and
	Photonics, 2003.
	
	\bibitem{foi2008practical}
	A.~Foi, M.~Trimeche, V.~Katkovnik, and K.~Egiazarian, ``{Practical
		Poissonian-Gaussian noise modeling and fitting for single-image raw-data},''
	{\em Image Processing, IEEE Transactions on Image Processing}, vol.~17,
	no.~10, pp.~1737--1754, 2008.
	
	\bibitem{li2015reweighted}
	J.~Li, Z.~Shen, R.~Yin, and X.~Zhang, ``A reweighted {L}$_2$ method for image
	restoration with poisson and mixed {P}oisson-{G}aussian noise,'' {\em Inverse
		Probl. Imaging (Springfield)}, vol.~9, no.~3, pp.~875--894, 2015.
	
	\bibitem{GoldsteinOsher2009}
	T.~Goldstein and S.~Osher, ``The split {B}regman method for {L}1-regularized
	problems,'' {\em SIAM Journal on Imaging Sciences}, vol.~2, no.~2,
	pp.~323--343, 2009.
	
	\bibitem{LinComparison}
	L.~Fu, T.~Lee, S.~M. Kim, A.~M. Alessio, P.~E. Kinahan, Z.~Chang, K.~D. Sauer,
	M.~K. Kalra, and B.~De~Man, ``{Comparison Between Pre-Log and Post-Log
		Statistical Models in Ultra-Low-Dose CT Reconstruction},'' {\em IEEE
		Transactions on Medical Imaging}, vol.~36, no.~3, pp.~707--720, 2017.
	
	\bibitem{xu2009electronic}
	J.~Xu and B.~M. Tsui, ``Electronic noise modeling in statistical iterative
	reconstruction,'' {\em IEEE Transactions on Image Processing}, vol.~18,
	no.~6, pp.~1228--1238, 2009.
	
	\bibitem{HsiehCTBook}
	J.~Hsieh, {\em {Computed tomography: principles, design, artifacts, and recent
			advances,2nd Edition}}.
	\newblock SPIE press, 2003.
	
	\bibitem{Gualtieri99orderedsubsets}
	G.~Gualtieri and J.~A. Fessler, ``Ordered subsets algorithms for transmission
	tomography,'' {\em Med. Biol}, vol.~44, pp.~2835--2851, 1999.
	
	\bibitem{joseph1978method}
	P.~M. Joseph and R.~D. Spital, ``A method for correcting bone induced artifacts
	in computed tomography scanners.,'' {\em Journal of computer assisted
		tomography}, vol.~2, no.~1, pp.~100--108, 1978.
	
	\bibitem{elbakri2002statistical}
	I.~A. Elbakri and J.~A. Fessler, ``Statistical image reconstruction for
	polyenergetic {X}-ray computed tomography,'' {\em IEEE Transactions on
		Medical Imaging}, vol.~21, no.~2, pp.~89--99, 2002.
	
	\bibitem{green1984iteratively}
	P.~J. Green, ``Iteratively reweighted least squares for maximum likelihood
	estimation, and some robust and resistant alternatives,'' {\em Journal of the
		Royal Statistical Society. Series B (Methodological)}, pp.~149--192, 1984.
	
	\bibitem{fessler1996spatial}
	J.~A. Fessler and W.~L. Rogers, ``Spatial resolution properties of
	penalized-likelihood image reconstruction: space-invariant tomographs,'' {\em
		IEEE Transactions on Image Processing}, vol.~5, no.~9, pp.~1346--1358, 1996.
	
	\bibitem{bouman1993a}
	C.~A. Bouman and K.~D. Sauer, ``A generalized gaussian image model for
	edge-preserving map estimation,'' {\em IEEE Transactions on Image
		Processing}, vol.~2, no.~3, pp.~296--310, 1993.
	
	\bibitem{burden2011numerical}
	R.~L. Burden and J.~D. Faires, {\em Numerical analysis}.
	\newblock Cengage Learning, 2011.
	
	\bibitem{WrightNO}
	J.~Nocedal and s.~J. Wright, {\em {Numerical optimization 2nd Edition}}.
	\newblock Springer-Verlag New York, 2006.
	
	\bibitem{goldstein2014fast}
	T.~Goldstein, B.~O'Donoghue, S.~Setzer, and R.~Baraniuk, ``Fast alternating
	direction optimization methods,'' {\em SIAM Journal on Imaging Sciences},
	vol.~7, no.~3, pp.~1588--1623, 2014.
	
	\bibitem{boyd2004convex}
	S.~Boyd and L.~Vandenberghe, {\em Convex optimization}.
	\newblock Cambridge university press, 2004.
	
	\bibitem{sauer1993a}
	K.~D. Sauer and C.~A. Bouman, ``A local update strategy for iterative
	reconstruction from projections,'' {\em IEEE Transactions on Signal
		Processing}, vol.~41, no.~2, pp.~534--548, 1993.
	
	\bibitem{fessler2000statistical}
	J.~A. Fessler, ``Statistical image reconstruction methods for transmission
	tomography,'' {\em Handbook of medical imaging}, vol.~2, pp.~1--70, 2000.
	
	\bibitem{ramani2012a}
	S.~Ramani and J.~A. Fessler, ``{A Splitting-Based Iterative Algorithm for
		Accelerated Statistical X-Ray CT Reconstruction},'' {\em IEEE Transactions on
		Medical Imaging}, vol.~31, no.~3, pp.~677--688, 2012.
	
	\bibitem{segars2008realistic}
	W.~P. Segars, M.~Mahesh, T.~J. Beck, E.~C. Frey, and B.~M.~W. Tsui,
	``{Realistic CT simulation using the 4D XCAT phantom.},'' {\em Medical
		Physics}, vol.~35, no.~8, pp.~3800--3808, 2008.
	
	\bibitem{rui2015ultra-low}
	X.~Rui, L.~Cheng, Y.~Long, L.~Fu, A.~M. Alessio, E.~Asma, P.~E. Kinahan, and
	B.~De~Man, ``Ultra-low dose {CT} attenuation correction for {PET/CT}:
	analysis of sparse view data acquisition and reconstruction algorithms,''
	{\em Physics in Medicine and Biology}, vol.~60, no.~19, p.~7437, 2015.
\end{thebibliography}



{
	\twocolumn[
	\begin{center}
		\Huge Statistical Image Reconstruction Using Mixed Poisson-Gaussian Noise
		Model for X-Ray CT: Supplementary Material
	\vspace{0.4in}
	\end{center}]
}

In this supplementary material, we  provide the details  of SP reconstruction problem by ADMM and  shifted Poisson Algorithm.
\section{SP Algorithm}
\label{APP:SP}
Introducing auxiliary variables $\bm{u}\in \mathbb{R}^{N_d},\bm{v}\in \mathbb{R}^{N_r},\bm{w}\in \mathbb{R}^{N_p}$, we rewrite the SP
reconstruction problem \eqref{SPCost} as the following equivalent constrained problem:
\begin{align}
\arg\min_{\bm{x},\bm{u},\bm{v}}
&\langle Ie^{-\bm{u}}+\bm{\sigma}^2,\bm{1}\rangle-\langle \bm{z} +\bm{\sigma}^2, \log(Ie^{-\bm{u}}+\bm{\sigma}^2)\rangle\nonumber\\
&+\lambda\|\bm{v}\|_1+\chi_c(\bm{w})\nonumber\\
\mathrm{s.t.}\quad &\bm{u}=\bm{A}\bm{x},
                   \bm{v}=\bm{C}\bm{x},
                   \bm{w}=\bm{x}.
\label{constSP}
\end{align}
To simplify, we reformulate \eqref{constSP}  as the following constrained optimization problem, where the constraints are written as a linear transform,
 \begin{eqnarray}
&\arg\min_{\bm{x},\bm{u},\bm{v}, \bm{w}}\mathcal{D}_{SP}(\bm{u})+\lambda\|\bm{v}\|_1+\chi_c(\bm{w}) \nonumber\\
&\mathrm{s.t.}\quad \bm{P}\bm{x}=(\bm{A}\bm{x}, \bm{C}\bm{x}, \bm{x})^T=(\bm{u},\bm{v}, \bm{w})^T
\label{modelSP}
\end{eqnarray}
where
\begin{align}
\mathcal{D}_{SP}(\bm{u})=\langle Ie^{-\bm{u}}+\bm{\sigma}^2,\bm{1}\rangle-\langle \bm{z} +\bm{\sigma}^2, \log(Ie^{-\bm{u}}+\bm{\sigma}^2)\rangle.
\end{align}
 The augmented Lagrange function of the optimization problem \eqref{modelSP} is defined as:
\begin{align}
\mathcal{L}_{SP}(\bm{x},\bm{u},&\bm{v}, \bm{w},\bm{b})=\mathcal{D}_{SP}(\bm{u})+\lambda\|\bm{v}\|_1+\chi_c(\bm{w})\nonumber\\
+&\langle \bm{b},\bm{P}\bm{x} -(\bm{u},\bm{v}, \bm{w})^T\rangle
+\frac{1}{2}\|\bm{P}\bm{x} -(\bm{u},\bm{v}, \bm{w})^T\|_{\bm{\mu}}^2
\end{align}
where $\bm{b}=(\bm{b}_1,\bm{b}_2,\bm{b}_3)^T$, $\bm{b}_1\in
\mathbb{R}^{N_d},\bm{b}_2\in \mathbb{R}^{N_r}, \bm{b}_3\in \mathbb{R}^{N_p}$
have the same size as $ \bm{A}\bm{x}, \bm{C}\bm{x}, \bm{x}$ respectively, $\bm{\mu}>0$ is the penalty parameter.  ADMM updates the sequence $( \bm{x}^{(j)},\bm{u}^{(j)},\bm{v}^{(j)},\bm{w}^{(j)},\bm{b}^{(j)})$ as,
\begin{subequations}
 \begin{numcases}
{}\bm{x}^{(j+1)}=\langle \bm{b}^{(j)},\bm{P}\bm{x} -(\bm{u}^{(j)},\bm{v}^{(j)}, \bm{w}^{(j)})^T\rangle\nonumber\\
~~~~~~~~~~~~~+\frac{1}{2}\|\bm{P}\bm{x} -(\bm{u}^{(j)},\bm{v}^{(j)}, \bm{w}^{(j)})^T\|_{\bm{\mu}}^2,\label{equationSPa}\\
(\bm{u}^{(j+1)},\bm{v}^{(j+1)},\bm{w}^{(j+1)})\nonumber\\
~~~~~~~=\arg\min_{\bm{u},\bm{v},\bm{w}}\mathcal{L}_{SP}(\bm{x}^{(j+1)},\bm{u},\bm{v},\bm{w},\bm{b}^{(j)}),\label{equationSPb}\\
 \bm{b}^{(j+1)} =\bm{b}^{(j)} + \bm{\mu}(\bm{P}\bm{x}^{(j+1)}-(\bm{u}^{(j+1)},\bm{v}^{(j+1)},\bm{w}^{(j+1)})^T).\nonumber\\\label{equationSPc}
 \end{numcases}
 \label{SPIter}
 \end{subequations}

We solve \eqref{equationSPa}  by the same method as \eqref{solution1}.
We solve \eqref{equationSPb}   separately for $\bm{u},\bm{v},\bm{w}$ and in parallel.
Subproblem of $\bm{u}^{(j+1)}$ is
\begin{align}
\bm{u}^{(j+1)}=&\arg\min_{\bm{u}}\mathcal{D}_{SP}(\bm{u})+\langle \bm{b}_1^{(j)},A\bm{x}^{(j+1)}-\bm{u}\rangle\nonumber\\
&+\frac{\mu_1}{2}\|A\bm{x}^{(j+1)}-\bm{u}\|_2^2 .\label{subSP1}
\end{align}
It is a  smooth, differentiable  and separable problem for each $u_i$.
Subproblem of $\bm{v}^{(j+1)},\bm{w}^{(j+1)}$ are the same as \eqref{Subsolb} and \eqref{Subsolc} respectively.
The dual variable $\bm{b}$ are updated straightforwardly as given in \eqref{equationSPc}.
The primal and dual residual for ADMM updates in  \eqref{SPIter} as the stopping criteria are computed in \eqref{PrimalRes} and \eqref{DualRes}.
Algorithm 
\ref{algSP} summarizes the optimization algorithm of the SP method.
 \begin{algorithm}[H]
		\caption{Shifted Poisson Algorithm}
		\label{algSP}
		\begin{algorithmic}
			\STATE \footnotesize{ \textbf{Input.} $\bm{x}^{(0)}$, $\lambda$, $\mu_1$, $\mu_2$, $\mu_3$. }
			\STATE Initial $\bm{u}^{(0)}=\bm{A}\bm{x}^{(0)}$, $\bm{v}^{(0)}=\bm{C}\bm{x}^{(0)}$, $\bm{w}^{(0)}=\bm{x}^{(0)}$, $\bm{b}^{(0)}=( \bm{b}_1^{(0)},\bm{b}_2^{(0)}, \bm{b}_3^{(0)})=0$ , $\rm{Maxiter}$, $\rm{tol}$,  $j=1$.
			\WHILE{$\|r^{(j)}\|>\rm{tol}$, $\|d^{(j)}\|>\rm{tol}$, $j<\rm{Maxiter}$}
            \STATE Solve for  $\bm{x}^{(j+1)}$ by  applying CG iterations to \eqref{solution1}.
			\STATE Compute $\bm{u}^{(j+1)}$ by solving \eqref{subSP1}.
            \STATE Solve for $\bm{w}^{(j+1)}$ using \eqref{Subsolb}.
             \STATE Solve for $\bm{w}^{(j+1)}$ using \eqref{Subsolc}.
            \STATE $\bm{b}_1^{(j+1)}=\bm{b}_1^{(j)}+\mu_1(\bm{A}\bm{x}^{(j+1)}-\bm{u}^{(j+1)})$.
			\STATE $\bm{b}_2^{(j+1)}=\bm{b}_2^{(j)}+\mu_2(\bm{C}\bm{x}^{(j+1)}-\bm{v}^{(j+1)})$.
			\STATE $\bm{b}_3^{(j+1)}=\bm{b}_3^{(j)}+\mu_3(\bm{x}^{(j+1)}-\bm{w}^{(j+1)})$.
			\STATE $j=j+1$.
			\ENDWHILE
		\end{algorithmic}
\end{algorithm}

\end{document}